\documentclass[11pt,twoside]{article}
\usepackage{a4wide}
\usepackage[latin1]{inputenc}
\usepackage{amsmath,amssymb,bbm,mathrsfs}
\usepackage{axodraw,subfigure}
\usepackage{cite}
\usepackage[dvips]{epsfig}
\usepackage{color}
\usepackage{pstricks,psfrag,scalefnt}
\usepackage{fancyhdr}
\advance \headheight by 3.0truept       

\newlength{\nseparation}
\newenvironment{nfigure}[1][]
        {\begin{figure}[#1]\hrule\vspace{\nseparation}\par}
        {\vspace{\nseparation}\par \hrule \end{figure}}

\newlength{\textlength}
\newlength{\overlinelength}
\newcommand{\ovl}[2][.55]{\settowidth{\textlength}{$#2$}
  \setlength{\overlinelength}{0.1pt}
  \addtolength{\overlinelength}{0.75\textlength}
  \makebox[\textlength][s]{$#2$} \hspace{-.55\textlength}
  \hspace{-\overlinelength}\hspace{#1\overlinelength}
  \overline{\makebox[\overlinelength][s]{\vphantom{$#2$}}}
  \hspace{-#1\overlinelength}\hspace{.55\textlength}}

\DeclareMathOperator{\diag}{diag}

\DeclareMathOperator{\sgn}{sgn}

\newcommand{\abs}[1]{\ensuremath{\left| #1 \right|}}

\newcommand{\ket}[1]{\ensuremath{\left| #1 \right\rangle}}
\newcommand{\VEV}[1]{\ensuremath{\left\langle #1\right\rangle}}
\newcommand{\VEVsmall}[1]{\ensuremath{\langle #1\rangle}}

\newcommand{\kk}{\ensuremath{K\!-\!\ovl{\!K}\,}}
\newcommand{\kkm}{\kk\ mixing}
\newcommand{\bbs}{\ensuremath{B_s\!-\!\ovl{\!B}_s\,}}
\newcommand{\bbms}{\bbs\ mixing}
\newcommand{\bbd}{\ensuremath{B_d\!-\!\ovl{\!B}_d\,}}
\newcommand{\bbmd}{\bbd\ mixing}
\newcommand{\eq}[1]{Eq.~(\ref{#1})}
\newcommand{\eqsand}[2]{Eqs.~(\ref{#1}) and (\ref{#2})}
\newcommand{\eqsto}[2]{Eqs.~(\ref{#1}--\ref{#2})}
\newcommand{\lt}{\left}
\newcommand{\rt}{\right}
\newcommand{\gev}{\mbox{GeV}}

\def\journal#1#2#3#4{#1~{\bf #2}, #3 (#4)}

\def\NPB#1#2#3{\journal{Nucl.\ Phys. B}{#1}{#2}{#3}}

\begin{document}
\thispagestyle{empty}
\hspace{-2ex}TTP10-52 \hfill SFB/CPP-10-132\\

~\\[-10pt]

\begin{center}
  {\LARGE\bfseries Flavor Physics in an SO(10) Grand Unified
      Model}
  \\[20pt]
  \textsc{Jennifer Girrbach${}^1$, Sebastian J\"ager${}^2$, Markus Knopf${}^1$, Waldemar
    Martens${}^1$, 
    \\[5pt]
    Ulrich Nierste${}^1$, Christian Scherrer${}^1$ and S\"oren Wiesenfeldt${}^{1,3}$}

  \bigskip \bigskip
  
  \centerline{\sl ${}^1$
    \parbox[t]{0.4\textwidth}{\small
      Institut f{\"u}r Theoretische Teilchenphysik
      \\
      Karlsruhe Institute of Technology
      \\
      Universit{\"a}t Karlsruhe
      \\
      Engesserstra\ss e 7
      \\
      D-76128 Karlsruhe, Germany}~~~~ ${}^2$
    \parbox[t]{0.4\textwidth}{\small
      University of Sussex
      \\
      Department of Physics and Astronomy \\
      Falmer    
      \\
      Brighton BN1 9QH
      \\
      United Kingdom} }
~\\[1mm]
\centerline{\sl ${}^3$
  \parbox[t]{0.8\textwidth}{\small Helmholtz Association,
    Anna-Louisa-Karsch-Str.\ 2, 10178 Berlin, Germany} }

  \vspace*{1cm}
  \centerline{\bf Abstract}
  \vspace*{0.3cm}
  \noindent
  \parbox{0.85\textwidth}{
   In supersymmetric grand-unified models, the lepton mixing
    matrix can possibly affect flavor-changing transitions in the quark
    sector. We present a detailed analysis of a model proposed by Chang,
    Masiero and Murayama, in which the near-maximal atmospheric neutrino
    mixing angle governs large new $b\to s$ transitions.   Relating the
    supersymmetric low-energy parameters to seven new parameters
    of this SO(10) GUT model, we perform a correlated study of several
    flavor-changing neutral current (FCNC) processes.  We find the
    current bound on ${\cal B}(\tau \to \mu \gamma)$ more constraining
    than ${\cal B}(B \to X_s \gamma)$.  {The LEP limit on the
      lightest Higgs boson mass implies an important lower bound on
      $\tan\beta$, which in turn limits the size of the new FCNC
      transitions. Remarkably, the combined analysis} does not rule
    out large effects in \bbms\ and we can easily accomodate the
    {large CP phase in the \bbs\ system which has recently been
      inferred from a global analysis of CDF and D\O\ data.  The model
      predicts a particle spectrum which is different from the popular
      Constrained Minimal Supersymmetric Standard Model (CMSSM).
      ${\cal B}(\tau \to \mu \gamma)$ enforces heavy masses, typically
      above 1 TeV, for the sfermions of the degenerate first two
      generations.  However, the ratio of the third-generation and
      first-generation sfermion masses is smaller than in the CMSSM
      and a (dominantly right-handed) stop with mass below 500 GeV is
      possible.  }} 
\end{center}


\section{Introduction}
\label{se:intro}

Although the standard model (SM) is extremely successful, it is likely
that it is only an effective theory, subsumed by a more fundamental
theory at short distances.  Weak-scale supersymmetry (SUSY) supplies a
means to stabilize a hierarchy between the electroweak and more
fundamental scales.  Remarkably, with the renormalization group (RG)
equations of the minimal supersymmetric extension of the standard model
(MSSM) above the weak scale, the three gauge couplings meet at
$M_\text{GUT} = 2\times 10^{16}$~GeV \cite{Amaldi:1991cn}.  This
supports the idea that the strong, weak, and electromagnetic
interactions are unified into a grand unified theory (GUT) with a single
gauge coupling \cite{Pati:1973uk,Georgi:1974sy}.  It is striking that
the experimental evidence for small but non-vanishing neutrino masses
fits nicely in this framework, as $M_\text{GUT}$ is of the right order
of magnitude to generate small Majorana masses for the neutrinos.

SO(10) \cite{so10} is arguably the most natural GUT group: both the SM
gauge and matter fields are unified, introducing only one additional
matter particle, the right-handed neutrino.  It is an anomaly-free
theory and therefore explains the intricate cancellation of the
anomalies in the standard model \cite{Georgi:1972bb}.  Moreover, it
contains $B-L$ as a local symmetry, where $B$ and $L$ are baryon and
lepton number, respectively; the breaking of $B-L$ naturally provides
light neutrino masses via the seesaw mechanism \cite{Minkowski:1977sc}.

Despite its theoretical attractiveness, the experimental hints for
supersymmetric GUTs have been sparse, putting stringent constraints on
models.  In particular, the impressive agreement of the flavor precision
measurements with the standard model leads {to the widespread belief}
that the Yukawa couplings are the only source of flavor violation; this
concept is known as minimal flavor violation \cite{minimal-flavor}.

In the (supersymmetric) standard model, fermion mixing is only
measurable among the left-handed states and described by the quark and
lepton mixing matrices, $V_\text{CKM}$ and $U_\text{PMNS}$.  The mixing
angles of $V_\text{CKM}$ are small, corresponding to the strong mass
hierarchy, while two angles in $U_\text{PMNS}$ turn out to be large.
These are the neutrino solar and atmospheric mixing angles, where the
latter is close to maximal, $\theta_{23} \simeq \left(42.3 {+5.1 \atop
    -3.3}\right)^\circ$ at $1\sigma$ \cite{nu-analysis}.  The definition
of minimal flavor violation in Ref.~\cite{minimal-flavor} involves
independent flavor symmetry groups for quarks and leptons.  It confines
the effects of $V_\text{CKM}$ to the quark sector and that of
$U_\text{PMNS}$ to the lepton sector.
In GUTs, however, this separation of quark and lepton sector is
abrogated as quarks and leptons are unified
and thus their masses and mixing are related to each other.  While
different patterns are possible, it is natural to expect imprints of
$U_\text{PMNS}$ on the quark sector as well.  For instance, the Yukawa
couplings (and thus the masses) of down quarks and charged leptons
unify in SU(5) with \cite{Georgi:1974sy}
\begin{align}
  \label{eq:yukawa-unification-su5}
  \mathsf{Y}_d = \mathsf{Y}_e^\top \,.
\end{align}
This relation indicates that one might {encounter} small rotations
between left-handed down quarks and right-handed leptons {in
  connection with} large mixing among right-handed down quarks and
left-handed leptons.  The mixing of the right-handed fermions is
unobservable due to the absence of right-handed currents at the weak
scale.  With weak-scale supersymmetry, however, the mixing of the
corresponding scalar partners of quarks and leptons becomes physical.
Hence, one might ask whether the large mixing angles are observable in
the quark sector
\cite{Baek:2000sj,Moroi:2000tk,Akama:2001em,Chang:2002mq,Hisano:2003bd}\footnote{For
  an earlier study with $V_\text{CKM}$ being the universal mixing
  matrix, see Ref.~\cite{Barbieri:1995rs}.}.

The concept of minimal flavor violation suggests the assumption that the
supersymmetry breaking parameters are universal at some scale.  This
ansatz is realized in the minimal supergravity (mSUGRA) scenario
\cite{msugra} (or a popular variant of it, the CMSSM \cite{cmssm}),
where the scale, at which the relations hold, is usually taken to be
$M_\text{GUT}$. {FCNC processes in this framework have been calculated
  already 20 years ago \cite{bbmr}.}  A more natural choice for
high-scale supersymmetry breaking, however, is to impose flavor
universality at the Planck scale, $M_\text{Pl} = G_N^{-1/2} = 1\cdot
10^{19}$~GeV.\footnote{Alternatively, one might choose the
  \emph{reduced} Planck scale, $M_\text{Pl} = \left(8\pi
    G_N\right)^{-1/2} = 2\cdot 10^{18}$~GeV, because it compensates for
  the factor $8\pi$ in the Einstein field equations.}  The reason to
take {$M_\text{GUT}$} instead of {$M_\text{Pl}$} is simply
that while the use of the renormalization group equations of the MSSM
below $M_\text{GUT}$ is undisputed, the analysis of the region between
$M_\text{GUT}$ and $M_\text{Pl}$ requires knowledge about the
grand-unified model.  The errors made in neglecting these effects are
proportional to a loop suppression factor times
$\ln\left(M_\text{Pl}/M_\text{GUT}\right)$; however, since the evolution
of the parameters from $M_\text{GUT}$ down to low energies breaks the
universality of the SUSY breaking parameters, new effects in FCNC
processes occur, as we will analyze in this paper.

{Now, in the} LHC era, it is desirable to have a predictive
theory framework which links the results of a decade of precision flavor
physics to quantities probed in high-$p_T$ collider physics, such as the
masses of superpartners.  The mSUGRA and CMSSM models minimize flavor
effects in an ad-hoc way and lead to an MFV version in the sense of
Ref.~\cite{minimal-flavor} of the MSSM.  The purpose of this paper is to
establish a well-motivated alternative scenario to the widely-studied
MFV variants of the MSSM.  We consider an SO(10) model {laid out} by
Chang, Masiero and Murayama (CMM model) \cite{Chang:2002mq}, which
amounts to a version of the MSSM with a well-controlled source of new
flavor violation linking the atmospheric neutrino mixing angle to
transitions between right-handed $b$ and $s$ quarks.
We perform a correlated analysis of several flavor-changing processes
in the quark and lepton sector.  This analysis involves seven parameters
in addition to the parameters of the standard model (SM).  Since the
same parameters enter observables studied in the high-$p_T$ programs of
CMS and ATLAS, the CMM model may serve as a benchmark model connecting
quark and lepton flavor physics to collider physics.  As a first step in
this direction we study the masses of superpartners and of the lightest
neutral Higgs boson.  In view of the rich Higgs sector of GUTs we
emphasize a particular advantage of probing {these} with flavor physics:
While flavor physics observables probe the Yukawa interactions between
the Higgs and matter supermultiplets, they only depend very weakly on
the poorly known parameters of the Higgs potential.
  
Prior to this paper no exhaustive RG analysis of the CMM model has been
published.  A CMM-inspired study has addressed the important topic of
$b\to s$ penguin amplitudes: In Ref.~\cite{Harnik:2002vs} the MFV-MSSM
was complemented by a flavor-changing $\tilde b_R-\tilde s_R$ term in
the right-handed down-squark mass matrix, without implementing GUT
relations among the MSSM parameters.  This study was triggered by an
experimental anomaly in the combined data of mixing-induced CP
asymmetries in $b\to s$ penguin amplitudes, which pointed to a
discrepancy with the SM value {inferred from the mixing-induced CP
  asymmetry measured in the tree-level decay $B_d\to J/\psi K_S$}.
Since the new $b\to s$ transition of the CMM model involves right-handed
quarks, the sign of the deviations of the CP asymmetries from their SM
values should depend on the parity of the final state (\emph{Kagan's
  theorem} \cite{Kagan:2004ia,Endo:2004dc}), unless the new contribution
dominates over the SM amplitude \cite{Larson:2004ha}.  A first study
relating MSSM to GUT parameters was performed in 2003
\cite{Jager:2003xv}, showing that in the CMM model the ---at that time
unknown--- \bbs\ oscillation frequency can exceed its SM value by up to
a factor of {5}.  Then B-factory data seemed to show that the
mixing-induced CP asymmetries in $b\to s$ penguin amplitudes are,
irrespectively of the parity of the final state, consistently lower than
the SM value: The naive average of the CP asymmetries was reported to
lie below the SM expectation by 3.8$\sigma$ in winter 2005
\cite{aspen2005} and the interest in the CMM idea faded.  Today's
situation, however, is again favorable for the CMM model: CDF and D\O\
find the \bbms\ oscillation frequency in agreement with the SM
\cite{Abulencia:2006ze}, which still leaves the possibility of roughly
50\% corrections from new physics because of large hadronic
uncertainties.  The same experiments, however, find hints for a new
CP-violating phase in \bbms\
\cite{Aaltonen:2007he,:2008fj,CDFPhis,D0Phis,Abazov:2010hv,CDFdimuon},
which might imply a complex correction to the \bbms\ amplitude of
roughly half the size of the SM contribution.  While the popular MFV
scenarios of the MSSM cannot provide this correction, even if
{flavor-diagonal} parameters (such as $A_t$) are taken complex
\cite{alt08}, this situation is covered by the range found for the CMM
model in Ref.~\cite{Jager:2003xv}.  On the other hand the significance
of the experimental anomalies in $b\to s$ penguin amplitudes is steadily
shrinking and current data do not challenge the SM much
\cite{hfag08,TheHeavyFlavorAveragingGroup:2010qj}.  The observed pattern
of possible new ${\cal O}(1)$ effects in \bbms\ and small corrections to
$b\to s$ penguin amplitudes below the current experimental sensitivity
is natural in the CMM model, as we discuss below.
  
The paper is organized as follows: In the next section we specify the
theoretical framework of the CMM model focusing on its peculiarities in
the flavor sector.  In section \ref{se:rge} we describe the RGE analysis
for the determination of the soft breaking parameters at the weak scale,
followed by a presentation of observables that have been used to
constrain the model in section \ref{se:observables}. Finally, before
concluding, we present {our results in section~\ref{sec:results} and
compare our study with other analyses in section~\ref{sec:disussion}.}

\section{Framework}\label{se:model}
In this section we describe the CMM model and fill in some details which
were not specified in Ref.~\cite{Chang:2002mq}.
SO(10) is successively broken to $\text{SU(3)}_C \times
\text{U(1)}_\text{em}$ as 
\begin{align}
  \text{SO(10)} \; \;
  \xrightarrow{\VEV{\text{16}_H},{\VEV{\overline{\text{16}}_H}},
               {\VEV{\text{45}_H}}} \; \;
  \text{SU(5)}   \; \;
  \xrightarrow{\VEV{\text{45}_H}}  \;\; & \text{G}_\text{SM} \equiv
  \text{SU(3)}_C \times \text{SU(2)}_L \times \text{U(1)}_Y \nonumber\\
  & \qquad\qquad\qquad
  \xrightarrow{\VEV{\text{10}_H},\, \VEVsmall{\text{10}_H^\prime}} \; \;
  \text{SU(3)}_C \times \text{U(1)}_\text{em} \,.\!
  \label{eq:break}
\end{align}
The first breaking occurs at $M_\text{SO(10)} \sim 10^{17}$~GeV, while
the SU(5)-symmetry is broken at the MSSM unification scale,
$M_\text{GUT}$.  Actually, both the SU(5) singlet {$S$} and adjoint
{$\Sigma_{24}$} of $\text{45}_H$ have non-vanishing vevs: While the
vev of the SU(5) adjoint,
$\VEV{\Sigma_{24}\left(\text{45}_H\right)}\equiv\sigma$, breaks SU(5) to
the standard model group, the singlet component acquires a vev, when
SO(10) is broken, $\VEV{S\left(\text{45}_H\right)}\equiv{v_0}$.  This
latter vev will become important for the Yukawa couplings discussed
below.
The pair of spinors, $\text{16}_H + \ovl{\text{16}}_H$, breaks the
$\text{U(1)}_{B-L}$ subgroup of SO(10), reducing the rank of the group
from five to four.
With this setup, we restrict ourselves to small Higgs {multiplets},
where the threshold corrections at the various breaking scales are small
and which allows for a perturbative SO(10) gauge coupling at the Planck
scale $M_\text{Pl}$.\footnote{A complete model
requires a suitable Higgs superpotential, both to achieve the pattern
of VEVs assumed here and to give GUT-scale masses to all components
in $\text{10}_H,\text{10}_H^\prime, \text{45}_H$
but for the two MSSM doublets  (see below).
The Higgs potential was not specified in
\cite{Chang:2002mq}, and we do not address this problem here. Rather, 
our focus in this paper is on the consequences of the breaking pattern
and flavor structure on low-energy phenomenology. We feel our
findings, in turn, motivate further work on the symmetry breaking
dynamics, possibly along the lines of \cite{Babu:2010ej},
which discusses a somewhat similar Higgs sector.}

The three generations of standard model matter fields are unified
{into three} spinorial representations, together with {three}
right-handed neutrinos,
\begin{align}
  16_i = \left( Q, u^c, d^c, L, e^c, \nu^c \right)_i , \quad i=1,2,3
  \;.
\end{align}
Here $Q$ and $L$ denote the quark and lepton doublet superfields and
$u^c$, $d^c$, $e^c$, and $\nu^c$ the corresponding singlet fields of
the up and down antiquark as well as the positron and the
antineutrino, respectively.  

\smallskip

The Yukawa superpotential reads
\begin{align}
  \label{eq:cmm-yukawa}
  W_Y & =  \frac12 \text{16}_i\, \mathsf{Y}_1^{ij}\, \text{16}_j\,
  \text{10}_H \; +\;   \text{16}_i\, \mathsf{Y}_2^{ij}\,
  \text{16}_j\, \frac{\text{45}_H\, \text{10}_H^\prime}{2\, M_\text{Pl}} 
  \; + \;
  \text{16}_i\, \mathsf{Y}_N^{ij}\, \text{16}_j
  \frac{\overline{\text{16}}_H \ovl{\text{16}}_H}{2\, M_\text{Pl}} \ .
\end{align}
Let us discuss the individual terms in detail.  The MSSM Higgs doublets
$H_u$ and $H_d$ are contained in $\text{10}_H$ and $\text{10}_H^\prime$,
respectively.  Only the up-type Higgs doublet {$H_u$ in
  $\text{10}_H$}, acquires a weak-scale vev such that the first term
gives masses to the up quarks and neutrinos only.  The masses for the
down quarks and charged fermions are then generated through the vev of
the down-type Higgs doublet of a second Higgs field {$H_d$ in
$\text{10}_H^\prime$.}  (A second Higgs field is generally needed in
order to have a non-trivial CKM matrix.)  They are obtained from the
second term in Eq.~(\ref{eq:cmm-yukawa}) which is of mass-dimension
five.  In fact, this operator stands for various, nonequivalent effective
operators with both the SU(5)-singlet and the SU(5)-adjoint vevs of the
adjoint Higgs field such that the coupling matrix $\mathsf{Y}_2$ can
only be understood symbolically.  The operator can be constructed in
various ways, for example by integrating out SO(10) fields at the Planck
scale.  The corresponding couplings can be symmetric or antisymmetric
\cite{Wiesenfeldt:2005zx,Barr:2005ss}, resulting in an asymmetric
{\slshape effective} coupling matrix $\mathsf{Y}_2$, as opposed to the
symmetric matrices $\mathsf{Y}_1$ and $\mathsf{Y}_N$.  This asymmetric
matrix allows for significantly different rotation matrices for the left
and right-handed fields.  For more details see
Appendix~\ref{se:cmm-higgs-yukawa}.  The dimension-five coupling also
triggers a natural hierarchy between the up and down-type quarks,
corresponding to small values of $\tan\beta$, where $\tan\beta$ is the
ratio of the vacuum expectation values (vevs), $\tan\beta =
\VEV{H_u}/\VEV{H_d}$.  Finally, the third term in
Eq.~(\ref{eq:cmm-yukawa}), again a higher-dimensional operator,
generates Majorana masses for the right-handed neutrinos.

The Yukawa matrices are diagonalized as
\begin{align}
  \begin{split}
    \mathsf{Y}_1 & = L_1\, \mathsf{D}_1\, 
                     {L_1^\top} \,,
    \\
    \mathsf{Y}_2 & = L_2 \, \mathsf{D}_2\, R_2^\dagger \,,
    \\
    \mathsf{Y}_N & = R_N\, \mathsf{D}_N\, P_N\, R_N^\top \,,
  \end{split}
  \label{eq:yukawa-diagonalization}
\end{align}
where $L_i$ and $R_i$ are unitary matrices, {$P_N$ is a phase matrix},
and $\mathsf{D}_{1,2,N}$ are diagonal with positive entries.  In order
to work out the physically observable mixing parameters, we choose the
first coupling to be diagonal, i.e., we transform the matter field as
$\text{16} \to L_1^\ast\, \text{16}$ such that
\begin{align}
  \label{eq:yukawa-decomposed}
  W_Y & =  \frac12 \text{16}^\top \mathsf{D}_1 
    \text{16}\, \text{10}_H + \text{16}^\top L_1^\dagger L_2
    \mathsf{D}_2 R_2^\dagger L_1^\ast\, \text{16}\, \frac{\text{45}_H\,
      \text{10}_H^\prime}{2\, M_\text{Pl}} + \text{16}^\top L_1^\dagger R_N
    \mathsf{D}_N P_N R_N^\top L_1^\ast\, \text{16}\,
    \frac{\overline{\text{16}}_H \ovl{\text{16}}_H}{2\, M_\text{Pl}}
  .
\end{align}
Since the up-quarks have diagonal couplings, either of the
$\mathsf{Y}_2$ mixing matrices, $L_1^\dagger L_2$ or $R_2^\dagger
L_1^\ast$, must describe the quark mixing.  We will work in the SU(5)
basis, in which the Yukawa couplings read
\begin{align}
  \label{eq:yukawa-su5}
  W_Y & = \left[ \frac{1}{4} \Psi^\top \mathsf{D}_1 \Psi + N^\top
    \mathsf{D}_1 \Phi \right] H + \sqrt{2}\Psi^\top L_1^\dagger L_2
  \mathsf{D}_2^\prime R_2^\dagger L_1^\ast\, \Phi\, {H^\prime} 
  \nonumber\\ 
  & \qquad\qquad \qquad
  + \frac{M_N}{2}\, N^\top L_1^\dagger R_N \mathsf{D}_N P_N R_N^\top
  L_1^\ast\, N \,,
  \\
  & \mspace{330mu} \mathsf{D}_2^\prime = \mathsf{D}_2
  \frac{{v_0}}{M_\text{Pl}} \ , \qquad M_N =
  \frac{\VEV{\overline{\text{16}}_H}
    \VEV{\ovl{\text{16}}_H}}{M_\text{Pl}} \nonumber
\end{align}
Here, we denote the SU(5) matter fields by
$\Psi_i=\left(Q_i,u_i^c,e_i^c\right)$, $\Phi_i=\left(d_i^c,L_i\right)$
and $N_i=\nu_i^c$ and the SU(5) Higgs fields by
$H=\left(H_u,\ast\right)$ and ${H^\prime}=\left(\ast,H_d\right)$.  
The color-triplets in $H$ and {$H^\prime$} which acquire masses of
  order $M_\text{GUT}$ are denoted by $\ast$.
{The vev $v_0$ is defined after \eq{eq:break}.}
Now we identify the quark mixing matrix as
\begin{align}
  V_q & = L_1^\top L_2^\ast \,.
\end{align}
($V_q$ coincides with the SM quark mixing matrix $V_\text{CKM}$ up to
phases.)
We can always choose a basis where one of the three Yukawa matrices is
diagonal.  In the CMM model, however, one assumes that $\mathsf{Y}_1$
and $\mathsf{Y}_N$ are simultaneously diagonalizable, i.e.
\begin{align}
  \label{eq:sim-diag}
  L_1^\dagger R_N = \mathbbm{1} \,.
\end{align}
This assumption is motivated by the observed values for the fermion
masses and mixings and might be a result of family symmetries.  First,
we note that the up-quarks are {more strongly} hierarchical than the
down quarks, charged leptons, and neutrinos.  As a result, the
eigenvalues of $\mathsf{Y}_N$ must almost have a double hierarchy,
compared to $\mathsf{Y}_1$.  Then, given the Yukawa couplings in an
arbitrary basis, we expect smaller off-diagonal entries in $L_1$ than in
$L_2$ because hierarchical masses generically correspond to small
mixing.  Moreover, the light neutrino mass matrix implies that, barring
cancellations, the rotations in $L_1$ should rather be smaller than
those in $V_\text{CKM}$ \cite{rot-seesaw}.  Hence, even if the relation
(\ref{eq:sim-diag}) does not hold exactly, the off-diagonal entries in
$L_1^\dagger R_N$ will be much smaller than the entries in
$V_\text{CKM}$ and they cannot spoil the large effects generated by the
lepton mixing matrix, $U_\text{PMNS}$.

{Our assumption that $\mathsf{Y}_1$ and $\mathsf{Y}_N$ are
  simultaneously diagonalizable permits an arbitrary phase matrix on the
  right-hand side of \eq{eq:sim-diag}. However, this phase matrix can be
  absorbed into $P_N$ introduced earlier in
  \eq{eq:yukawa-diagonalization} (where this matrix could have been
  absorbed into $R_N$).} Now, with $\mathsf{Y}_1$ and $\mathsf{Y}_N$
being simultaneously diagonal, the flavor structure is (apart from
supersymmetry breaking terms, which we will discuss below) fully
contained in the remaining coupling, $\mathsf{Y}_2$, and
Eq.~(\ref{eq:yukawa-decomposed}) simply reads
\begin{align}
  W_Y & =  \frac12 \text{16}^\top \mathsf{D}_1 
    \text{16}\, \text{10}_H + \text{16}^\top V_q^\ast \mathsf{D}_2
    R_2^\dagger L_1^\ast\, \text{16}\, \frac{\text{45}_H\,
      \text{10}_H^\prime}{2 M_\text{Pl}} + \text{16}^\top \mathsf{D}_N
    P_N\, \text{16}\, \frac{\overline{\text{16}}_H
      \ovl{\text{16}}_H}{2\,M_\text{Pl}} .
\end{align}
It is clear that this coupling has to account for both the quark and
lepton mixing.  Hence, $\mathsf{Y}_2$ cannot be symmetric.

As mentioned above, the higher-dimensional operator can be generated in
various ways, generically resulting in the asymmetric effective coupling
matrix $\mathsf{Y}_2$.  The dominant contributions come from the singlet
vev, $v_0 \sim M_\text{SO(10)}$, which is an order of magnitude higher
than $\sigma \sim M_\text{GUT}$.  In this case, the contributions are
{approximately} the same for down quarks and charged leptons; a more
detailed discussion is given in Appendix~\ref{se:cmm-higgs-yukawa}.
Then we can identify the lepton mixing matrix as
\begin{align}
  U_D & = P_N^\ast R_2^\dagger L_1^\ast \,.
\end{align}
Again, $U_D$ coincides with the lepton mixing matrix
$U_\text{PMNS}^\ast$ up to phases.
In this paper, the Majorana phases contained in $P_N$ are
irrelevant and can therefore be neglected.  We can then express the
Yukawa coupling of the down quarks and charged leptons as
\begin{align}
  \label{eq:yukawa-down-decomposed}
  \mathsf{Y}_2 & = V_q^\ast\, \mathsf{D}_2\,U_D \;.
\end{align}

The relation (\ref{eq:yukawa-down-decomposed}) holds in the CMM model as
long as we concentrate on the heaviest generation, namely the bottom
quarks and the tau lepton.  The masses of the lighter generations do not
unify, so the higher-dimensional operators must partially contribute
differently to down quarks and charged leptons (see
Appendix~\ref{se:cmm-higgs-yukawa}).  Now one might wonder whether these
corrections significantly modify the relation
(\ref{eq:yukawa-down-decomposed}); however, the approximate bottom-tau
unification and the good agreement between the SM predictions and the
experimental data for $B_d-\ovl{B}_d$ mixing, $\Delta{M_K}$ and
$\epsilon_K$ severely constrain these potential modification, as
discussed in Ref.~\cite{Trine:2009ns}. {A corresponding analysis in
  the lepton sector (in a wider SU(5) framework) exploiting $\mu\to e
  \gamma$ can be found in Ref.~\cite{Girrbach:2009uy}.}
We can therefore safely {neglect corrections to} 
Eq.~(\ref{eq:yukawa-down-decomposed}).

In terms of MSSM fields, the couplings simply read
\begin{align}
  W_Y & = Q_i\, \mathsf{D}_1^{ij}\, u^c_j\, H_u + Q_i \left( V_q^\ast\,
    \mathsf{D}_2^\prime\, U_D \right)^{ij} d^c_j\, H_d 
  \nonumber \\ 
  & \qquad + L_i\, \mathsf{D}_1^{ij}\, \nu^c_j\, H_u + L_i \left( U_D^\top\,
    \mathsf{D}_2^\prime\, V_q^\dagger \right)^{ij} e^c_j\, H_d +
  \frac{1}{2}\, \nu^c_i\, \mathsf{D}_N^{ij}\, \nu^c_j \,.
  \label{eq:cmm-yukawa-sm}
\end{align}
Here $Q_i\, \mathsf{D}_1^{ij}\, u^c_j\, H_u$ is short-hand for
\mbox{$\epsilon_{mn}\, Q^{\alpha m}_i\, \mathsf{D}_1^{ij}\, u^c_{\alpha
    j}\, H_u^n$} with the $\text{SU(3)}_C$ and $\text{SU(2)}_L$ indices
$\alpha=1,2,3$ and $m,n=1,2$, respectively, and similarly for the other
couplings. {\eq{eq:cmm-yukawa-sm} holds for exact SO(10) symmetry;
  below $M_\text{SO(10)}$ the Yukawa couplings $D_1^{ij}$
  in the first and third terms will be different, as well as those in
  the second and fourth term.} 

Both $V_q$ and $U_D$ are unitary matrices, which generically have nine
parameters each, namely three mixing angles and six phases.  In the SM,
we can eliminate five of the six phases in $V_\text{CKM}$ by making
phase rotations of the quark fields.  Due to the Majorana nature of the
neutrinos, we are left with three phases in $U_\text{PMNS}$.  In the CMM
model, however, we cannot rotate the quark and lepton fields separately
{without violating the implicit GUT constraint}.  Once we eliminate
all but one phase in $V_q$, we are left with the full set of phases in
$U_D$.  To see the additional phases explicitly, let us write down the
mixing matrix for the tri-bimaximal solution, corresponding to
$\theta_{12} = \arcsin\left(1/\sqrt{3}\right) \simeq 35^\circ$,
$\theta_{13} = 0^\circ$, and $\theta_{23} = 45^\circ$,
\begin{align}
  \label{eq:mnx-phases}
  U_D^\text{TBM} & =\Theta_L U_\text{PMNS}^{\text{TMB}\ast}
\Theta_R =
    \begin{pmatrix}
    \sqrt{\frac{2}{3}}\, e^{-ia_1} & \frac{1}{\sqrt{3}}\,
    e^{-ia_2} & 0
    \\[3pt]
    -\frac{1}{\sqrt{6}}\, e^{-ia_4} & \frac{1}{\sqrt{3}}\,
    e^{-i\left(-a_1+a_2+a_4\right)} & \frac{1}{\sqrt{2}}\,
    e^{-i\left(-a_1+a_3+a_4\right)}
    \\[3pt]
    \frac{1}{\sqrt{6}}\, e^{-ia_5} & -\frac{1}{\sqrt{3}}\,
    e^{-i\left(-a_1+a_2+a_5\right)} & \frac{1}{\sqrt{2}}\,
    e^{-i\left(-a_1+a_3+a_5\right)}.
  \end{pmatrix}
  .
\end{align}
The sixth phase (the `standard' phase $\delta$) drops out due to
$\theta_{13} = 0^\circ$.  In Eq.~(\ref{eq:mnx-phases}), we choose a
parametrization, where the phases could be absorbed via the phase
matrices
\begin{align}\label{eq:ThetaLR}
\Theta_L =
 \text{diag}(e^{-ia_1},e^{-ia_4},e^{-ia_5}),\quad \Theta_R
=
\text{diag}(1,
 e^{i(a_1-a_2)}, e^{i(a_1-a_3)}),\quad U_D =\Theta_L
U_\text{PMNS}^\ast \Theta_R. 
\end{align}
acting on the fields on the left and right, respectively.  However, we
only have this freedom for either $V_q$ or $U_D$.  We choose $V_q \equiv
V_\text{CKM}$ to be in its standard parametrization, so $U_D$ will have
the structure indicated in Eq.~(\ref{eq:mnx-phases}).  These phases are
important constituents of our observables (see
Section~\ref{se:observables}). If we restrict to transitions between the
second and third generation as in \bbs mixing then only one phase
(difference) enters the observables. Then we can write\footnote{The
  corrections to the diagonalization matrix of the right-handed down
  quarks, $U_D$, are studied in \cite{Trine:2009ns}.}
\begin{align}
 U_D = \text{diag}(1,\, e^{i\xi},\,1)\,U_\text{PMNS}^\ast, \qquad\qquad \xi
=a_5-a_4.
\end{align}

\smallskip

Let us now add the supersymmetry breaking terms,
\begin{align}
  \mathscr{L}_\text{soft} & = - \widetilde{\text{16}}^\ast_i \,
  \mathsf{m}^{2\,ij}_{\widetilde{\text{16}}}\, \widetilde{\text{16}}_j -
  m^2_{\text{10}_H}\, \text{10}^\ast_{H} \text{10}_H - m^2_{\text{10}_H^\prime}\,
  \text{10}^\ast_{H^\prime} \text{10}_{H^\prime} \nonumber
  \\
  & \quad - m^2_{\overline{\text{16}}_H}\,
  \ovl{\text{16}}^\ast_H \ovl{\text{16}}_H - m^2_{\text{16}_H} \text{16}^\ast_H
  \text{16}_H - m^2_{\text{45}_H}\, \text{45}^\ast_H \text{45}_H \nonumber
  \\
  & \quad - \left( \frac{1}{2} \widetilde{\text{16}}_i\,
    \mathsf{A}_1^{ij}\, \widetilde{\text{16}}_j\, \text{10}_H +
    \widetilde{\text{16}}_i\, \mathsf{A}_2^{ij}\,
    \widetilde{\text{16}}_j\, \frac{\text{45}_H\,
      \text{10}_{H^\prime}}{2\, M_\text{Pl}} + 
    \widetilde{\text{16}}_i\, \mathsf{A}_N^{ij}\,
    \widetilde{\text{16}}_j \frac{\ovl{\text{16}}_H
      \ovl{\text{16}}_H}{2\, M_\text{Pl}} + \text{h.c.}  \right) ,
\end{align}
where $\mathsf{m}$ are the soft scalar mass matrices and
$\mathsf{A}_i$ the (dimensionful) coefficients of the scalar trilinear
couplings.  In addition, there are $B$-terms for the Higgs fields as
well as gaugino mass terms.  As discussed above, we assume universal
parameters at $M_\text{Pl}$,
\begin{subequations}
  \label{eq:cmm-universal}
  \begin{align}
    \label{eq:cmm-universal-mass}
    \mathsf{m}^2_{\widetilde{\text{16}}_i} & = m_0^2\ \mathbbm{1} \,,
    {\qquad m^2_{\text{10}_H} = m^2_{\text{10}_H^\prime} = m^2_{\text{16}_H} = m^2_{\overline{\text{16}}_H} =
    m^2_{\text{45}_H}} = m_0^2 \,,
    \\
    \label{eq:cmm-universal-aterm}
    \mathsf{A}_1 & = a_0\, \mathsf{Y}_1 \,, \qquad \mathsf{A}_2 =
    a_0\, \mathsf{Y}_2 \,, \qquad \mathsf{A}_N = a_0\, \mathsf{Y}_N
    \,,
  \end{align}
\end{subequations}
as well as one universal gaugino mass, $m_{\tilde{g}}$.
Thus at $M_\text{Pl}$, the soft masses are diagonal in any flavor basis.
At lower energies, this universality is broken.  In particular, it is
broken at $M_\text{GUT}$, which leads to a different phenomenology than
the CMSSM \cite{cmssm} or mSUGRA \cite{bbmr}.  The renormalization group
evolution is conveniently performed in a flavor basis in which the
up-type Yukawa couplings are diagonal (up basis).

For completeness we also give the soft breaking terms for the
CMM model in terms of SU(5) fields:
\begin{align}
  \mathscr{L}_\text{soft} & = - \widetilde{\Psi}^\ast_i \,
  \mathsf{m}^{2\,ij}_{\widetilde{\Psi}}\, \widetilde{\Psi}_j 
- \widetilde{\Phi}_i^{{\ast}} \,
  \mathsf{m}^{2\,ij}_{\widetilde{\Phi}}\, \widetilde{\Phi}_j 
- \left[ {\frac12}  \widetilde{N}_i \,
  \mathsf{m}^{2\,ij}_{\widetilde{N}}\, \widetilde{N}_j 
   { \; +\;  \mbox{h.c.}} \right] \nonumber
\\
  &\quad - m^2_{H}\, H^\ast H - m^2_{H^\prime}\,
  H^{\prime\ast} H^\prime
   - m^2_{\text{24}_H}\, \text{24}^\ast_H \text{24}_H \nonumber
  \\
  & \quad {-} \left[ 
     \left( \frac{1}{4} \widetilde{\Psi}{}^\top \mathsf{A}_1\, 
    \widetilde{\Psi} + \widetilde{N}{}^\top
    \mathsf{A}_\nu \widetilde{\Phi} \right) H + \sqrt{2}\widetilde{\Psi}{}^\top 
  \mathsf{A}_2\, \widetilde{\Phi}\, H^\prime +
  \frac{M_N}{2}\, \widetilde{N}^\top  \mathsf{A}_N \, \widetilde{N} 
               \,  { \; + \; \mbox{h.c.}} \right] .
\end{align}
The fields $\Psi_i$, $\Phi_i$, $N_i$, $H$ and $H^\prime$ live in
the representations $\text{10}$, $\overline{\text{5}}$, $\text{1}$,
$\text{5}$ and $\overline{\text{5}}$ of SU(5), respectively.

In leading order, the soft mass matrix for the right-handed down
squarks, $\mathsf{m}^2_{\tilde d}$, keeps its diagonal form but the
third generation gets significant corrections from the large top Yukawa
coupling, which are parametrized by the real parameter $\Delta_{\tilde
  d}$,
\begin{align}
  \mathsf{m}^2_{\tilde d}\left(M_Z\right) & = \diag \left( m^2_{\tilde
      d}, \, m^2_{\tilde d}, \, m^2_{\tilde d} - \Delta_{\tilde d}
  \right) .\label{smm}
\end{align}
Here and in the following, the small Yukawa couplings of the first two
generations are set to zero in the renormalization group equations.
Now choosing the super-CKM basis\footnote{For the soft-terms and rotation matrices we will
always use the convention of \cite{Rosiek:1995kg}} where the down quarks are mass
eigenstates, this matrix is no longer diagonal,
\begin{align}  \label{eq:vmv}
 \mathsf{m}_D^2 = U_D \mathsf{m}_{\tilde{d}}^2 U_D^\dagger=
\left(\begin{array}{ccc}
m_{\tilde{d}}^2 & 0 & 0\\ 
0 & m_{\tilde{d}}^2-\frac{1}{2}\Delta_{\tilde{d}} &
-\frac{1}{2}\Delta_{\tilde{d}}\mathrm{e}^{i\xi}\\
0 & -\frac{1}{2}\Delta_{\tilde{d}}\mathrm{e}^{-i\xi}&
m_{\tilde{d}}^2-\frac{1}{2}\Delta_{\tilde{d}}     
                                \end{array}
\right), \qquad\quad \xi \equiv a_5 -a_4 ,
\end{align}
allowing flavor-changing quark-squark-gluino and quark-squark-neutralino
vertices (Fig.~\ref{fig:fc-vertices}). Similarly, we get for the
sleptons $\mathsf{m}_L^2 = U_D \mathsf{m}_{\tilde{l}}^2 U_D^\dagger$.
The CP phase\footnote{In \cite{Trine:2009ns} the phase $\xi$ corresponds
  to $\phi_{B_s}$ in absence of Yukawa corrections to the first two
  generations. Note that in \cite{Trine:2009ns} a different convention
  for the soft terms of $\tilde{d}^c,\, \tilde{u}^c,\,\tilde{e}^c$ is
  used: $\tilde{d^c}\mathsf{m}_{\tilde{d}}^2\tilde{d^c}^*$ and not
  $\tilde{d^c}^\ast \mathsf{m}_{\tilde{d}}^2\tilde{d^c}$ such that
  $\mathsf{m}_{\tilde{d}}^2 =
  \left(\mathsf{m}_{\tilde{d}}^2\right)_{\mbox{\cite{Trine:2009ns}}}^\ast $. }
$\xi$ is of utmost importance for the phenomenology of $b\to s$
transitions.  It is worthwhile to compare the situation at hand with the
usual MSSM with generic flavor structure: In the latter model all
off-diagonal elements of the squark mass matrices are ad-hoc complex
parameters, constrained only by the hermiticity of the squark mass
matrices.  In the CMM model, the phase factor $e^{i\xi}$ originates from
the Yukawa matrix $\mathsf{Y}_2$ in \eq{eq:yukawa-down-decomposed} and
enters \eq{eq:vmv} through a rotation of right-handed superfields.

\begin{nfigure}
  \centering
  \hspace{-20pt}
  \subfigure[]{
    \label{fig:fc-vertices:a}
    \centering 
    \scalebox{1}{
      \begin{picture}(100,75)(0,-15)
        \DashArrowLine(50,10)(15,10){2}
        \Text(0,10)[l]{$\tilde{{d}}_{i\alpha}$}
        \ArrowLine(50,45)(50,10)
        \Text(50,60)[t]{${d}_{j\beta}$}
        \Line(50,10)(85,10)
        \Gluon(50,10)(85,10){2}{4}
        \Text(100,10)[r]{$\tilde{g}^a$}
 	\Text(50,-10)[c]{\begin{footnotesize}$i\sqrt{2}
T^a_{\alpha\beta}(U_D)_{ji} P_R$\end{footnotesize}} 
      \end{picture}
    } }
  \hspace{100pt}
  \subfigure[]{
    \label{fig:fc-vertices:b}
    \centering 
    \scalebox{1}{
      \begin{picture}(100,75)(0,-15)
        \DashArrowLine(50,10)(15,10){2}
        \Text(0,10)[l]{$\tilde{{d}}_{i\alpha}$}
        \ArrowLine(50,45)(50,10)
        \Text(50,60)[t]{${d}_{j{\beta}}$}
        \Line(50,10)(85,10)
        \Text(100,10)[r]{$\tilde{\chi}^0_k$}
	\Text(50,-10)[c]{
\begin{footnotesize}$ i\left(Y_j^D (U_D)_{ji} Z_N^{3k} P_L
   -\frac{\sqrt{2}e}{3\cos\theta_W}(U_D)_{ji}
   Z_N^{1k\ast}P_R\right){\delta_{\alpha\beta}}$ 
\mbox{{~~~for $i\neq j$}}  
\end{footnotesize}}
      \end{picture}
    } }
  \caption{Quark-squark-gluino and quark-squark-neutralino
      vertices {for $i,j=2,3$.} 
      {Here $d_{j\beta}$ is the Dirac field of the 
      down-quark mass eigenstate of the $j$-th generation. 
      $\tilde{d}_{i\alpha}$ is the $i$-th-generation right-handed
      down-squark mass eigenstate (coinciding with the interaction
      eigenstate in the basis with $\mathsf{Y}_1=\mathsf{D}_1$).}}
    \label{fig:fc-vertices}
\end{nfigure}

Similarly, relation (\ref{eq:cmm-universal-aterm}) holds at the Planck
scale.  Running {the MSSM trilinear terms} $\mathsf{A}_d$ and
$\mathsf{A}_e$ down to the electroweak scale, off-diagonal entries
appear in the super-CKM basis due to the large mixing matrix $U_D$.
These entries yield additional flavor violating effects.
The running of the parameters in the various regions will be discussed
in the following section. In our notation, we denote trilinear breaking 
terms that are defined in the super-CKM basis by a hat (e.g. $\hat{\mathsf{A}}_d$).

\smallskip

Let us finally discuss two important aspects of the analysis which
originate from the model's group structure.
One, when the SU(5) singlet component of the spinorial Higgs field,
$\text{16}_H$, acquires a vev, SO(10) is not broken to its maximal
subgroup $\text{SU(5)} \times \text{U(1)}_X$ (where
$X=5\left(B-L\right)-4Y$) but to SU(5).  The SO(10) spinor decomposes as
$\text{16} \to \text{10}_1 + \ovl{5}_{-3} + \text{1}_5$ with respect to
$\text{SU(5)} \times \text{U(1)}_X$, so we see that the SU(5) singlet
has a non-trivial $\text{U(1)}_X$ charge.  Acquiring its vev, it breaks
$\text{U(1)}_X$ and reduces the rank of the group from five to four.
Now, because of this rank reduction, additional D-term contributions to
the soft masses appear, which are associated with the spontaneously
broken diagonal generator of $\text{U(1)}_X$ \cite{d-term}.  They are
proportional to the $\text{U(1)}_X$ charge of the SU(5)-fields but do
not depend on the precise form of the $\text{U(1)}_X$ breaking
superpotential, nor on the scale where it is broken.  In contrast, they
depend on the soft masses and are of the same size as the other SUSY
breaking terms, even though the scale of the $\text{U(1)}_X$ breaking is
many orders of magnitude larger.  Hence, these contributions can be
thought of as corrections to the relations
(\ref{eq:cmm-universal-mass}).

The SO(10) vector field decomposes as $10 \to \text{5}_{-2} +
\ovl{5}_2$ with respect to $\text{SU(5)} \times \text{U(1)}_X$.
Hence, the soft masses of the SU(5) fields are given by
\begin{align}
  m^2_{\widetilde{\Psi}_i} \left(t_{\text{SO(10)}}\right) & =
  m^2_{\widetilde{\text{16}}_i} \left(t_{\text{SO(10)}}\right) + D \,, &
  m^2_{H} \left(t_{\text{SO(10)}}\right) & = {m^2_{\text{10}_H}}
  \left(t_{\text{SO(10)}}\right) - 2\, D \,, \nonumber
  \\
  m^2_{\widetilde{\Phi}_i} \left(t_{\text{SO(10)}}\right) & =
  m^2_{\widetilde{\text{16}}_i} \left(t_{\text{SO(10)}}\right) - 3\, D
  \,, &
  {m^2_{H^\prime}} \left(t_{\text{SO(10)}}\right) & = {m^2_{\text{10}_H^\prime}}
  \left(t_{\text{SO(10)}}\right) + 2\, D \,, \nonumber
  \\
  m^2_{\widetilde{N}_i} \left(t_{\text{SO(10)}}\right) & =
  m^2_{\widetilde{\text{16}}_i} \left(t_{\text{SO(10)}}\right) + 5\, D
  \,,
  \label{eq:d-terms}
\end{align}
where $D$ denotes the additional D-term contribution {and
  $t=\ln\mu_{\rm r}$ with the renormalization scale $\mu_{\rm r}$.}  $D$ is
another parameter which enters our analysis when we relate weak scale
observables to universal parameters at $M_\text{Pl}$.  Since $D$ affects
all fermion generations in the same way, its effect on flavor physics is
small.



Two, we have to check whether the fields of the unbroken subgroups are
properly normalized.  Decomposing the vector and adjoint of SO(10) in
SU(5) representations, we see that both the fundamental and adjoint
SU(5)-fields need to be rescaled by a factor of $\sqrt{2}$
\cite{normierung}.  In order to have a continuous gauge coupling,
however, we should instead rescale the SO(10) generators by a factor
$1/\sqrt{2}$,
\begin{align}
  T_{ij} & = \frac{1}{\sqrt{2}} \mathcal{T}_{ij} \,,
  \intertext{where $\mathcal{T}_{ij}$ are the SO(10) generators in the
    usual normalization, satisfying}
  \left(\mathcal{T}_{ij}\right)_{mn} & = i \left( \delta_{im}
    \delta_{jn} - \delta_{in} \delta_{jm} \right) , &
  \left[ \mathcal{T}_{ij}, \mathcal{T}_{kl} \right] & = i \left(
    \delta_{jk} \mathcal{T}_{il} - \delta_{il} \mathcal{T}_{jk} -
    \delta_{jl} \mathcal{T}_{ik} + \delta_{ik} \mathcal{T}_{jl} \right)
  .
\end{align}
At the same time, this redefinition of the SO(10) generators avoids a
rescaling of the top Yukawa coupling by a factor $\sqrt{2}$
\cite{nath-syed}.

\smallskip In summary, the CMM model is a simple but {well-motivated}
SO(10) model, which allows for large mixing among right-handed down
quarks and therefore interesting effects in flavor changing processes.
Actually, these effects are a consequence of the underlying GUT
structure (evident in the relation $Y_d=Y_e^\top$), the large top
coupling and weak-scale supersymmetry.
Compared to the SM, we have only a small number of additional parameters
affecting the low-energy physics we plan to study: So far we have
encountered the SUSY breaking parameters $m_0$, $m_{\tilde{g}}$ and
$a_0$, the D-term correction $D$ and the CP phase $\xi$.  We will
need two more parameters, $\tan\beta$ and the phase of the Higgs
  mass parameter $\mu$.

This small set of parameters makes the model very predictive.

\section{Renormalization Group Equations}\label{se:rge}
\subsection{Top Yukawa Coupling and its Infrared Fixed Point}\label{sec:yt}
For small values of $\tan\beta$, the top Yukawa $y_t$ coupling is of
order unity.  In this case, the coupling can become non-perturbative
below the Planck scale, in particular in GUT scenarios which generically
include larger representations than the MSSM.  The SO(10) RGE for the
gauge and top Yukawa coupling have an infrared quasi-fixed point at one
loop for $g^2/y_t^2=56/55$
\cite{Hill:1980sq,Pendleton:1980as,Carena:1993bs}.  Thus, for larger
values of $y_t$ at $M_\text{SO(10)}$, its value may become
non-perturbative below the Planck scale. In the CMM model the main
driver of the FCNC effects is the RG revolution between $M_{{\rm Pl}}$
and $M_{{\rm SO(10)}}$. Therefore, with increasing $\tan\beta$ the model
specific $b\rightarrow s$ transitions quickly die out.

In the CMM model, the infrared fixed point corresponds to $\tan\beta
\simeq 2.7$ as one can see in Fig.~\ref{fig:fp}.  Our analysis will be
located close to this fixed point, hence a precise a knowledge of $y_t$
is important.  For this reason we will use the two-loop RGE in the MSSM.
The default values in our analysis are $\tan\beta=3$ and $\tan\beta =
6$.

\begin{center}
\begin{nfigure}[!tb]
\centering
\psfrag{t}{\begin{large}$t=\ln(\mu_{\rm r})$\end{large}}
\psfrag{yt}{\begin{large}${y}_t(t)$\end{large}}
\psfrag{label1}{\begin{small}$\ \tan \beta=2.2$\end{small}}
\psfrag{label2}{\begin{small}$\ \tan \beta_c\approx 2.7$\end{small}}
\psfrag{label3}{\begin{small}$\ \tan \beta=4.5$\end{small}}
\psfrag{tMZ}{\begin{large}$t_{M_Z}$\end{large}}
\psfrag{t5}{\begin{large}$\hspace{-0.8cm}t_{\text{GUT}}$\end{large}}
\psfrag{t10}{\begin{large}$t_{\text{SO(10)}}$\end{large}}
\psfrag{tPl}{\hspace{0.3cm}\begin{large}$t_{\text{Pl}}$\end{large}}
\includegraphics[width=10cm]{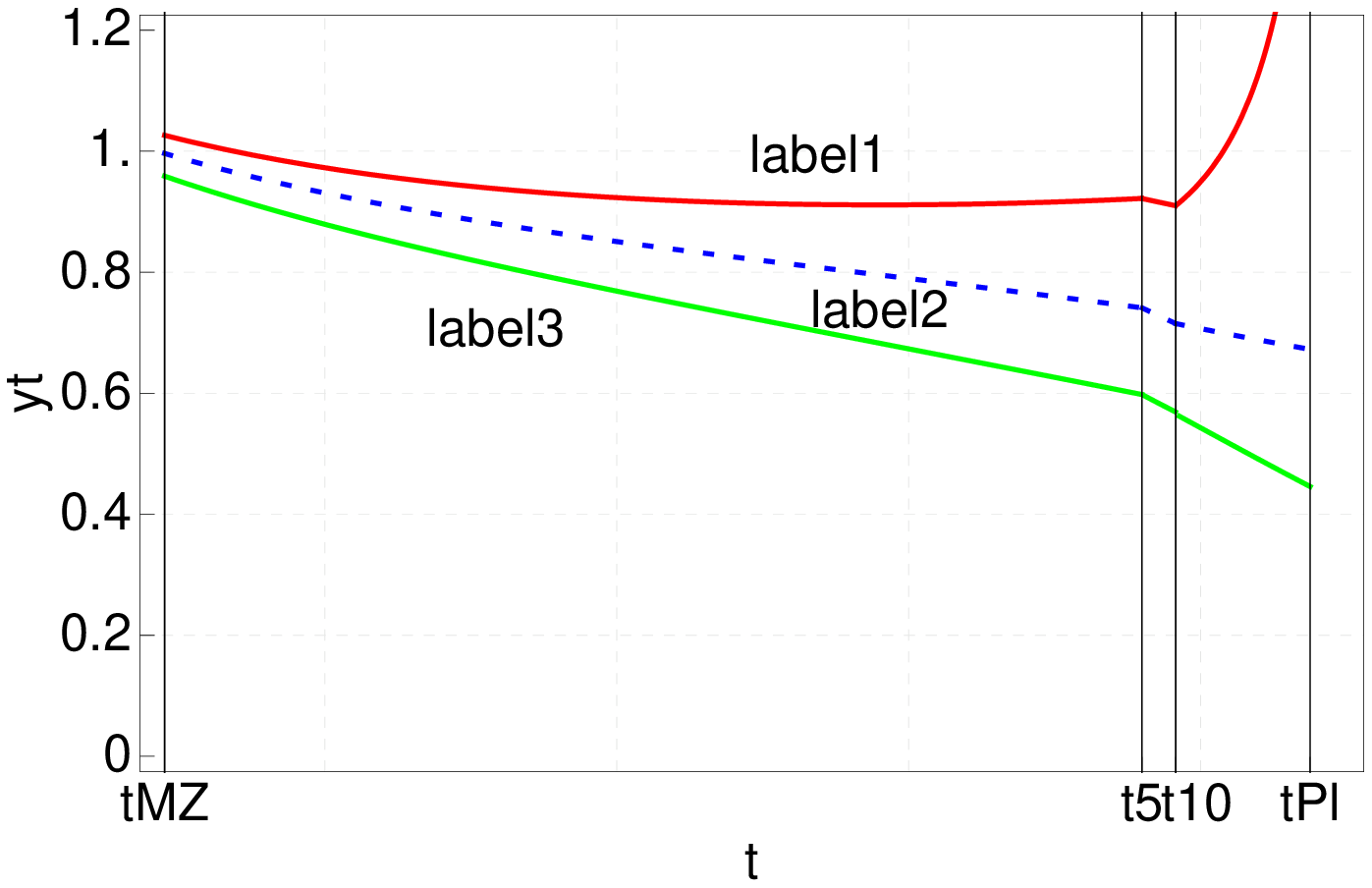}
\caption{
If $\tan\beta$ is too small, $y_t$ becomes non-perturbative below the Planck scale. 
The dotted line corresponds to a value of $\tan\beta$, where $g/y_t$ reaches its fixed point at $t_{\text{SO(10)}}$. The kinks in the functions are due to the change of the gauge group.}
\label{fig:fp}
\end{nfigure}
\end{center}

\subsection{Threshold correction and conversion to $\overline{\text{\slshape DR}}$ Scheme}
  
We use the two-loop RG equations for the gauge and Yukawa couplings in the
$\overline{\text{\slshape DR}}$ scheme with one-loop SUSY threshold
corrections at the electroweak scale \cite{Martin:1993zk,Pierce:1996zz}.  
The reason for NLO accuracy here is the delicate dependence of the FCNC effects
on $y_t(M_Z)$ shown in Fig.~\ref{fig:fp}.
For scheme consistency the one-loop threshold corrections must be included with two-loop RGEs.
Above $t_{{\rm GUT}}$ one-loop accuracy is sufficient.
In the MSSM we use the approximated formula from Ref.~\cite{Pierce:1996zz}
that include only potentially large corrections. For simplicity the decoupling scale 
is set to $M_Z$.
The initial values for the gauge couplings $\widetilde\alpha_i \equiv \alpha_i/(4\pi)$ are then
given as
\begin{align}
  \widetilde\alpha_1(M_Z) & = \frac{5}{3}\frac{\alpha_e(M_Z)}{4\pi
    \cos^2\theta_W}  \,, \nonumber
  \\[2pt]
  \widetilde\alpha_2(M_Z) & = \frac{\alpha_e(M_Z)}{4\pi \sin^2\theta_W}
  \,,
  \label{eq:rge-gauge}
  \\
  \widetilde\alpha_3(M_Z) & = \frac{1}{4\pi}
  \frac{\alpha_s(M_Z)}{1-\Delta\alpha_s} , & \Delta\alpha_s & =
  \frac{\alpha_s(M_Z)}{2 \pi} \left[ \frac{1}{2} - \frac{2}{3}
    \ln\frac{m_t}{M_Z} - 2 \ln\frac{m_{\tilde{g}_3}}{M_Z} - \frac{1}{6}
    \sum_{i=1}^{12} \ln\frac{M_{\tilde{q}_i}}{M_Z} \right] , \nonumber
\end{align}
where stands $M_{\tilde{q}_i}$ for the mass eigenvalues of the 12 up and down squarks
and $m_{\tilde{g}_3}$ is the gluino mass. Here and in the following
a tilde on a quantity always means that it has been divided by $4\pi$.

For the Yukawa couplings, we take both complex SUSY parameters and large
off-diagonal elements in $\mathsf{m}^2_{d}$ and $\mathsf{A}_d$ into
account.
Then the top Yukawa coupling including threshold corrections is given by
\begin{align}
  \tilde y_t(M_Z) & = \frac{m_t}{4\pi v \sin\beta \left(1+\frac{\Delta
        m_t}{m_t}\right)} \,,
  \\[3pt]
  \frac{\Delta m_t}{m_t} & = \widetilde\alpha_3(M_Z) \left[ 4
    \ln\frac{M_Z^2}{m_t^2} + \frac{20}{3} - \frac{4}{3} \left(
      B_1(0,m_{\tilde g_3},m_{\tilde t_1}) + B_1(0,m_{\tilde
        g_3},m_{\tilde t_2}) \right) \right. \nonumber
  \\
  & \mspace{90mu} \left. + \frac{4}{3}  e^{i \delta_{\tilde t}} \sin\left(2
        \theta_{\tilde t}\right) \frac{m_{\tilde g_3}}{m_t} \left(
        B_0(0,m_{\tilde g_3},m_{\tilde t_1}) - B_0(0,m_{\tilde
          g_3},m_{\tilde t_2}) \right)  \right] , \nonumber
\end{align}
where $\theta_{\tilde t}$ and $\delta_{\tilde t}$ denote the stop mixing parameters
defined later in this paragraph. The electroweak vev is denoted as 
$v=\sqrt{\VEV{H_u}^2+\VEV{H_d}^2}\approx 174\ \text{GeV}$.
The loop functions $B_0$ and $B_1$ are given as follows:
\begin{subequations}
  \label{eq:loop-fct-b}
  \begin{align}
    B_0(0,m_1,m_2) & = -\ln\frac{M^2}{M_Z^2} + 1 + \frac{m^2}{m^2-M^2}
    \ln\frac{M^2}{m^2} \,,
    \\
    B_1(0,m_1,m_2) & = \frac{1}{2} \left[ -\ln\frac{M^2}{M_Z^2} +
      \frac{1}{2} +\frac{1}{1-x} +\frac{\ln x}{(1-x)^2} - \theta(1-x)
      \ln x \right] ,
  \end{align} 
\end{subequations}
with $M=\max\left(m_1,m_2\right)$, $m=\min\left(m_1,m_2\right)$, and
$x=m_2^2/m_1^2$.

The corrections for the bottom coupling are slightly more involved.
We include these corrections to account for CP phases. In the
end, however, they turn out to be not relevant for small $\tan\beta$.
\begin{align}
  \tilde y_b(M_Z) & = -\frac{\widehat{m}_b^{\text{SM}}(M_Z)}{4\pi v
    \cos\beta \left(1+\frac{\Delta m_b}{m_b}\right)} \,,
  \\[3pt]
  \frac{\Delta m_b}{m_b} & = \left( \frac{\Delta m_b}{m_b}
  \right)^{\tilde t \tilde\chi^+} + \left( \frac{\Delta m_b}{m_b}
  \right)^{\tilde b \tilde g_3} , \nonumber
  \\[6pt]
  \left(\frac{\Delta m_b}{m_b} \right)^{\tilde t \tilde\chi^+} & =
  \tilde y_t \mu^* \frac{\tilde{A}^*_t \tan\beta + \mu \tilde
    y_t}{m_{\tilde t_1}^2 - m_{\tilde t_2}^2} \left[
    B_0(0,\abs{\mu},m_{\tilde t_1}) - B_0(0,\abs{\mu},m_{\tilde t_2})
  \right] \nonumber
  \\
  & \quad - \widetilde\alpha_2 \frac{\mu^* m_{\tilde g_2}
    \tan\beta}{\abs{\mu}^2-m_{\tilde g_2}^2} \left[\cos^2\theta_{\tilde
      t}\, B_0(0,m_{\tilde g_2},m_{\tilde t_1}) +\sin^2\theta_{\tilde
      t}\, B_0(0,m_{\tilde g_2},m_{\tilde t_2}) \right.  \nonumber
  \\
  & \mspace{160mu} \left. -\cos^2\theta_{\tilde t}\,
    B_0(0,\abs{\mu},m_{\tilde t_1}) - \sin^2\theta_{\tilde t}\,
    B_0(0,\abs{\mu},m_{\tilde t_2}) \right] , \nonumber
  \\
  \left(\frac{\Delta m_b}{m_b} \right)^{\tilde b \tilde g_3} & =
  -\frac{4}{3} \tilde\alpha_3(M_Z) \left[ B_1(0,m_{\tilde g_3},m_{\tilde
      b_1}) + B_1(0,m_{\tilde g_3},m_{\tilde b_2}) - 2 \frac{m_{\tilde
        g_3}}{m_b} \sum_{i=1}^{6} Z_D^{6 i *} Z_D^{3 i} B_0(0,m_{\tilde
      g_3},m_{\tilde d_i}) \right] . \nonumber
\end{align}
with $\widehat{m}_b^{\text{SM}}(M_Z) = 2.92$ GeV.  The matrix $Z_D$ is
the $6\times 6$ mixing matrix for the down squarks defined in Ref.~\cite{Rosiek:1995kg}; 
$m_t$ UN $m_b$ denote the pole masses of the top and bottom quarks, respectively; and
the loop functions are given in Eqs.~(\ref{eq:loop-fct-b}).
$\tilde A_t$ is the $(3,3)$ entry of the trilinear soft breaking term
for the up squarks. $\mu$ is the SUSY Higgs parameter
and $m_{\tilde t_i}$, $m_{\tilde b_i}$ are the eigenvalues of the stop
and sbottom mass matrix. Furthermore, we denote the mass of the $\text{SU(2)}_L$ gaugino
by $m_{\tilde g_2}$.
Finally, the initial condition for the tau coupling reads
\begin{align}
  \tilde y_{\tau}(M_Z) & = -\frac{m_{\tau}}{4\pi v \cos\beta} \,.
\end{align}
The $2 \times 2$ mass matrix of the scalar top quarks,
\begin{align}
  \mathcal M_{\tilde t}^2 & =
  \begin{pmatrix}
    m_{\tilde q_3}^2 + m_t^2 + \left(\frac{1}{2}-\frac{2}{3}
      \sin^2{\theta_W}\right) M_Z^2 \cos(2\beta) & -m_t \left(\frac{\tilde
        A_t}{\tilde y_t}+\frac{\mu^*}{\tan\beta}\right) \cr -m_t
    \left(\frac{\tilde A_t^*}{\tilde y_t}+\frac{\mu}{\tan\beta}\right) &
    m_{\tilde u_3}^2 + m_t^2 + \frac{2}{3} \sin^2{\theta_W} M_Z^2 \cos(2\beta)
  \end{pmatrix}
  ,
\end{align}
is diagonalized by the unitary matrix $\widetilde Z^T_U$,
\begin{align}
  \widetilde Z^T_U \mathcal M_{\tilde t}^2 \widetilde Z^*_U & =
  \begin{pmatrix}
    m_{\tilde t_1}^2 & 0 \cr 0 & m_{\tilde t_2}^2
  \end{pmatrix}
  \ , \mspace{60mu} \widetilde Z^T_U =
  \begin{pmatrix}
    \cos\theta_{\tilde t} & e^{i \delta_{\tilde t}} \sin\theta_{\tilde
      t} \cr -e^{-i \delta_{\tilde t}} \sin\theta_{\tilde t} &
    \cos\theta_{\tilde t}
  \end{pmatrix}
  ,
\end{align}
which is the $(3,6)$-submatrix of $Z^T_U$, the analogon of $Z^T_D$ for the
up squarks \cite{Rosiek:1995kg}.  The mixing angle and phase are
computed via
\begin{align}
  \tan\theta_{\tilde t} & = \frac{2 m_t \abs{\frac{\widetilde
        A_t}{\tilde y_t} + \frac{\mu^\ast}{\tan\beta}}}{m^2_{\tilde q_3}
    - m^2_{\tilde u_3} + \left( \frac{1}{2} - \frac{4}{3}\sin^2\theta_W
    \right) M_Z^2 \cos(2\beta)} \,, &
  \delta_{\tilde t} & = \arg \left[ -m_t \left( \frac{\widetilde
        A_t}{\tilde y_t} + \frac{\mu^*}{\tan\beta} \right) \right] ,
\end{align}
where the $(3,3)$ elements of the (diagonal) soft breaking masses
have been denoted by $m^2_{\tilde u_3}$ and $m^2_{\tilde q_3}$.
Note that we do not include threshold corrections in the mixing
matrices, because they appear only in expressions that are of
one-loop order already. The resulting effect would be one more order
higher, which can safely be neglected.

\subsection{Gauge and Yukawa Couplings}

As discussed above, we use the two-loop RGEs in the MSSM.  They can be
found in Ref.~\cite{Martin:1993zk} and are listed in our notation
below. We do not include Higgs self-interactions in the RGEs
because we do not specify the couplings of the Higgs superfields
to each other. Qualitatively they would not change the outcome of our analysis
since Higgs self-interactions are always flavor blind. Including them
would only lead to an absolute shift in the allowed parameter space of the model.
 We neglect
both the small Yukawa couplings of the lighter generations as well as
the CKM matrix, as its flavor violating entries are small compared to
those in $U_D$. Here and in the following,
$t=\ln{\mu_{\rm r}}$, where $\mu_{\rm r}$ is the renormalization scale. 

\label{rge-mssm}

{\allowdisplaybreaks
  \begin{align}
    \frac{d}{d t} \widetilde\alpha_1 & = 2 \widetilde\alpha_1^2 \left(
      \frac{33}{5} + \frac{199}{25} \widetilde\alpha_1
      +\frac{27}{5}\widetilde\alpha_2 + \frac{88}{5}
      \widetilde\alpha_3 -\frac{26}{5} \abs{\tilde y_t}^2 -
      \frac{14}{5} \abs{\tilde y_b}^2 - \frac{18}{5} \abs{\tilde
        y_{\tau}}^2 \right)
    \\
    \frac{d}{d t} \widetilde\alpha_2 & = 2 \widetilde\alpha_2^2
    \left(1 + \frac{9}{5} \widetilde\alpha_1 +25 \widetilde\alpha_2 +
      24 \widetilde\alpha_3 -6 \abs{\tilde y_t}^2 - 6 \abs{\tilde
        y_b}^2 - 2 \abs{\tilde y_{\tau}}^2 \right)
    \\
    \frac{d}{d t} \widetilde\alpha_3 & = 2 \widetilde\alpha_3^2
    \left(-3 + \frac{11}{5} \widetilde\alpha_1 +9 \widetilde\alpha_2 +
      14 \widetilde\alpha_3 - 4 \abs{\tilde y_t}^2 - 4 \abs{\tilde
        y_b}^2 \right)
    \\
    \frac{d}{d t} \tilde y_t & = \tilde y_t \left( 6 \abs{\tilde
        y_t}^2 + \abs{\tilde y_b}^2 - \frac{16}{3} \widetilde\alpha_3
      - 3 \widetilde\alpha_2 - \frac{13}{15} \widetilde\alpha_1
    \right) \nonumber
    \\
    & \quad + \tilde y_t \left( - 22 \abs{\tilde y_t}^4 - 5
      \abs{\tilde y_b}^4 -5 \abs{\tilde y_b \tilde y_t}^2 -
      \abs{\tilde y_b \tilde y_{\tau}}^2 \right.  \nonumber
    \\
    & \mspace{70mu} + \left.  16 \widetilde\alpha_3 \abs{\tilde y_t}^2
      + \frac{6}{5} \widetilde\alpha_1 \abs{\tilde y_t}^2 +6
      \widetilde\alpha_2 \abs{\tilde y_t}^2 + \frac{2}{5}
      \widetilde\alpha_1 \abs{\tilde y_b}^2 \right. \nonumber
    \\
    & \mspace{70mu} - \left. \frac{16}{9} \widetilde\alpha_3^2 +
      \frac{15}{2} \widetilde\alpha_2^2 + \frac{2743}{450}
      \widetilde\alpha_1^2 + 8 \widetilde\alpha_3 \widetilde\alpha_2 +
      \frac{136}{45} \widetilde\alpha_3\widetilde\alpha_1 +
      \widetilde\alpha_1\widetilde\alpha_2 \right)
    \\
    \frac{d}{d t} \tilde y_b & = \tilde y_b \left( 6 \abs{\tilde
        y_b}^2 + \abs{\tilde y_t}^2 + \abs{\tilde y_{\tau}}^2 -
      \frac{16}{3} \widetilde\alpha_3 - 3 \widetilde\alpha_2 -
      \frac{7}{15} \widetilde\alpha_1 \right) \nonumber
    \\
    & \quad +\tilde y_b \left( - 22 \abs{\tilde y_b}^4 - 5 \abs{\tilde
        y_t}^4 - 3 \abs{\tilde y_{\tau}}^4 - 5 \abs{\tilde y_b \tilde
        y_t}^2 - 3 \abs{\tilde y_b \tilde y_{\tau}}^2 \right.
    \nonumber
    \\
    & \mspace{70mu} + \left. 16 \widetilde\alpha_3 \abs{\tilde y_b}^2
      + \frac{2}{5} \widetilde\alpha_1 \abs{\tilde y_b}^2 + 6
      \widetilde\alpha_2 \abs{\tilde y_b}^2 + \frac{6}{5}
      \widetilde\alpha_1 \abs{\tilde y_{\tau}}^2 + \frac{4}{5}
      \widetilde\alpha_1 \abs{\tilde y_t}^2 \right.  \nonumber
    \\
    & \mspace{70mu} + \left. \frac{16}{9} \widetilde\alpha_3^2 +
      \frac{15}{2} \widetilde\alpha_2^2 + \frac{287}{90}
      \widetilde\alpha_1^2 + 8 \widetilde\alpha_3 \widetilde\alpha_2 +
      \frac{8}{9} \widetilde\alpha_3 \widetilde\alpha_1 +
      \widetilde\alpha_1 \widetilde\alpha_2 \right)
    \\
    \frac{d}{d t} \tilde y_{\tau} & = \tilde y_{\tau} \left( 4
      \abs{\tilde y_{\tau}}^2 + 3 \abs{\tilde y_b}^2 - 3
      \widetilde\alpha_2 - \frac{9}{5} \widetilde\alpha_1 \right)
    \nonumber
    \\
    & \quad +\tilde y_{\tau} \left( - 19 \abs{\tilde y_{\tau}}^4 - 9
      \abs{\tilde y_{\tau} \tilde y_b}^2 - 3 \abs{\tilde y_b \tilde
        y_t}^2 + 16 \widetilde\alpha_3 \abs{\tilde y_b}^2 -
      \frac{2}{5} \widetilde\alpha_1 \abs{\tilde y_b}^2 \right.
    \nonumber
    \\
    & \mspace{70mu} + \left. \frac{6}{5} \widetilde\alpha_1
      \abs{\tilde y_{\tau}}^2 + 6 \widetilde\alpha_2 \abs{\tilde
        y_{\tau}}^2 + \frac{15}{2} \widetilde\alpha_2^2 + \frac{9}{5}
      \widetilde\alpha_1 \widetilde\alpha_2 + \frac{27}{2}
      \widetilde\alpha_1^2 \right)
  \end{align}
}%

\subsubsection*{SU(5)}

At $M_\text{GUT}$, the gauge couplings unify.  As is well known, this
unification is not exact in the MSSM at the two-loop level but will be
compensated by threshold effects, caused by the GUT particle spectrum.
Due to the larger uncertainties of the strong coupling, we use the
criterion $\widetilde\alpha_1(t_{\text{GUT}}) =
\widetilde\alpha_2(t_{\text{GUT}}) \equiv \widetilde\alpha$.  Similarly,
we choose the bottom coupling as input for $\mathsf{Y}_2$.

The singlet neutrinos are integrated out at their mass scales, the
heaviest of which is an order of magnitude smaller than $M_\text{GUT}$.
However, we do not take the effect of the neutrino coupling $\tilde
y_{\nu_3}$ between $M_{N3}$ and $M_\text{GUT}$ into account.  At
$M_\text{GUT}$, we identify $\tilde y_{\nu_3}=\tilde y_t$ according to
Eq.~(\ref{eq:cmm-yukawa-sm}).

We use one-loop RGE as given in \cite{Hisano:1998fj}.  In our notation,
they read
\begin{align}
  \frac{d}{d t} \tilde\alpha & = - 6 \tilde\alpha^2 \,,
  \\
  \frac{d}{d t} \tilde y_t & = \tilde y_t \left( -\frac{96}{5}
    \tilde\alpha + 9 \abs{\tilde y_t}^2 + 4 \abs{\tilde y_b}^2 +
    \abs{\tilde y_{\nu_3}}^2 \right) ,
  \\
  \frac{d}{d t} \tilde y_b & = \tilde y_b \left( - \frac{84}{5}
    \tilde\alpha + 10 \abs{\tilde y_b}^2 + 3 \abs{\tilde y_t}^2 +
    \abs{(U_D)_{33}}^2 \abs{\tilde y_{\nu_3}}^2 \right) ,
  \\
  \frac{d}{d t} \tilde y_{\nu_3} & = \tilde y_{\nu_3} \left(
    -\frac{48}{5} \tilde\alpha + 7 \abs{\tilde y_{\nu_3}}^2 + 3
    \abs{\tilde y_t}^2 + 4 \abs{(U_D)_{33}}^2 \abs{\tilde y_b}^2 \right)
  .
\end{align}

\subsubsection*{SO(10)}

The Yukawa couplings for the down quarks are generated via the
non-renormalizable term.  To derive its RGE, we generalize the equations
from Ref.~\cite{Martin:1993zk} to a dimension-five coupling.  Here we
make use 
of the non-renormalization theorem in
supersymmetry, i.e.~that only wave-function renormalization
contributes to the beta functions. To verify that this theorem
  is applicable to the dimension-5 term at the one-loop level, note
  that each vertex diagram is equivalent to a vertex correction of a 
 dimension-four interaction: E.g.\ diagrams in which the two matter 
  supermultiplets are part of the loop are identical to the sum of 
  corresponding diagrams with $45_H 10_H^\prime$ replaced by  
  single Higgs superfields transforming as $\underline{10}$,   
  $\underline{120},\ldots$.
The RGE for $\widetilde{\mathsf{Y}}_2$ reads:
\begin{align}
  \frac{d}{d t} \widetilde{\mathsf{Y}}_2 = -\frac{95}{2}
  \tilde{\alpha} \widetilde{\mathsf{Y}}_2 + 10 \left(
    \widetilde{\mathsf{Y}}_1 \widetilde{\mathsf{Y}}_1^\dagger
    \widetilde{\mathsf{Y}}_2 + \widetilde{\mathsf{Y}}_2
    \widetilde{\mathsf{Y}}_1^\dagger
    \widetilde{\mathsf{Y}}_1
     \right),\label{rgy}
\end{align} 
where again $\tilde{\alpha}=\alpha/(4\pi)$,
$\tilde{\mathsf{Y}}_i=\mathsf{Y}_i/(4\pi)$ and $t=\ln{\mu_{\rm r}}$.
In practice, however, we will only need the {RGE} for the
bottom-coupling,
\begin{align}
  \frac{d}{d t} \tilde{y}_b & = \tilde{y}_b \left( -\frac{95}{2}
    \tilde{\alpha} + 10 \left( 1 + \abs{\left(U_D\right)_{33}}^2 \right)
    \abs{\tilde{y}_t}^2 \right) .
\end{align}

Note that {$\mathsf{Y}_2$} and $\tilde{y}_b$ are the SO(10)
  couplings, which will be rescaled
at the SO(10) breaking scale (see Eq.~(\ref{eq:cmm-yukawa-sm})),
e.g.
\begin{align}
  \tilde{y}_b^\prime\left(t_\text{SO(10)}\right) & =
  \frac{{v_0}}{M_{\text{Pl}}}\ \tilde{y}_b
  \left(t_\text{SO(10)}\right) \,,
\end{align}
where the prime denotes the SU(5) coupling. The prime, however,
is omitted in our SU(5) RGEs.

The equations for the top coupling and the gauge coupling read
\begin{align}
  \frac{d}{d t} \tilde{y}_t & = \tilde{y}_t \left( -\frac{63}{2}
    \tilde{\alpha} + 28 \abs{\tilde{y}_t}^2 \right)
  \\
  \frac{d}{d t} \tilde{\alpha} & = -8 \tilde{\alpha}^2 \ .
\end{align}

\subsection{Supersymmetry Breaking Parameters}
The soft masses and $A$-terms at the scale $M_Z$  are fixed by
the universal terms $a_0$, $m_0^2$, and $D$ through
the renormalization group equations (RGE).  Instead
of guessing their values at $M_\text{Pl}$, we will consider three
parameters at $M_Z$ which are allowed by theoretical and
experimental constraints.  These are the soft masses of the
first generation of right-handed up and down squarks and the
(11)-element of the trilinear coupling of the down squarks,
\begin{align}
  m_{\tilde{u}_1}^2(M_Z) \,, \qquad m_{\tilde{d}_1}^2(M_Z) \,, \qquad
  a_1^d(M_Z) \equiv \lt[ a^d(M_Z) \rt]_{11}
\,.
\end{align}
We work in the weak basis with diagonal {$\mathsf{Y}_1$} and the
  trilinear term $a_1^d$ is defined with the corresponding Yukawa
  coupling factored out, in analogy to $a_0$ in 
  \eq{eq:cmm-universal-aterm}.
  With these initial conditions we can evolve the soft terms 
up to $M_\text{GUT}$, where
the MSSM fields are unified into the SU(5) multiplets
$\Phi$ and $\Psi$ with
\begin{align}
  m_{\widetilde{\Psi}_1}^2\left(t_\text{GUT}\right) & =
  m_{\tilde{u}_1}^2\left(t_\text{GUT}\right) , &
  m_{\widetilde{\Phi}_1}^2\left(t_\text{GUT}\right) & =
  m_{\tilde{d}_1}^2\left(t_\text{GUT}\right) .
\end{align}
After running from $M_\text{GUT}$ to $M_\text{SO(10)}$ we can
calculate $D$ by means of Eqs.~(\ref{eq:d-terms}),
\begin{align}
  D & = \frac{1}{4} \left[
    m^2_{\widetilde{\Psi}_1}\left(t_\text{SO(10)}\right) -
    m^2_{\widetilde{\Phi}_1}\left(t_\text{SO(10)}\right) \right] ,
\end{align}
and determine
\begin{align}
  m^2_{\widetilde{\text{16}}_1}\left(t_\text{SO(10)}\right) =
  \frac{1}{4}\left[ 3
    m^2_{\widetilde{\Psi}_1}\left(t_\text{SO(10)}\right) +
    m^2_{\widetilde{\Phi}_1}\left(t_\text{SO(10)}\right) \right] .
\end{align} 
Then the universal scalar soft  mass at the Planck scale is found:
\begin{align}
  m_0^2 = m^2_{\tilde{\text{16}}_1}\left(t_\text{Pl}\right)
\end{align}
The determination of the universal gaugino mass $m_{\tilde{g}}$
  is much simpler: At leading order the ratio
  $\kappa\equiv m_{\tilde{g}_i}(t)/\tilde{\alpha}_i(t)$ is RG invariant,
  independent of $i$ and equal to its SU(5) and SO(10) GUT values,
  $\kappa=m_{\tilde{g}}(t)/\tilde{\alpha}(t)$ \cite{Martin:1993zk}.
We determine $\kappa$ from the gluino mass and the QCD coupling:
\begin{align}
  m_{\tilde{g}_i}(t) & = \kappa\, \tilde\alpha_i(t) \;,
  \intertext{where}
  \kappa & \equiv \frac{m_{\tilde{g}_3}(M_Z)}{\tilde{\alpha}_3(M_Z)} \
  .
\end{align} 
The RGE needed to determine the Planck scale parameters are
\begin{eqnarray}
\text{MSSM:} && \frac{d}{d t} {a_1^d} = -\left(\frac{32}{3}
                \tilde\alpha_3^2 + 6 \tilde\alpha_2^2 +
                \frac{14}{15}\tilde\alpha_1^2\right) \kappa \nonumber\\ 
\text{SU(5):} && \frac{d}{d t} {a_1^d} = 
        -\frac{168}{5} \tilde\alpha^2\kappa \nonumber\\ 
\text{SO(10):} && \frac{d}{d t} {a_1^d} = -95 \tilde\alpha^2\kappa
      \quad \Rightarrow \quad a_0=a_1^D(t_{\text{Planck}})
 \label{rgeplanck}
\end{eqnarray} 
and
\begin{eqnarray}
\text{MSSM:} && \frac{d}{d t} m^2_{\tilde u_1} =
-\frac{32}{3}\kappa^2\tilde\alpha_3^3 -
\frac{32}{15}\kappa^2\tilde\alpha_1^3 - \frac{4}{5}
\frac{S_{\text{GUT}}}{\tilde\alpha_{\text{GUT}}}\tilde\alpha_1^2 \nonumber\\ &&
\frac{d}{d t} m^2_{\tilde d_1} = -\frac{32}{3}\kappa^2\tilde\alpha_3^3
- \frac{18}{15}\kappa^2\tilde\alpha_1^3 + \frac{2}{5}
\frac{S_{\text{GUT}}}{\tilde\alpha_{\text{GUT}}}\tilde\alpha_1^2 \nonumber
\\ \text{SU(5):} && \frac{d}{d t} {m^2_{\tilde\Psi_1}} = -\frac{144}{5}
\kappa^2\tilde\alpha^3 \nonumber\\ && \frac{d}{d t} {m^2_{\tilde\Phi_1}} =
-\frac{96}{5} \kappa^2\tilde\alpha^3 \nonumber\\ \text{SO(10):} && \frac{d}{d
  t} m^2_{\tilde{\mathbf{16}}_1} = -45 \kappa^2\tilde\alpha^3 \quad
\Rightarrow \quad m^2_0=m^2_{\tilde{\mathbf{16}}_1}(t_{\text{Planck}})
 \label{rgeplanck2}
\end{eqnarray} 
Here we have used the quantity
\begin{equation}
\label{EqDefSGUT}
 S_{\text{GUT}} \equiv m^2_{H_u}(t_{\text{GUT}})-m^2_{H_d}(t_{\text{GUT}})
\end{equation}
which is defined in a more general way in Eq.~(4.27) of
\cite{Martin:1993zk}. We exploit the leading-order RG invariance of
the ratio
$\frac{S}{\tilde\alpha_1}=\frac{S_{\text{GUT}}}{\tilde\alpha_{\text{GUT}}}$
to eliminate several soft masses from the RGE.

In summary, as inputs for the CMM model we need the soft masses
of $\tilde u_R$ and $\tilde d_R$ of the first generations $m_{\tilde
  u_1}^2$, $m_{\tilde d_1}^2$ and $a_1^d$, the mass $m_{\tilde g_3}$
as well as the phase of $\mu$.  Additionally, $\tan\beta$ and the
phase $\xi$ can be chosen as free input parameters, 
but $\tan\beta$ cannot be large because of the bottom Yukawa
  coupling is suppressed by a factor of $M_\text{SO(10)}/M_{\text{Pl}}$.
Initially, we
set {$m_{\tilde u_1}^2=m_{\tilde d_1}^2=M_{\tilde q}$} at the weak scale
and use a three-dimensional polynomial fit for the quantity $S_{\text{GUT}}$.
This fit is computed by initially setting $S_{\text{GUT}}=0$ and obtaining
well convergent values after two runs depending on the variables 
$M_{\tilde q}(M_Z)$, $a_1^d(M_Z)$ and $m_{\tilde g_3}(M_Z)$.

We run up to the Planck scale using the RGE and the unification
conditions specified above.  Then we evolve back from
$M_\text{Pl}$ through SO(10), SU(5) and the MSSM to the electroweak
scale and determine the remaining relevant parameters like soft
masses. We can further now determine the magnitude of the MSSM
  Higgs parameter $\mu$ from the condition of electroweak symmetry
  breaking: With $ m^2_{H_u}$ and $m^2_{H_d}$ from the first
  run we determine $|\mu(M_Z)|$ using
\begin{align}
 \left|\mu\right| = \frac{m^2_{H_u}\sin^2\beta - 
         m^2_{H_d}\cos\beta^2}{\cos(2\beta)}-\frac{1}{2}M_Z,
\end{align}
which is used as input for the second run of the RGE.  The phase of
$\mu$ is left as a free input. With the first run also
  $S_{\text{GUT}}/\tilde\alpha_{\text{GUT}}$ is determined {anew}.  To
stabilize our solution we repeat the RG evolution to the Planck
  scale and back with the input values refined through the first run.
We find good convergence already after two complete runs.

The RGE for the soft SUSY-breaking terms of the first generation
  are given in \eqsand{rgeplanck}{rgeplanck2}. The RGE governing the
  soft terms {of the third generation that are needed
  for the running from the Planck scale back to the electroweak scale}
  are more complicated because of the {flavor} mixing
  stemming from $U_D$ and the involvement of $\tilde y_t$. These
  equations are listed and are discussed in the following
  Secs.~\ref{sec:a} and \ref{sec:m}.

\subsection{RGE of trilinear terms}\label{sec:a}
At the Planck scale we have
\begin{eqnarray}
\widetilde{\mathsf{A}}_1 = a_0 
         \widetilde{\mathsf{Y}}_1 \ , & \qquad 
\widetilde{\mathsf{A}}_2 = a_0 \widetilde{\mathsf{Y}}_2, 
\end{eqnarray} 
so that the trilinear terms are diagonal in the same basis as the
Yukawa couplings. In our basis with diagonal $\widetilde{\mathsf{Y}}_1$,
$\widetilde{\mathsf{Y}}_u$ the matrix $\widetilde{\mathsf{A}}_1$,
$\widetilde{\mathsf{A}}_u$ stays diagonal down to the scale $M_Z$.  It
is therefore sufficient to consider $\tilde A_t := (\tilde
A_u)_{33}$. However, the large atmospheric mixing angle induces a
non-negligible (3,2) element in $\widetilde{\mathsf{A}}_2$, 
$\widetilde{\mathsf{A}}_d$ at $M_Z$. This corresponds to a non-negligible (2,3)
element in $\widetilde{\mathsf{A}}_e$. $(\widetilde{\mathsf{A}}_d)_{32}$
induces novel $\tilde b_L \rightarrow \tilde s_R$ transitions.
\subsubsection*{SO(10)}
{The RGE for $\tilde A_t= (\widetilde{\mathsf{A}}_1)_{33}$ 
is easily obtained from \cite{Martin:1993zk}.
We derive the RGE for $\hat{\tilde{\mathsf{A}}}_2$ in the same
  way as those for $\hat{\tilde{\mathsf{Y}}}_2$ in \eq{rgy}, by generalizing 
  Eqs.~(2.7)--(2.10) of \cite{Martin:1993zk}. The group factors are
  calculated in a straightforward way and can be found e.g.\ in 
  \cite{Jager:2003zz}. The desired equations read
\begin{eqnarray}
\frac{d}{d t} \tilde A_t &=& -\frac{63}{2} \tilde\alpha \left(2
\tilde\alpha \kappa \tilde y_t + \tilde A_t\right) +84 \tilde A_t
|\tilde y_t|^2 \,,\nonumber\\[0.5cm]
\label{EqRGESO10DownATerm} 
\frac{d}{d t} \hat{\tilde{\mathsf{A}}}_2 & = & -\frac{95}{2} \tilde\alpha
\left(2 \tilde\alpha \kappa \hat{\tilde{\mathsf{Y}}}_2 +
  \hat{\tilde{\mathsf{A}}}_2\right) 
 \nonumber\\ && \quad + 10\,  
 \Big(\hat{\tilde{\mathsf{Y}}}_1 \hat{\tilde{\mathsf{Y}}}_1^{\dagger} 
      \hat{\tilde{\mathsf{A}}}_2
  + \hat{\tilde{\mathsf{A}}}_2 U_D \hat{\tilde{\mathsf{Y}}}_1 
    \hat{\tilde{\mathsf{Y}}}_1^{\dagger}
U_D^{\dagger}  + 2\, \hat{\tilde{\mathsf{A}}}_1 \hat{\tilde{\mathsf{Y}}}_1^{\dagger} 
 \hat{\tilde{\mathsf{Y}}}_2 + 2\, \hat{\tilde{\mathsf{Y}}}_2 U_D
\hat{\tilde{\mathsf{Y}}}_1^{\dagger} \hat{\tilde{\mathsf{A}}}_1 U_D^{\dagger}\Big)
\end{eqnarray}} 
\subsubsection*{SU(5)}
Using the RGEs from \cite{Hisano:1998fj} and the rescaling conditions at
the SO(10) scale analogously to the Yukawa couplings, the relevant
equations read
\begin{eqnarray}
  \hat{\tilde{\mathsf{A}}}^{\nu}(t_{\text{SO(10)}}) = 
  \hat{\tilde{\mathsf{A}}}^U(t_{\text{SO(10)}}) \ , 
  && 
 \qquad  (\hat{\tilde{\mathsf{A}}}_2(t_{\text{SO(10)}}))_{\text{SU(5)}}
  = \frac{{v_0} }{M_{\text{Pl}}}(\hat{\tilde{\mathsf{A}}}_2(t_{\text{SO(10)}}))_{\text{SO(10)}}
\end{eqnarray} 
\begin{eqnarray}
\frac{d}{d t} \tilde A_t &=& -\frac{96}{5}\tilde\alpha
\left(2\tilde\alpha\kappa \tilde y_t + \tilde A_t\right) + 2 \tilde
y_t \left(\tilde y_{\nu_3}^* \tilde A_{\nu_3} + 4 \tilde y_b^* \tilde
A_b\right) \nonumber\\
&& + \tilde A_t \left(27 |\tilde y_t|^2 + |\tilde y_{\nu_3}|^2 + 3
|\tilde y_b|^2 \right) \,, \nonumber
\end{eqnarray}
\begin{eqnarray}
\frac{d}{d t} \hat{\tilde{\mathsf{A}}}_2 & = & -\frac{84}{5} \tilde\alpha
\left(2 \tilde\alpha\kappa \hat{\tilde{\mathsf{Y}}}_2 +
\hat{\tilde{\mathsf{A}}}_2\right) + \left( 4 |\tilde y_b|^2 + 10
\hat{\tilde{\mathsf{Y}}}_2 \hat{\tilde{\mathsf{Y}}}_2^{\dagger} +3 \hat{\tilde{\mathsf{Y}}}_1
\hat{\tilde{\mathsf{Y}}}_1^{ \dagger} \right) \hat{\tilde{\mathsf{A}}}_2 \nonumber\\ 
&& + 8 \hat{\tilde{\mathsf{A}}}_2 \hat{\tilde{\mathsf{Y}}}_2^{\dagger} \hat{\tilde{\mathsf{Y}}}_2
+\hat{\tilde{\mathsf{A}}}_2 U_D \hat{\tilde{\mathsf{Y}}}_\nu^{\dagger}
\hat{\tilde{\mathsf{Y}}}_{\nu} U_D^{\dagger} + 8 \tilde y_b^*\tilde A_b
\hat{\tilde{\mathsf{Y}}}_2 \nonumber\\ 
&& + 6 \hat{\tilde{\mathsf{A}}}_1 \hat{\tilde{\mathsf{Y}}}_1^{\dagger} \hat{\tilde{\mathsf{Y}}}_2
+ 2 \hat{ \tilde{\mathsf{Y}}}_2 U_D \hat{\tilde{\mathsf{Y}}}_\nu^{\dagger}
\hat{\tilde{\mathsf{A}}}_{\nu} U_D^{\dagger } \,,\nonumber\\[0.5cm] 
\frac{d}{d t} \hat{\tilde{\mathsf{A}}}_{\nu} & = & -\frac{48}{5} \tilde\alpha
\left(2 \tilde\alpha \kappa \hat{\tilde{\mathsf{Y}}}_{\nu} +
\hat{\tilde{\mathsf{A}}}_{\nu} \right) + \left(3 |\tilde y_t|^2 + |\tilde
y_{\nu_3}|^2 + 7 \hat{\tilde{\mathsf{Y}}}_{\nu} \hat{\tilde{\mathsf{Y}}}_\nu^{\dagger}\right)
 \hat{\tilde{\mathsf{A}}}_{\nu} \nonumber\\
&& +6 \tilde y_t^* \tilde A_t \hat{\tilde{\mathsf{Y}}}_{\nu} + 2 \tilde
y_{\nu_3}^* \tilde A_{\nu_3} \hat{\tilde{\mathsf{Y}}}_{\nu} + 4 \hat{\tilde{\mathsf{A}}}^{\nu}
 U_D^{\dagger} \hat{\tilde{\mathsf{Y}}}_2^{\dagger} \hat{\tilde{\mathsf{Y}}}_2
U_D \nonumber\\
&& + 11 \hat{\tilde{\mathsf{A}}}_{\nu} \hat{\tilde{\mathsf{Y}}}_\nu^{\dagger} 
\hat{\tilde{\mathsf{Y}}}_{\nu} + 8 \hat{\tilde{\mathsf{Y}}}_{\nu} U_D^{\dagger}
\hat{\tilde{\mathsf{Y}}}_2^{\dagger} \hat{\tilde{\mathsf{A}}}_2 U_D
\end{eqnarray} 
Here again $\tilde A_t$, $\tilde A_b$ and $\tilde A_{\nu_3}$ are the
(33) entries of the matrices $\hat{\tilde{\mathsf{A}}}_1$,
$\hat{\tilde{\mathsf{A}}}_2$ and $\hat{\tilde{\mathsf{A}}}_{\nu}$.
\subsubsection*{MSSM}
We integrate out the righthanded neutrino
at the GUT scale and use the RGEs from \cite{Martin:1993zk}. Furthermore, we employ the SU(5) relation 
$\mathsf{A}_e(t_{\text{GUT}})=(\mathsf{A}_d(t_{\text{GUT}}))^T$ 
and evolve the trilinear terms down to the scale $M_Z$.
\begin{eqnarray}
\frac{d}{d t} \tilde A_t & =& \tilde A_t \left(8 |\tilde y_t|^2 + |\tilde y_b|^2 - \frac{16}{3} \tilde\alpha_3 - 3 \tilde\alpha_2 - \frac{13}{15} \tilde\alpha_1\right) \nonumber\\
&& +\tilde y_t \left(10 \tilde y_t^* \tilde A_t + 2 \tilde y_b^* \tilde A_b - \frac{32}{3} \tilde\alpha_3^2\kappa - 6 \tilde\alpha_2^2\kappa -\frac{26}{15}\tilde\alpha_1^2\kappa\right) \,,\nonumber\\[0.5cm]
\frac{d}{d t} \hat{\tilde{\mathsf{A}}}_d & = & \left(3 |\tilde y_b|^2 + |\tilde y_{\tau}|^2 + 5 \hat{\tilde{\mathsf{Y}}}_d^{*} (\hat{\tilde{\mathsf{Y}}}_d)^T + \hat{\tilde{\mathsf{Y}}}_{u}^* (\hat{\tilde{\mathsf{Y}}}_u)^T -\frac{16}{3} \tilde\alpha_3 - 3\tilde\alpha_2 - \frac{7}{15} \tilde\alpha_1 \right)\hat{\tilde{\mathsf{A}}}_d \nonumber\\
&& +\left(6 \tilde y_b^* \tilde A_b + 2 \tilde y_{\tau}^* \tilde A_{\tau} +4 \hat{\tilde{A}}_d \hat{\tilde{\mathsf{Y}}}_d^{\dagger} + 2 \hat{\tilde{\mathsf{A}}}_u \hat{\tilde{\mathsf{Y}}}_u^{\dagger} -\frac{32}{3} \tilde\alpha_3^2\kappa - 6 \tilde\alpha_2^2\kappa - \frac{14}{15} \tilde\alpha_1^2\kappa \right)\hat{\tilde{\mathsf{Y}}}_d \,,\nonumber\\[0.5cm]
\frac{d}{d t} \hat{\tilde{\mathsf{A}}}_e & = & \left(3 |\tilde y_b|^2 + |\tilde y_{\tau}|^2 + 5 \hat{\tilde{\mathsf{Y}}}_e^{*} (\hat{\tilde{\mathsf{Y}}}_e)^T - 3 \tilde\alpha_2 - \frac{9}{5} \tilde\alpha_1 \right)\hat{\tilde{\mathsf{A}}}_e \nonumber\\
&& +\left(6 \tilde y_b^* \tilde A_b + 2 \tilde y_{\tau}^* \tilde A_{\tau} + 4 \hat{\tilde{\mathsf{A}}}_e \hat{\tilde{\mathsf{Y}}}_e^{\dagger} - 6 \tilde\alpha_2^2\kappa - \frac{18}{10}\tilde\alpha_1^2\kappa \right) \hat{\tilde{\mathsf{Y}}}_e \,.
\end{eqnarray}
\subsection{RGE for soft masses}\label{sec:m}
Employing the universality conditions of \eq{eq:cmm-universal-mass} at
the Planck scale, the soft masses stay diagonal in the basis {with
  diagonal $\widetilde{\mathsf{Y}}_u$}.  We list the RGEs for the first
and second generation (index 1) and the third generation (index 3), which
is {separates due to the} large top Yukawa coupling. 
\subsubsection*{SO(10)}
We use the RGE from appendix B.1 of \cite{Jager:2003zz}.
\begin{eqnarray}
\frac{d}{d t}m^2_{\widetilde{\text{16}_1}}&=&-45\kappa^2\tilde\alpha^3 \,,\nonumber\\
\frac{d}{d t}m^2_{\widetilde{\text{16}_3}}&=&-45\kappa^2\tilde\alpha^3 +20 |\tilde y_t|^2\left[2m^2_{\widetilde{\text{16}_3}}+m^2_{\text{10}}\right] + 20|\tilde A_t|^2 \,,\nonumber\\
\frac{d}{d t}m^2_{\text{10}_H}&=&-36\kappa^2\tilde\alpha^3 +16|\tilde y_t|^2 \left[2 m^2_{\widetilde{\text{16}_3}} + m^2_{\text{10}} \right] + 16 |\tilde A_t|^2 \,,\nonumber\\
\frac{d}{d t}m^2_{\text{10}_H^\prime}&=&-36\kappa^2\tilde\alpha^3 \,.
\end{eqnarray} 
\subsubsection*{SU(5)}
After taking into account the D-term splitting in
Eq.~(\ref{eq:d-terms}), we evolve the soft masses down to the GUT scale
using the RGEs from \cite{Hisano:1998fj}.  For the numerical solution we
can safely set $\tilde y_{\nu_3}=\tilde y_t$.
\begin{eqnarray}
\frac{d}{d t} m^2_{\tilde\Phi_1}&=&-\frac{96}{5}\kappa^2\tilde\alpha^3 + 8(U_D^\dagger \hat{\tilde{\mathsf{A}}}_2^{\dagger} \hat{\tilde{\mathsf{A}}}_2 U_D)_{11} \nonumber\\
&&+ 8 |(U_D)_{31}|^2 |\tilde y_b|^2\left[m^2_{\tilde\Phi_1} + m^2_{H'} + m^2_{\tilde \Psi_3}\right]\,,\nonumber\\
\frac{d}{d t} m^2_{\tilde\Phi_3}&=&-\frac{96}{5}\kappa^2\tilde\alpha^3 + 8(U_D^\dagger \hat{\tilde{\mathsf{A}}}_2^{\dagger} \hat{\tilde{\mathsf{A}}}_2 U_D)_{33} + 2|\tilde A_{\nu_3}|^2 + 2|\tilde y_{\nu_3}|^2\left[m^2_{\tilde\Phi_3} + m^2_{H} + m^2_{\tilde N_3}\right]\nonumber\\
&&+ 8 |(U_D)_{33}|^2 |\tilde y_b|^2\left[m^2_{\tilde\Phi_3} + m^2_{H'} + m^2_{\tilde \Psi_3}\right]\,,\nonumber \\
\frac{d}{d t} m^2_{\tilde\Psi_1}&=&-\frac{144}{5}\kappa^2\tilde\alpha^3 \,,\nonumber\\
\frac{d}{d t} m^2_{\tilde\Psi_3} &=&-\frac{144}{5}\kappa^2\tilde\alpha^3 + 4|\tilde y_b|^2\left[m^2_{\tilde\Psi_3} + m^2_{H'} + (U_D m^2_{\tilde\Phi} U_D^\dagger)_{33}\right]\nonumber\\
&&+6|\tilde y_t|^2\left[2 m^2_{\tilde\Psi_3}+m^2_H\right] + 4(|( \hat{\tilde{\mathsf{A}}}_2)_{32}|^2+|\tilde{A}_{b}|^2) + 6 |\tilde{A}_{t}|^2 \,,\nonumber\\
\frac{d}{d t} m^2_{\tilde N_1} &=& 0 \,,\nonumber\\
\frac{d}{d t} m^2_{\tilde N_3} &=& 10|\tilde y_{\nu_3}|^2\left[m^2_{\tilde N_3} + m^2_H + m^2_{\tilde\Phi_3}\right] + 10(|(\hat{\tilde{\mathsf{A}}}_{\nu})_{31}|^2+ (\hat{\tilde{\mathsf{A}}}_{\nu})_{32}|^2+|\tilde A_{\nu_3}|^2) \,,\nonumber\\
\frac{d}{d t} m^2_{H} &=& -\frac{96}{5} \kappa^2\tilde\alpha^3 + 6|\tilde y_t|^2\left[2 m^2_{\tilde\Psi_3} + m^2_{H}\right] + 2 |\tilde y_{\nu_3}|^2\left[m^2_{\tilde\Phi_3} + m^2_{\tilde N_3} +  m^2_H\right] \nonumber\\
&& + 2(|(\hat{\tilde{\mathsf{A}}}_{\nu})_{31}|^2+|(\hat{\tilde{\mathsf{A}}}_{\nu})_{32}|^2+|\tilde A_{\nu_3}|^2) + 6|\tilde A_t|^2) \,,\nonumber\\
\frac{d}{d t} m^2_{H'} &=& -\frac{96}{5} \kappa^2\tilde\alpha^3 + 8|\tilde y_b|^2\left[m^2_{\tilde\Psi_3}+ m^2_{H'} + (U_D m^2_{\tilde\Psi} U_D^\dagger)_{33}\right] \nonumber\\
&&+ 8(|( \hat{\tilde{\mathsf{A}}}_2)_{32}|^2+|\tilde{A}_{b}|^2) \,.
\end{eqnarray} 
\subsubsection*{MSSM}
In the last step, we evolve the soft masses down to $M_Z$ using the RGE
from \cite{Martin:1993zk}.
\begin{eqnarray}
\frac{d}{d t} m^2_{\tilde q_1} &=& -\frac{32}{3}\kappa^2\tilde\alpha_3^3 - 6\kappa^2\tilde\alpha_2^3-\frac{2}{15}\kappa^2\tilde\alpha_1^3 +\frac{1}{5}\frac{S_{\text{GUT}}}{\tilde\alpha_{\text{GUT}}}\tilde\alpha_1^2\,,\nonumber\\
\frac{d}{d t} m^2_{\tilde q_3} &=&-\frac{32}{3}\kappa^2\tilde\alpha_3^3 - 6\kappa^2\tilde\alpha_2^3-\frac{2}{15}\kappa^2\tilde\alpha_1^3 +\frac{1}{5}\frac{S_{\text{GUT}}}{\tilde\alpha_{\text{GUT}}}\tilde\alpha_1^2 \nonumber\\
&& + 2|\tilde y_t|^2\left[m^2_{\tilde q_3} + m^2_{H_u} + m^2_{\tilde u_3}\right] + 2|\tilde y_b|^2\left[m^2_{\tilde q_3} + m^2_{H_d} + (U_D m^2_{\tilde d} U_D^\dagger)_{33}\right]\nonumber\\
&& + 2(|\tilde A_t|^2 + |( \hat{\tilde{\mathsf{A}}}_d)_{32}|^2+|\tilde{A}_{b}|^2) \,,\nonumber\\
\frac{d}{d t} m^2_{\tilde u_1} &=& -\frac{32}{3}\kappa^2\tilde\alpha_3^3 - \frac{32}{15}\kappa^2\tilde\alpha_1^3-\frac{4}{5}\frac{S_{\text{GUT}}}{\tilde\alpha_{\text{GUT}}}\tilde\alpha_1^2\,,\nonumber\\
\frac{d}{d t} m^2_{\tilde u_3} &=& -\frac{32}{3}\kappa^2\tilde\alpha_3^3 - \frac{32}{15}\kappa^2\tilde\alpha_1^3-\frac{4}{5}\frac{S_{\text{GUT}}}{\tilde\alpha_{\text{GUT}}}\tilde\alpha_1^2 \nonumber\\
&& +4|\tilde y_t|^2\left[m^2_{\tilde u_3} + m^2_{\tilde q_3} +
  m^2_{H_u}\right] + 4\tilde |A_t|^2\,,\nonumber 
\end{eqnarray}
\begin{eqnarray}
\frac{d}{d t} m^2_{\tilde d_1} &=& - \frac{32}{3}\kappa^2\tilde\alpha_3^3-\frac{8}{15}\kappa^2\tilde\alpha_1^3+\frac{2}{5}\frac{S_{\text{GUT}}}{\tilde\alpha_{\text{GUT}}}\tilde\alpha_1^2 \nonumber\\
&& +4|\tilde y_b|^2 |(U_D)_{31}|^2\left[m^2_{\tilde d_1} + m^2_{\tilde q_3} + m^2_{H_d}\right] + 4(U_D^\dagger \hat{\tilde{\mathsf{A}}}_d^{\dagger}\hat{\tilde{\mathsf{A}}}_d U_D)_{11}\,,\nonumber\\
\frac{d}{d t} m^2_{\tilde d_3} &=& - \frac{32}{3}\kappa^2\tilde\alpha_3^3-\frac{8}{15}\kappa^2\tilde\alpha_1^3+\frac{2}{5}\frac{S_{\text{GUT}}}{\tilde\alpha_{\text{GUT}}}\tilde\alpha_1^2 \nonumber\\
&& +4|\tilde y_b|^2 |(U_D)_{33}|^2\left[m^2_{\tilde d_3} + m^2_{\tilde
    q_3} + m^2_{H_d}\right] + 4(U_D^\dagger
\hat{\tilde{\mathsf{A}}}_d^{\dagger}\hat{\tilde{\mathsf{A}}}_d
U_D)_{33}\,,\nonumber \\
\frac{d}{d t} m^2_{\tilde l_1} &=& -6\kappa^2\tilde\alpha_2^3 - \frac{6}{5}\kappa^2\tilde\alpha_1^3-\frac{3}{5}\frac{S_{\text{GUT}}}{\tilde\alpha_{\text{GUT}}}\tilde\alpha_1^2 \nonumber\\
&& +2 |\tilde y_\tau|^2 |U_{31}|^2\left[m^2_{\tilde l_1} + m^2_{H_d} + m^2_{\tilde l_3}\right] + 2(U^\dagger \hat{\tilde{\mathsf{A}}}_e \hat{\tilde{\mathsf{A}}}_e^{\dagger} U)_{11}\,,\nonumber\\
\frac{d}{d t} m^2_{\tilde l_3} &=& -6\kappa^2\tilde\alpha_2^3 - \frac{6}{5}\kappa^2\tilde\alpha_1^3-\frac{3}{5}\frac{S_{\text{GUT}}}{\tilde\alpha_{\text{GUT}}}\tilde\alpha_1^2 \nonumber\\
&& 2 |\tilde y_\tau|^2 |U_{33}|^2\left[2 m^2_{\tilde l_3} + m^2_{H_d}\right] + 2(U^\dagger \hat{\tilde{\mathsf{A}}}_e \hat{\tilde{\mathsf{A}}}_e^{\dagger} U)_{33}\,,\nonumber\\
\frac{d}{d t} m^2_{\tilde e_1} &=& - \frac{24}{5}\kappa^2\tilde\alpha_1^3 + \frac{6}{5}\frac{S_{\text{GUT}}}{\tilde\alpha_{\text{GUT}}}\tilde\alpha_1^2\,,\nonumber\\
\frac{d}{d t} m^2_{\tilde e_3} &=& - \frac{24}{5}\kappa^2\tilde\alpha_1^3 + \frac{6}{5}\frac{S_{\text{GUT}}}{\tilde\alpha_{\text{GUT}}}\tilde\alpha_1^2 \nonumber\\
&& +4|\tilde y_\tau|^2\left[m^2_{\tilde e_3} + m^2_{H_d} + (U m^2_{\tilde l} U^\dagger)_{33}\right] + 4(|(\hat{\tilde{\mathsf{A}}}_e)_{23}|^2+|\tilde{A}_{\tau}|^2)\,,\nonumber\\
\frac{d}{d t} m^2_{H_u}&=&-6\kappa^2\tilde\alpha_2^3-\frac{6}{5}\kappa^2\tilde\alpha_1^3+\frac{3}{5}\frac{S_{\text{GUT}}}{\tilde\alpha_{\text{GUT}}}\tilde\alpha_1^2 \nonumber\\
&&+ 6|\tilde y_t|^2 \left[m^2_{H_u} + m^2_{\tilde q_3} + m^2_{\tilde u_3}\right] + 6|\tilde A_{t}|^2+\,,\nonumber\\
\frac{d}{d t}m^2_{H_d} &=& -6\kappa^2\tilde\alpha_2^3 -\frac{6}{5}\kappa^2\tilde\alpha_1^3 -\frac{3}{5}\frac{S_{\text{GUT}}}{\tilde\alpha_{\text{GUT}}}\tilde\alpha_1^2 \nonumber\\
&&+ 6|\tilde y_b|^2\left[m^2_{H_d} + m^2_{\tilde q_3} + (U_D m^2_{\tilde d} U_D^\dagger)_{33}\right] + 2 |\tilde y_\tau|^2\left[m^2_{H_d} + m^2_{\tilde l_3} + (U m^2_{\tilde l} U^\dagger)_{33}\right] \nonumber\\
&& +6(|\tilde A_{b}|^2+|(\hat{\tilde{\mathsf{A}}}_d)_{32}|^2) + 
2(|\tilde A_{\tau}|^2+|(\hat{\tilde{\mathsf{A}}}_e)_{23}|^2) \,.
\end{eqnarray} 
\boldmath
\subsection{Parameters at $M_\text{GUT}$}\label{se:parMGUT}
\unboldmath
The philosophy of the CMM model is somewhat different from that of the
CMSSM.  Although both need only a few input parameters and are in a
sense minimal flavor violating, the CMSSM assumes flavor universality at
the GUT scale with quark and lepton flavor structures being
unrelated. By contrast, the CMM model invokes universality (see
Eq.~(\ref{eq:cmm-universal})) at a more natural scale, namely
$M_\text{Pl}$. All flavor violation stems from an non-renormalizable
term related to $\mathsf{Y}_d$ due to the assumption that the Majorana
mass matrix and the up Yukawa coupling are simultaneously diagonalizable.
Furthermore, the CMM model is minimal in the sense that it is only
constructed with Higgs representations that are needed for symmetry
breaking anyway.

Contrary to the CMSSM, at the GUT scale universality is already broken
in the CMM model due to the running $M_\text{Pl}\to M_\text{SO(10)}\to
M_\text{GUT}$. We illustrate the difference with the input parameters
$M_{\tilde q} = 1500~$GeV, $m_{\tilde g_3} = 500~$GeV,
$a_1^d(M_Z)/M_{\tilde q} = 1.5$, $\arg(\mu) = 0$ and $\tan\beta = 6$.
With our running procedure the universal parameters at the Planck
scale have the values:
\begin{align}
  a_0 = 1273~\text{GeV},\qquad m_0 = 1430~\text{GeV}, 
 \qquad {m_{\tilde{g}} =184~\text{GeV}.}
\end{align}
Using the super-CKM basis (as denoted by the hat) for the trilinear
terms and the up basis for masses,
we already arrive at the following non-universal parameters at the GUT
scale:
{\allowdisplaybreaks
\begin{subequations}
\begin{align}
& \mathsf{\hat{\tilde{A}}}_u(M_\text{GUT}) = \begin{pmatrix}
		  0& 0& 0\\	
		  0& 0& 0\\ 
		  0 & 0 & 46
		\end{pmatrix}\,{\gev}, \quad
\mathsf{\hat{\tilde{A}}}_d(M_\text{GUT}) = \begin{pmatrix}
		  0& 0& 0\\	
		  0& 0& 0\\ 
		  0 & 0.3 & -3.5
		\end{pmatrix}\,{\gev},\\
&\mathsf{\hat{\tilde{A}}}_\nu(M_\text{GUT}) = \begin{pmatrix}
		  0& 0& 0\\	
		  0& 0& 0\\ 
-0.0013 & 	0.0023 & 	43.4 
		\end{pmatrix}\,{\gev},\quad\\
& m_{\tilde \Phi}(M_\text{GUT}) = \diag\left(1426, 1426, 1074\right)~\text{GeV},\\
& m_{\tilde \Psi}(M_\text{GUT}) = \diag\left(1444,1444,1077\right)~\text{GeV},\\
& m_{\tilde N}(M_\text{GUT}) = \diag\left(1459,1459,1078\right)~\text{GeV},\\ 
& m_{H_u}(M_\text{GUT}) = 1126~\text{GeV},\quad m_{H_d}(M_\text{GUT}) =
1446~\text{GeV}, \\
& {m_{\tilde{g}}(M_\text{GUT})=211~\text{GeV}.} 
\end{align}
\end{subequations}}
With $\tilde y_t(M_\text{GUT}) = 0.046$ and $\tilde y_b(M_\text{GUT})
= -0.0026$ we can now no longer write $\mathsf{A} = a_0 \mathsf{Y}$,
especially $\mathsf{A}_d$ has already developed an off-diagonal entry
inducing $\tilde s_R\to \tilde b_L$-transitions.  Moreover, the third
generation masses already separate significantly from those of the
first two generations at the GUT scale.

{The idea of universal soft breaking terms at $M_\text{Pl}$ and
  flavor-violation from $y_t$-driven RG running above $M_\text{GUT}$
  has been studied by many authors, both in SU(5) and SO(10) scenarios
  \cite{Barbieri:1995rs,Moroi:2000tk,Chang:2002mq, 
        Harnik:2002vs,Hisano:2003bd,Jager:2003xv,Cheung:2007pj,Hisano:2008df,
        kpy,Buras:2010pm}. 
 A detailed comparison of our results with the literature will be 
 given in Sec.~\ref{sec:disussion}.}

\section{Observables}
\label{se:observables}
In this Section, we briefly summarize the observables that are used to
constrain the CMM model parameter space.

\subsection{\boldmath{$B_s-\ovl{B}_s$} Mixing}
\bbs\ oscillations are governed by the Schr\"odinger equation
\begin{align}
  i \frac{d}{dt}
  \begin{pmatrix}
    \ket{B_s(t)} \cr \ket{\bar{B}_s (t)}
  \end{pmatrix}
  = \left( \mathsf{M}^s - \frac{i}{2} \mathsf{\Gamma}^s \right)
  \begin{pmatrix}
    \ket{B_s(t)} \cr \ket{\bar{B}_s (t)} 
  \end{pmatrix}
\end{align}
with the mass matrix $\mathsf{M}^s$ and the decay matrix
$\mathsf{\Gamma}^s$.  The physical eigenstates $\ket{B_{H,L}}$ with the
masses $M_{H,L}$ and the decay rates $\Gamma_{H,L}$ are obtained by
diagonalizing $\mathsf{M}^s-i\mathsf{\Gamma}^s/2$.  The physical
observables are the mass and width differences as well as the CP phase,
\begin{align}
  \Delta M_s & = M^s_H - M^s_L = 2 \abs{\mathsf{M}_{12}^s} , \nonumber
  \\
  \Delta\Gamma_s & = \Gamma^s_L-\Gamma^s_H = 2
  \abs{\mathsf{\Gamma}_{12}^s} \cos\phi_s \,, \nonumber
  \\
  \phi_s & = \text{arg} \left(
    -\frac{\mathsf{M}^s_{12}}{\mathsf{\Gamma}^s_{12}} \right) .
\label{eq:defphi}
\end{align}

In {the CMM model}, there are two operators contributing to the
oscillations,
\begin{subequations}
  \begin{align}
    \label{eq:bs-mixing-left}
    \mathcal{O}_L & = \ovl{s}_{L,\alpha}\, \gamma_\mu\, b_{L,\alpha}\,
    \ovl{s}_{L,\beta}\, \gamma^\mu\, b_{L,\beta}
    \\
    \label{eq:bs-mixing-right}
    \mathcal{O}_R & = \ovl{s}_{R,\alpha}\, \gamma_\mu\, b_{R,\alpha}\,
    \ovl{s}_{R,\beta}\, \gamma^\mu\, b_{R,\beta} \ .
  \end{align}
\end{subequations}
In the standard model, only the left-handed operator
(\ref{eq:bs-mixing-left}) is present due to the absence of the
right-handed vector bosons.  With weak-scale supersymmetry, however, the
vertices in Fig.~\ref{fig:fc-vertices} contribute to both
$\mathcal{O}_L$ and $\mathcal{O}_R$ with the quark-squark-gluino vertex
in Fig.~\ref{fig:fc-vertices:a} dominating.

{The $B_s-\ovl{B}_s$ oscillations are governed by}
\begin{align}
  \label{eq:ms12}
  \mathsf{M}^s_{12,\,\text{CMM}} & = \frac{G_F^2 M_W^2 M_{B_s}}{12\pi^2}
  \left(f^2_{B_s} \hat{B}_{B_s}\right) \left(V_{ts}^\ast V_{tb}\right)^2
  \left(C_L(\mu_b) + C_R(\mu_b)\right) 
  \;.
\end{align}
Here $G_F$ is the Fermi constant, $M_{B_s}$ and $M_W$ are the masses of
$B_s$ meson and $W$-boson, respectively. {The renormalization scale
  entering the Wilson coefficients $C_{L,R}$ is $\mu_b\sim m_b$.}  The
long-distance QCD effects are contained in the {equal} hadronic
matrix element of ${\cal O}_{L,R}$ and are parametrized
by 
\begin{align}\label{equ:fBs}
  f_{B_s} \sqrt{\hat{B}_{B_s}} = \left(0.2580 \pm 0.0195\right) \,\text{GeV}
  \,,
\end{align}
where we use the values listed in \cite{Lenz:2010gu}: $f_{B_s} =
228\pm3\pm17~$MeV and $\hat{B}_{B_s} = 1.28\pm0.02\pm0.03$.  Finally,
the coefficients $C_L$ and $C_R$ read\footnote{Note, that in
  \cite{Trine:2009ns} $C_{L,R}$ include the factor r = 0.985 which removes
  the NLO QCD corrections to $S_0(x_t)$ in the SM.}
\begin{align}
  C_L(\mu_b) & = \eta_B F_{tt} \,,\\
C_R(\mu_b) & = \frac{\left(U_D^{23\ast}
U_D^{33}\right)^2}{\left(V^*_{ts} V_{tb}\right)^2}
  \frac{8\pi^2\alpha_s^2(M_Z)}{G_F^2 M_W^2 m^2_{\tilde g_3}}
 \eta_B
  S^{(\tilde g)}(x,y),
\label{eq:cr}
\end{align}
where $\eta_B=0.55$ \cite{bjw}, the function $F_{tt}$ is given {e.g.}\ 
in Eq.~(4.5) of \cite{Buras:2000qz}
and $S^{(\tilde g)}(x,y)$ denotes the loop function
\begin{align}
  S^{(\tilde g)}(x,y) & = \frac{11}{18}\left[G(x,x) + G(y,y) -
    2G(x,y))\right] - {\frac{2}{9}}\left[F(x,x) + F(y,y) -2F(x,y)\right] ,
  \nonumber
  \\
  F(x,y) & = \frac{1}{y-x} \left[\frac{x \ln x}{(x-1)^2} - \frac{1}{x-1}
    - (x \leftrightarrow y)\right] , \nonumber
  \\
  G(x,y) & = \frac{1}{x-y}\left[\frac{x^2 \ln x}{(x-1)^2} -
    \frac{1}{x-1} - (x \leftrightarrow y)\right] , \mspace{60mu} x =
  \frac{m^2_{\tilde d_2}}{m^2_{\tilde g_3}} \,, \quad y =
  \frac{m^2_{\tilde d_3}}{m^2_{\tilde g_3}} \ .
\end{align}
{Next we insert $U_D^{i3}$ from \eq{eq:ThetaLR} into 
\eq{eq:cr} to make the dependence on the new CP phase $\xi$ explicit:}
\begin{equation}\label{equ:WilsonBsBsbar}
 {C = C_L +} e^{-2i\xi}\left|C_R^\text{CMM}\right|,
\end{equation}
{In \eqsand{eq:cr}{equ:WilsonBsBsbar} we have, in the spirit of
  this paper, concentrated on the dominant new effect involving large
  parameters (namely $\xi$ and the atmospheric neutrino mixing
  angle). Among the neglected effects are the MFV-like contributions
  proportional to $V_{ts}^{*2}$ involving left-handed squarks and
  gluinos.  These contributions are not only small in magnitude
  compared to the second term in \eq{equ:WilsonBsBsbar} (a few percent
  of the SM coefficient), they are also in phase with the SM
  contribution and do not alter the CP asymmetries in \bbms. The MFV
  boxes involving charged Higgs bosons and those with charginos and
  squarks could be neglected as well, but are nevertheless included in
  our analysis through the function $F_{tt}$ of \cite{Buras:2000qz}. }
The free phase $\xi$ is essential: First, it is the source of a
  {possibly} large CP phase $\arg C$ and second, it {may tame} the CMM
  contribution to $\Delta M_s$, which for $\xi= 0$ can easily exceed
the experimental bound. But with a non-zero $\xi$ {the two
  contributions in \eq{equ:WilsonBsBsbar} can be arranged to keep
  $|C|$ in the range complying within the allowed region for $\Delta
  M_s$.}  {Since} $\xi$ and $\xi + \pi$
cannot be distinguished in \bbms,  {we} only consider the case $\xi\in[0,
  \pi]$, {noting that} $b\to s\gamma$ depends only weakly on this
phase. Mixing-induced CP asymmetries {in $b\to s$ penguin decays}
constitute a possibility to distinguish between $\xi$ and $\xi + \pi$,
with $a_\text{CP}^\text{mix}\left(B_d\to\phi
K_S\right)<a_\text{CP}^\text{mix,SM}\left(B_d\to\phi K_S\right)$
favoring $\xi\in[0,\pi]$.

The current experimental status is as follows.  The CDF experiment
measured the mass difference to be \cite{Abulencia:2006ze},
\begin{align}
  \Delta M_s = \left( 17.77 \pm 0.10 \left(\text{stat.}\right) \pm 0.07
    \left(\text{syst.}\right) \right) \text{ps}^{-1} ,
\end{align}
in agreement with the D\O\ range and the SM prediction \cite{Lenz:2006hd},
\begin{align}
  \Delta M_s^{\text{SM}} & = \left(19.30\pm6.68\right) \text{ps}^{-1} .
\end{align}
Combining both experiments gives \cite{PDG}
\begin{align}\label{equ:DeltaMsPDG}
 \Delta M_s^\text{PDG}  & = \left(17.77\pm0.12\right)~\text{ps}^{-1}.
\end{align}
The SM CP phase {in \eq{eq:defphi}} is small \cite{Lenz:2006hd,Lenz:2010gu},
\begin{align}
  \phi_s^{\text{SM}} & = \left(4.3^{+3.5}_{-3.1}\right) \times 10^{-3} .
\end{align}
The CP phase has been constrained by both the CDF and
D\O\ collaborations {in different ways. The angular analysis of tagged 
$B_s\to J/\psi\phi$ decays determines $2\beta_s$, with SM   
value $\beta_s^\text{SM} = -\arg
\left(-\frac{V_{ts}^\ast V_{tb}}{V_{cs}^\ast V_{cb}}\right)\linebreak =
0.01811^{+0.0085}_{-0.00082}$ \cite{Lenz:2010gu}. Neglecting the 
tiny $ \phi_s^{\text{SM}}$, new physics in $\mathsf{M}_{12}^s$ will lead to
$2\beta_s = 2 \beta_s^\text{SM} - \phi_s$ and $\phi_s$ in 
\eq{eq:defphi} can a-priori be of order 1.   
The new results for $2\beta_s$} presented in summer 2010 are
given as \cite{CDFPhis,D0Phis}
\begin{subequations}
  \begin{align}
    -2 \beta_s^\text{CDF}  \equiv - 2 \beta_s^\text{SM} +
    {\phi_s} &\in \left[-1.04,-0.04\right] \cup \left[-3.10,-2.16\right]\quad (68\%\ \text{CL}) ,
    \\
    & \in \left[-\pi,-1.78\right]\cup \left[-1.36,0.26\right]\cup\left[2.88,\pi\right]\quad(95\%\ \text{CL})\label{equ:CDF95CL}\\
    \phi_s^\text{D\O}  \equiv -2 \beta_s^\text{SM} + {\phi_s} 
    =& -0.76^{+0.38}_{-0.36}(\text{stat})\pm 0.02(\text{syst})\\
    &\in \left[-1.65, 0.24\right]\cup \left[1.14,2.93\right]\quad (95\%\ \text{CL}) \label{equ:D095CL}
     \;.
  \end{align}
\end{subequations}
So far there is no combination of the CDF and D\O\ results available.
Recently D\O\ has measured the inclusive dimuon asymmetry $A^b =
\frac{N_b^{++} - N_b^{--}}{N_b^{++} + N_b^{--}}$ using $6.1~fb^{-1}$ of
integrated luminosity where $N_b^{++}$ counts the number of
$\left(B^0(t), \bar B^0(t)\right)\to \left(\mu^+, \mu^+\right)$ and
$N_b^{--}$ decays into $ \left(\mu^-, \mu^-\right)$
\cite{Abazov:2010hv}.  The same asymmetry can also be obtained from
semileptonic decays $a_\text{fs} = \frac{\Gamma\left(\bar B^0\to X\ell^+
    \nu_\ell\right)-\Gamma\left( B^0\to \bar X\ell^-
    \bar\nu_\ell\right)}{\Gamma\left(\bar B^0\to X\ell^+
    \nu_\ell\right)+\Gamma\left( B^0\to \bar X\ell^-
    \bar\nu_\ell\right)} =A^b$.  {The two measurements combine} to
\cite{Abazov:2010hv}
\begin{align}
 a_\text{fs} = -0.00957 \pm 0.00251 \pm 0.00146
\end{align}
for a mixture between $B_d$- and $B_s$-mesons with
\begin{align}
 a_\text{fs} = \left(0.506\pm 0.043\right)a_{fs}^d+ \left(0.494\pm0.043\right) a_{fs}^s.
\end{align}
Comparison with the predicted SM value $a_\text{fs}^\text{SM} =
\left(-0.23^{+0.05}_{-0.06}\right)\cdot 10^{-3}$ \cite{Lenz:2006hd}
yields a $3.2\sigma$ discrepancy. Averaging with the CDF result
$a_\text{fs} = 0.008\pm 0.0090 \pm 0.0068$ \cite{CDFdimuon} results in a
$2.9\sigma$ {deviation from the SM}:
\begin{align}
 a_\text{fs} = -0.0085\pm 0.0028\qquad\qquad \text{at 68\% CL}.
\end{align}
The relation with the CP phase $\phi_s$ is given by $a_\text{fs}^s = \frac{\left|\mathsf{\Gamma}_{12}^s\right|}{\left|\mathsf{M}_{12}^s\right|}\sin\phi_s$. Assuming there is no new physics in $a_\text{fs}^d$ the experimental value translates into $a_\text{fs}^s = -0.017\pm {0.056}$ which corresponds to
\begin{align}\label{equ:dimuonSinphis}
 \sin\phi_s = -2.2\pm 1.4\,\qquad\qquad\text{at 95\% CL},
\end{align}
with a central value in the unphysical region. 
For our numerical analysis we naively use a weighted average of the experimental values 
for $\sin\phi_s$ only employing the second interval in (\ref{equ:CDF95CL}) and the first 
in (\ref{equ:D095CL}), as well as eq. (\ref{equ:dimuonSinphis}). At 95\% CL we obtain 
\begin{align}\label{equ:expSinphis}
 \sin\phi_s = -0.77\pm 0.47.
\end{align}
{The global analysis in \cite{Lenz:2010gu} found also hints of new
  physics in \bbmd, which alleviates the problem in \eq{equ:dimuonSinphis}.
  The best-fit value for the corresponding CP phase $\phi_d$ is much
  smaller in magnitude than $\phi_s$. In \cite{Trine:2009ns} it has been
  shown that a non-zero $\phi_d$ and a phenomenologically equally
  welcome contribution to $\epsilon_K$ can arise in the CMM model from 
  dimension-5 Yukawa terms. In this paper we do not consider 
  these sub-dominant terms which would introduce new parameters to the 
  analysis.}

\subsection{\boldmath{$b \to s \gamma$}}

The atmospheric mixing angle in $U_D$ has a strong impact in $b\to
s\gamma$.  In the SM it is mediated via a $W$ boson in which the the
$b_L\to s_R$-transition is proportional to the strange quark mass
$\propto m_s$ and thus negligible compared to the $b_R\to
s_L$-transition $\propto m_b$.  
{In the CMM model amplitudes with both chiralities occur:}
\begin{align}
& {\cal A} (b_L\to s_R\gamma)\quad  \propto\;\; \left(U_D
m^2_{\tilde{d}}U_D^\dagger\right)_{32} 
\qquad\qquad\quad\,\,\rightarrow C_7^\prime \label{blsr} \\
& {\cal A} (b_R \to s_L\gamma)\quad  \propto\;\; \left(V_q^\dagger
m^2_{\tilde{q}}V_q\right)_{32} \qquad\qquad\qquad 
\rightarrow C_7  . \label{brsl}
\end{align}
{\eq{blsr} is the effect of the genuine $\tilde b_R$--$\tilde
  s_L$ transition of the CMM model. It contributes to $C_7^\prime$ and
  therefore yields a positive contribution to ${\cal B} (b\to
  s\gamma)$.  The term in \eq{brsl} constitutes an MFV-like
  (i.e.\ CKM-driven) gluino-squark contribution to $C_7$. We will see
  later that in the ballpark of the viable parameter region of the
  model the second contribution is larger and actually reduces ${\cal
    B}(b \to s \gamma)$. This is the only place where we find a
  formally subdominant (namely CKM-suppressed) contribution important. Its
  relevance stems partially from the interference of the term in
  \eq{brsl} with the SM term.  Therefore the contribution in \eq{brsl}
  enters ${\cal B} (b\to s\gamma)$ linearly, while the one in
  \eq{blsr} modifies this branching ratio quadratically.}
   
The branching ratio for $b \to s \gamma$ {is usually written as}
\begin{align}
  \label{eq:bsgamma}
  \mathcal B(b\rightarrow s\gamma) = \mathcal{B}_{SL} \frac{6
    \abs{V_{tb} V_{ts}^\ast}^2}{\pi \abs{V_{cb}}^2
    g\left(m_c^2/m_b^2\right)} \left( \abs{\hat{C}_7(\mu_b)}^2 +
    \abs{\hat{C}_7^\prime(\mu_b)}^2 \right) ,
\end{align}
where $\mathcal{B}_{SL} = 0.1033 \pm 0.0028$ \cite{PDG} is semileptonic
branching ratio and $g(z)=1-8z+8z^3-z^4-12z^2\ln(z)$.  The effective
Wilson coefficients are given by  \cite{Besmer:2001cj,Borzumati:1999qt}
\begin{align}
  \hat{C}_7(\mu_b) & = C_7^{\text{eff}}(\mu_b) - \left[
    C_{7b\tilde{g}}(\mu_b) + \frac{1}{m_b}
    C_{7\tilde{g}\tilde{g}}(\mu_b) \right] \frac{16 \sqrt{2} \pi^3
    \alpha_s(\mu_b)}{G_F V_{tb} V_{ts}^\ast} \ , \nonumber
  \\[2pt]
  \hat{C}_7^\prime(\mu_b) & = C_7^\prime(\mu_b) - \left[
    C_{7b\tilde{g}}^\prime(\mu_b) + \frac{1}{m_b}
    C_{7\tilde{g}\tilde{g}}^\prime(\mu_b) \right] \frac{16 \sqrt{2}
    \pi^3 \alpha_s(\mu_b)}{G_F V_{tb} V^\ast_{ts}} \ ,
\end{align}
where $\alpha_s$ is the strong gauge coupling. The RGE evolution to the scale $\mu_b$ is given by
\begin{align}
  C_{7b\tilde{g}}(\mu_b) & = \eta^{\frac{39}{23}} C_{7b\tilde{g}}(\mu_W)
  + \frac{8}{3} \left(\eta^{\frac{37}{23}}-\eta^{\frac{39}{23}}\right)
  C_{8b\tilde{g}}(\mu_W) \,,
  \\
  C_{7\tilde{g}\tilde{g}}(\mu_b) & = \eta^{\frac{27}{23}}
  C_{7\tilde{g}\tilde{g}}(\mu_W) + \frac{8}{3}
  \left(\eta^{\frac{25}{23}}-\eta^{\frac{27}{23}}\right)
  C_{8\tilde{g}\tilde{g}}(\mu_W) 
  \intertext{(for the running of the primed coefficients, substitute
    $C_i^\prime$ for $C_i$);}
  C_7^{\text{eff}}(\mu_b) & = C_7^{\text{SM}}(\mu_b) +
  \eta^{\frac{16}{23}} C_7(\mu_W) + \frac{8}{3}
  \left(\eta^{\frac{14}{23}}-\eta^{\frac{16}{23}}\right) C_8(\mu_W) \,,
  \\
  C_7'(\mu_b) & = \eta^{\frac{16}{23}} C_7^\prime(\mu_W) + \frac{8}{3}
  \left(\eta^{\frac{14}{23}}-\eta^{\frac{16}{23}}\right)
  C_8^\prime(\mu_W)
\end{align}
and
\begin{equation}
  \eta \equiv \frac{\alpha_s(\mu_W)}{\alpha_s(\mu_b)}.
\end{equation}
For the SM contribution we use
\begin{equation}
  C_7^{\text{SM}}(\mu_b) = {-0.335}.
\end{equation}
Without new physics contribution this value reproduces the SM NNLO result \cite{Misiak:2006ab}:
\begin{align}
  \mathcal{B}(b\rightarrow s\gamma)^{\text{SM}}_{E_\gamma > 1.6\,
    \text{GeV}} = \left( 3.15 \pm 0.23 \right) \times 10^{-4} \,.
\end{align} 
An average of the experimental data of BABAR, Belle and CLEO yields
\cite{Barberio:2006bi}:
\begin{align}
  \mathcal{B}(b\rightarrow s\gamma)^{\text{exp}}_{E_\gamma > 1.6\,
    \text{GeV}} = \left( 3.55 \pm 0.24^{+0.09}_{-0.10} \pm 0.03\right)
  \times 10^{-4} \,,
\end{align}
where the errors are combined statistical and systematic, systematic
due to the shape function, and the $b\to d\gamma$ fraction. The SM
prediction lies within the $3\sigma$ range, but since the central
values differ from each other there is still room for new physics.
\\ The MSSM contributions are computed with the following formulas
\cite{Besmer:2001cj,Borzumati:1999qt} (using the abbreviation $V
\doteq (4 G_F \, V_{tb} V_{ts}^\ast)/\sqrt{2}$): The chargino-,
neutralino- and Higgs contributions read:
{\allowdisplaybreaks
  \begin{align}
    C_{7}(\mu_W) & = - \frac{1}{2} \left[\cot^2\beta\,
      x_{tH}(Q_{u}F_1(x_{tH})+F_2(x_{tH}))+
      x_{tH}(Q_uF_3(x_{tH})+F_4(x_{tH}))\right] \nonumber
    \\
    & \quad + \frac{1}{2V}\sum_{j=1}^6\sum_{l=1}^2\frac{1}{m^2_{\tilde
        u_j}}B^d_{2j\ell}B^{d*}_{3j\ell}\left[F_1(x_{\tilde{\chi}^{\pm}_\ell
        \tilde{u}_j})+ Q_{u}F_2(x_{\tilde{\chi}^{\pm}_{\ell} \tilde{u}_j})\right]
\nonumber
    \\
    & \quad + \frac{1}{2V}\sum_{j=1}^6\sum_{l=1}^2\frac{1}{m^2_{\tilde
        u_j}}\frac{m_{\tilde{\chi}^{\pm}_\ell}}{m_b}B^d_{2j\ell}A^{d*}_{3j\ell}
    \left[F_3(x_{\tilde{\chi}^{\pm}_\ell \tilde{u}_j})+Q_u F_4(x_{\tilde{\chi}^{\pm}_\ell
        \tilde{u}_j})\right] \nonumber
    \\
    & \quad + \frac{Q_d}{2V}\sum_{j=1}^6\sum_{l=1}^4\frac{1}{m^2_{\tilde
        d_j}}\left[D^d_{2j\ell}D^{d*}_{3j\ell}F_2(x_{\tilde{\chi}^0_\ell \tilde
        d_j})+\frac{m_{\tilde{\chi}^0_\ell}}{m_b}D^d_{2j\ell}
      C^{d*}_{3j\ell}F_4(x_{\tilde{\chi}^0_\ell \tilde d_j})\right] \label{C7}
    \\
    C_{8}(\mu_W) & = - \frac{1}{2} \left[\cot^2\beta\,
      x_{tH}F_1(x_{tH})+x_{tH}F_3(x_{tH}) \right] \nonumber
    \\
    & \quad + \frac{1}{2V}\sum_{j=1}^6\sum_{l=1}^2\frac{1}{m^2_{\tilde
        u_j}}\left[B^d_{2j\ell}B^{d*}_{3j\ell}F_2(x_{\tilde{\chi}^{\pm}_\ell
        \tilde{u}_j}) +\frac{m_{\tilde{\chi}^{\pm}_\ell}}{m_b}B^d_{2j\ell}A^{d*}_{3j\ell}
      F_4(x_{\tilde{\chi}^{\pm}_\ell \tilde{u}_j}) \right] \nonumber
    \\
    & \quad + \frac{1}{2V}\sum_{j=1}^6\sum_{l=1}^4\frac{1}{m^2_{\tilde
        d_j}}\left[D^d_{2j\ell}D^{d*}_{3j\ell}F_2(x_{\tilde{\chi}^0_\ell \tilde{d}_j})
      +\frac{m_{\tilde{\chi}^0}}{m_b}D^d_{2j\ell}C^{d*}_{3j\ell}F_4(x_{\tilde{\chi}^0_\ell
        \tilde{d}_j}) \right] \label{C8}
    \\
    C_{7}^{\prime}(\mu_W) & = -\frac{1}{2} \,
    \frac{m_sm_b}{m_t^2}\tan^2\beta\,
    x_{tH}(Q_{u}F_1(x_{tH})+F_2(x_{tH})) \nonumber
    \\
    & \quad + \frac{1}{2V}\sum_{j=1}^6\sum_{l=1}^2\frac{1}{m^2_{\tilde
        u_j}}A^d_{2j\ell}A^{d*}_{3j\ell}\left[F_1(x_{\tilde{\chi}^{\pm}_\ell
        \tilde{u}_j})+Q_{u}F_2(x_{\tilde{\chi}^{\pm}_\ell \tilde{u}_j})\right] \nonumber
    \\
    & \quad + \frac{1}{2V}\sum_{j=1}^6\sum_{l=1}^2\frac{1}{m^2_{\tilde u_j}}
    \frac{m_{\tilde{\chi}^{\pm}_\ell}}{m_b}A^d_{2j\ell}B^{d*}_{3j\ell}
    \left[F_3(x_{\tilde{\chi}^{\pm}_\ell \tilde{u}_j})+Q_u F_4(x_{\tilde{\chi}^{\pm}_\ell
        \tilde{u}_j})\right] \nonumber
    \\
    & \quad + \frac{Q_d}{2V}\sum_{j=1}^6\sum_{l=1}^4\frac{1}{m^2_{\tilde
        d_j}}\left[C^d_{2j\ell}C^{d*}_{3j\ell}F_2(x_{\tilde{\chi}^0_\ell \tilde{d}_j})
      +\frac{m_{\tilde{\chi}^0_\ell}}{m_b}C^d_{2j\ell}
      D^{d*}_{3j\ell}F_4(x_{\tilde{\chi}^0_\ell \tilde{d}_j}) \right] \label{C7'}
    \\
    C_{8}^{\prime}(\mu_W) & = -\frac{1}{2} \,
    \frac{m_sm_b}{m_t^2}\tan^2\beta\, x_{tH}F_1(x_{tH}) \nonumber
    \\
    & \quad + \frac{1}{2V}\sum_{j=1}^6\sum_{l=1}^2\frac{1}{m^2_{\tilde
        u_j}}\left[A^d_{2j\ell}A^{d*}_{3j\ell}F_2(x_{\tilde{\chi}^{\pm}_\ell
        \tilde{u}_j}) +\frac{m_{\tilde{\chi}^{\pm}_\ell}}{m_b}A^d_{2j\ell}B^{d*}_{3j\ell}
      F_4(x_{\tilde{\chi}^{\pm}_\ell \tilde{u}_j}) \right] \nonumber
    \\
    & \quad + \frac{1}{2V}\sum_{j=1}^6\sum_{l=1}^4\frac{1}{m^2_{\tilde
        d_j}}\left[C^d_{2j\ell}C^{d*}_{3j\ell}F_2(x_{\tilde{\chi}^0_\ell \tilde{d}_j})
      +\frac{m_{\tilde{\chi}^0_\ell}}{m_b}C^d_{2j\ell}D^{d*}_{3j\ell}
      F_4(x_{\tilde{\chi}^0_\ell \tilde{d}_j}) \right] \, ,
  \end{align}}
where $Q_u=2/3$, $Q_d=-1/3$, $x_{\tilde{\chi}^{0,\pm}_\ell \tilde q_j} =
m_{\tilde{\chi}^{0,\pm}_\ell}^2/m_{\tilde q_j}^2 $ and
$x_{tH}=m_t^2/m_{H^\pm}^2 $. 
The gluino contributions read:
{\allowdisplaybreaks
\begin{align}
  C_{7b,\tilde{g}}(\mu_W) & = \ \ -\frac{Q_d}{16 \pi^2} \, \frac{4}{3}
  \sum_{k=1} ^6 \frac{1}{m_{\tilde{d}_k}^2} \left( \Gamma_{DL}^{kb} \,
    \Gamma_{DL}^{\ast\,ks} \right) F_2(x_{gd_k}) \, , \nonumber
  \\
  C_{7\tilde{g},\tilde{g}}(\mu_W) & = m_{\tilde g_3}\, \frac{Q_d}{16
    \pi^2} \, \frac{4}{3} \sum_{k=1} ^6 \frac{1}{m_{\tilde{d}_k}^2}
  \left( \Gamma_{DR}^{kb} \, \Gamma_{DL}^{\ast\,ks} \right)
  F_4(x_{gd_k})\, , \nonumber
  \\
  C_{8b,\tilde{g}}(\mu_W) & = \ \ -\frac{1}{16 \pi^2} \sum_{k=1} ^6
  \frac{1}{m_{\tilde{d}_k}^2} \left( \Gamma_{DL}^{kb} \,
    \Gamma_{DL}^{\ast\,ks} \right) \, \left[ - \frac{1}{6} F_2(x_{gd_k})
    - \frac{3}{2} F_1(x_{gd_k}) \right] \, , \nonumber
  \\
  C_{8\tilde{g},\tilde{g}}(\mu_W) & = m_{\tilde g_3}\, \frac{1}{16 \pi^2}
  \sum_{k=1} ^6 \frac{1}{m_{\tilde{d}_k}^2} \left( \Gamma_{DR}^{kb} \,
    \Gamma_{DL}^{\ast\,ks} \right) \, \left[- \frac{1}{6} F_4(x_{gd_k})
    - \frac{3}{2} F_3(x_{gd_k}) \right] \,.
\end{align}}
The ratios $x_{gd_k}$ are defined as $x_{gd_k} \equiv m_{\tilde
  g}^2/m_{\tilde{d}_k}^2$.  The Wilson coefficients of the corresponding
primed operators are obtained through the interchange $\Gamma_{DR}^{ij}
\leftrightarrow \Gamma_{DL}^{ij} $.
Finally, we define the functions $F_i$ appearing in the Wilson coefficients
listed above:
\begin{align}
  F_1(x) & = \frac{1}{12 \left(x-1\right)^4} \left( x^3 -6x^2 +3x +2
    +6x\log x\right) \, , \nonumber
  \\
  F_2(x) & = \frac{1}{12 \left(x-1\right)^4} \left(2x^3 +3x^2 -6x +1
    -6x^2\log x\right) \, , \nonumber
  \\
  F_3(x) & = \frac{1}{\phantom{1} 2 \left(x-1\right)^3} \left( x^2 -4x
    +3 +2\log x\right) \, , \nonumber
  \\
  F_4(x) & = \frac{1}{ \phantom{1} 2 \left(x-1\right)^3} \left( x^2 -1
    -2x\log x\right) \, .
  \label{loopfunc}
\end{align}
The matrices appearing in the above expressions are now expressed in
terms of the mixing matrices according to the convention of
\cite{Rosiek:1995kg} except for the vacuum expectation values:
\begin{align}
  v_1^{\mbox{\cite{Besmer:2001cj}}} & = \frac{1}{\sqrt{2}}
  v_1^{\mbox{\cite{Rosiek:1995kg}}} \;, & v_2^{\mbox{\cite{Besmer:2001cj}}} & =
  \frac{1}{\sqrt{2}} v_2^{\mbox{\cite{Rosiek:1995kg}}}
\end{align}
The mixing matrices of up and down quarks are
\begin{align}
  (\Gamma_{DL})_{iI} & =Z_D^{Ii} & (\Gamma_{DR})_{iI} & =Z_D^{(I+3)i}
  \\
  (\Gamma_{UL})_{iI} & =Z_U^{Ii*} & (\Gamma_{UR})_{iI} & =Z_U^{(I+3)i*} \,.
\end{align} 
Other abbreviations that appear are:
\begin{align}
  A^d_{ijl} & = \frac{e}{\sqrt{2}\sin\theta_W M_W\cos\beta} M_d^{ik}
  Z_U^{kj}Z_-^{2l}
  \\
  B^d_{ijl} & = \frac{e}{\sqrt{2}\sin\theta_W M_W\sin\beta} \left(
    K^\dagger M_u\right)^{ik} Z_U^{(k+3)j} Z_+^{2l*} -
  \frac{e}{\sin\theta_W} Z_U^{ij} Z_+^{1l*}
  \\
  C^d_{ijl} & = \frac{e}{\sqrt{2}\sin\theta_W M_W\cos\beta} M_d^{ik}
  Z_D^{kj*} Z_N^{3l} - \frac{\sqrt{2}e}{\cos\theta_W} Q_d Z_D^{(i+3)j*}
  Z_N^{1l}
  \\
  D^d_{ijl} & = \frac{e}{\sqrt{2}\sin\theta_W M_W\cos\beta} M_d^{ik}
  Z_D^{(k+3)j*} Z_N^{3l*} + \frac{1}{\sqrt{2}}Z_D^{ij*}\left[ \left(2
      Q_d+1\right) \frac{e}{\cos\theta_W} Z_N^{1l*} -
    \frac{e}{\sin\theta_W} Z_N^{2l*} \right]
\end{align}
where $M_u$ and $M_d$ are diagonal $3\times3$-matrices that contain the
masses of up and down quarks respectively in their diagonal elements.
All mixing matrices are according to \cite{Rosiek:1995kg}.
For completeness we also list the conversion of conventions for the
mixing matrices of charginos, neutralinos and charged Higgs bosons:
\begin{align}
  U & = Z_-^\dagger \,, & V & = Z_+^\dagger \,, &
  N & = Z_N^\dagger \,, & Z_E & = Z_H^\dagger \,.
\end{align}

\subsection{\boldmath$\tau \to \mu \gamma$}

So far, large transitions in the observables we have looked at stem from a large
mixing among the right-handed down-type squarks, induced by GUT relations. 
Therefore, it is important to correlate those results with the results from a decay in the lepton sector
where the PMNS matrix is directly responsible for the transition: $\tau\to \mu\gamma$.
In the SM with massive neutrinos this decay is unobservably small, 
such that any signal would be a clear proof for new physics. The experimental upper bounds are:
\begin{align}
  &\mathcal B(\tau\rightarrow\mu\gamma)^{\text{exp}} < 4.5 \times 10^{-8}\quad\text{at 90\% CL (Belle)\cite{Hayasaka:2007vc} }\\
&\mathcal B(\tau\rightarrow\mu\gamma)^{\text{exp}} < 4.4 \times 10^{-8}\quad\text{at 90\% CL (BaBar)\cite{Benitez:2010gm} }.
\end{align}
In the CMM model the atmospheric mixing angle enters $Z_L$ and the PMNS matrix itself 
in slepton-neutralino and chargino-sneutrino vertices. We use the one-loop result of \cite{Hisano:1995cp} 
but employ the notation of \cite{Jager:2003zz} and correction of a factor $\cos\theta_W$ .
Furthermore, we consider a limit which is suitable for the CMM model:  
Setting $y_\mu = 0$, we consider only $\tau_R\to\mu_L\gamma$ transitions.
The branching ratio reads:
\begin{align}
  \mathcal{B}(\tau\rightarrow\mu\gamma) = \frac{\tau_\tau m^5_\tau}{4
    \pi} \abs{C_7^{\tilde{\chi}^\pm} + C_7^{\tilde{\chi}^0}}^2
\end{align}
with the $\tau$ lifetime $\tau_\tau = 290.6 \times 10^{-15}$~s and the
$\tau$ mass $m_\tau = 1.77699$~GeV \cite{PDG}. The Wilson coefficients
are given by:
\begin{align}
  C_7^{\tilde{\chi}^\pm} & = \frac{e^3}{32 \pi^2 \sin^2\theta_W}
  \sum_{J=1}^{3} \sum_{i=1}^{2} U_D^{2J} U_D^{3J\ast}  \left[
Z_+^{1i*}
    Z_+^{1i} \frac{H_1(x_{Ji})}{m^2_{\chi^+_i}} - Z_+^{1i*}Z_-^{2i*}
    \frac{H_2(x_{Ji})}{\sqrt{2}\cos\beta \ m_{\chi^+_i} M_W} \right]
  \\
  C_7^{\tilde{\chi}^0} & = \frac{e^3}{32 \pi^2 \sin^2\theta_W}
  \sum_{J=1}^{6} \sum_{i=1}^{4} \frac{1}{m^2_{\chi^0_i}} \left[
    Z_{L}^{2J*} Z_{L}^{3J} \abs{Z_N^{1i} \sin\theta_W + Z_N^{2i}
      \cos\theta_W}^2 \frac{H_3(y_{Ji})}{2 \cos^2\theta_W} \right.
  \nonumber
  \\[2pt]
  & \mspace{120mu} - \left. Z_L^{2J*} Z_L^{6J} Z_N^{3i} \left( Z_N^{1i*}
      \sin\theta_W + Z_N^{2i*} \cos\theta_W \right) \frac{m_\tau
      H_3(y_{Ji})}{2\cos\theta_W M_W \cos\beta} \right. \nonumber
  \\[2pt]
  & \mspace{120mu} 
+ \left. Z_L^{2J*} Z_L^{3J} Z_N^{3i*} \left(
      Z_N^{1i*} \sin\theta_W + Z_N^{2i*} \cos\theta_W \right)
    \frac{m_{\chi^0_i}H_4(y_{Ji})}{2 \cos\theta_W M_W \cos\beta} \right.
  \nonumber
  \\[2pt]
  & \mspace{120mu} + \left. Z_L^{2J*} Z_L^{6J} Z_N^{1i*} \left(
      Z_N^{1i*} \sin\theta_W + Z_N^{2i*} \cos\theta_W \right)
    \frac{m_{\chi^0_i} \sin\theta_W H_4(y_{Ji})}{m_\tau \cos^2\theta_W}
  \right] ,
\end{align}
where  in the convention of \cite{Rosiek:1995kg} $Z_+$ and $Z_-$ are the chargino mixing
matrices, $Z_N$ is the
neutralino mixing matrix, $Z_L$ is the lepton mixing matrix, $Z_\nu =U_D$ is the sneutrino
mixing matrix
and
\begin{align}
  x_{Ji} & = \frac{m^2_{\tilde \nu_J}}{m^2_{\chi^+_i}} \ , \mspace{60mu}
  y_{Ji} = \frac{m^2_{\tilde l_J}}{m^2_{\chi^0_i}} \ .
\end{align} 
The loop functions are given by:
\begin{align}
  \begin{split}
    H_1(x) & = \frac{1-6x+3x^2+2x^3-6x^2\ln x}{12(x-1)^4}
    \\
    H_2(x) & = \frac{-1+4x-3x^2+2x^2 \ln x}{2(x-1)^3} 
    \\
    H_3(x) & = \frac{-2-3x+6x^2-x^3-6x \ln x}{12(x-1)^4}
    \\
    H_4(x) & = \frac{1-x^2+2x \ln x}{2(1-x)^3}
  \end{split}
\end{align}
Neglecting left-right mixing in the slepton sector, the rotation matrix
is given as
\begin{equation}
 Z_L =
\left(\begin{array}{cc}
       U_D^\ast & 0\\
  0 & V_\text{CKM}^\top\\
      \end{array}
\right).
\end{equation}
From this we can read off that in the neutralino contribution the two terms proportional
to $Z_L^{2J*}Z_L^{3J}\approx U_D^{2J}U_D^{3J*}$ dominates whereas the terms
$\propto Z_L^{2J*}Z_L^{6J}$ need LR-mixing.

\subsection{The neutral Higgs mass}

Another observable that is quite restrictive for the CMM model is the
{mass of the lightest neutral,} CP-even Higgs boson of the MSSM.  At 
tree level its mass is bounded from above by the $Z$ boson mass.
However, radiative corrections shift the mass to higher values.  An
approximate formula at $\mathcal O (\alpha\alpha_s)$ is given by
\cite{Heinemeyer:2004ms}
\begin{align}
  M_h^2 & = M_{h}^{2,\text{tree}} + \frac{3}{2}\frac{G_F\sqrt{2} \
    \overline m_t^4}{\pi^2} \left\{ -\ln\left(\frac{\overline
        m^2_t}{M_S^2}\right) + \frac{\abs{X_t}^2}{M_S^2}
    \left(1-\frac{\abs{X_t}^2}{12 M_S^2}\right) \right\} \nonumber
  \\
  & \mspace{75mu} -3\frac{G_F\sqrt{2}\alpha_s \overline m_t^4}{\pi^3}
  \left\{ \ln^2\left(\frac{\overline m^2_t}{M_S^2}\right) + \left[
      \frac{2}{3}-2\frac{\abs{X_t}^2}{M_S^2}
      \left(1-\frac{\abs{X_t}^2}{12 M_S^2}\right) \right]
    \ln\left(\frac{\overline m^2_t}{M_S^2}\right) \right\} ,\label{equ:Mh2corr}
\end{align}
where
\begin{align}
  X_t & = -\frac{A_t}{y_t} - \frac{\mu^*}{\tan\beta} \,,
\end{align}
{$\overline m_t=165\pm 2$} GeV is the $\overline{\rm MS}$ mass of
the top quark and
\begin{equation}
  M_S^2 = \sqrt{m^2_{\tilde q_3} m^2_{\tilde u_3}} \,.
\end{equation}
The tree level Higgs mass is given by
\begin{equation}
  M_{h}^{2,\text{tree}} = \frac{1}{2} \left[M^2_A + M^2_Z -
    \sqrt{(M^2_A+M^2_Z)^2 - 4 M^2_Z M^2_A \cos^2(2\beta)}\right] 
\end{equation}
where the mass of the CP odd Higgs boson can be computed by:
\begin{equation}
  M^2_A = \frac{m^2_{H_u}-m^2_{H_d}}{\cos(2\beta)}-M^2_Z
\end{equation}
The experimental lower bound  (for large $\tan\beta$) is $ M_{h}^{\text{exp}}
\geqslant 89.8~\text{GeV}$ \cite{PDG}. Since the coupling strength of the
$Z$ boson to $h^0$ depends on the MSSM Higgs mixing angles, especially on
$\sin(\beta-\alpha)$, the experimental lower bound for small $\tan\beta$, relevant for our analysis,
is close to the Higgs mass bound in the SM \cite{Barate:2003sz}:
\begin{equation}
  M_{h}^{\text{exp}} \geqslant 114.4~\text{GeV}
\end{equation}
In the next section we will see that for $\tan\beta = 3$ the constraints from the
lightest Higgs mass are much more stringent than the FCNC bounds. This is
due to the fact that the large top Yukawa coupling drives the masses of the third
squark generation to smaller values such that the corrections to the tree level Higgs mass 
cannot compensate for the difference between $M_h^\text{tree}$ and the experimental
lower bound.

\subsection{Further experimental input parameters}
For our analysis we used the following experimental input :\\

\begin{tabular}{ll}
$\alpha_e(M_Z) =   1/128.129 \quad $\cite{Martens:2010nm,Teubner:2010ah}  &\qquad
    $\sin^2\theta_W = 0.23138\quad$\cite{Martens:2010nm,PDG}  \nonumber\\
$\alpha_s(M_Z) =   0.1184 \quad $\cite{Bethke:2009jm} &\qquad
 $G_F = 1.16637\times10^{-5}\ \text{GeV}^{-2}\quad$\cite{PDG} \nonumber\\
$M_W = 80.398\ \text{GeV} \quad$\cite{PDG} &\qquad
   $m_t = 173.3\ \text{GeV} \quad \text{(pole mass)} \quad$\cite{:1900yx} \nonumber\\
$M_Z = 91.1876 \ \text{GeV} \quad$\cite{PDG} &\qquad
   $m_b(m_b) = 4.163 \ \text{GeV}\quad $\cite{Chetyrkin:2009fv} \nonumber\\ 
$m_\tau = 1.777\ \text{GeV} \quad$\cite{PDG} &\qquad
   $m_b = 4.911 \ \text{GeV} \quad \text{(pole mass)}$\,.
\end{tabular}\\
The pole mass of the bottom quark was obtained
using the above value for $m_b(m_b)$ and the program {\tt RunDec}~\cite{Chetyrkin:2000yt}.


For the MNS matrix we use the tri-bimaximal mixing \cite{Harrison:2002er}, i.e.
a parametrization with $\theta_{12} = 30^\circ$, $\theta_{23} = 45^\circ$, 
and $\theta_{13} = 0^\circ$. The CKM matrix is constructed via the
Wolfenstein parametrization \cite{Wolfenstein:1983yz} using the latest parameters from the 
CKMfitter group \cite{Charles:2004jd}:
\begin{eqnarray}
\lambda &=& 0.22543 \nonumber\\
A &=& 0.812 \nonumber\\
\overline{\rho} &=& 0.144  \nonumber\\
\overline{\eta} &=& 0.342 \,.
\end{eqnarray}

\section{Results}\label{sec:results}

The correlation of observables in Sec.~\ref{se:observables} allows us
to constrain the parameter space of the CMM model.  In order to test
the model, we first choose a scenario {in which the specific
  signatures of the model are enhanced and flavor-violating effects
  are maximal:} As discussed in Sec.~\ref{sec:yt} with $\tan\beta = 3$
the top Yukawa coupling is near its infrared fixed point such that the
mass splitting between {the first two generations and the third
  one} is maximal without losing the perturbativity of $y_t$.  The
rotation into the super-CKM basis (see Eq.~(\ref{eq:vmv})) translates
this into maximal flavor violation. Whereas the FCNC constraints still
allow some regions in the CMM model parameter space, the model is
challenged by the experimental lower bound on the Higgs mass.
However, this can be reconciled in a relaxed scenario with $\tan\beta
= 6$. We discuss both the $\tan\beta = 3$ and the $\tan\beta = 6$
cases. In the first case we get maximal effects in the flavor sector,
because of the large intergenerational squark mass splitting. In this
scenario we {explore the viable parameter space of the CMM
  model.}  If we find that the model is not excluded, then this will
also be true for larger values of $\tan\beta$.  The $\tan\beta = 6$
case corresponds to a consistent scenario.  {We further take
  $\mu$ real to avoid problems with electric dipole moments.}

\subsection*{Vacuum stability and positive soft {bilinear terms}}
Since the trilinear $A$-terms can lead to charge- and color-breaking
minima of the scalar potential, the CMM input parameter $a^d_1$ is
restricted to fulfill the stability bound \cite{Casas:1997ze}
\begin{align}
 \left|a_1^d(M_Z)\right|<\sqrt{3\left(m_{\tilde q_1}^2 + m_{\tilde d_1}^2 + m_{H_d}^2\right)}.
\end{align}
We have checked that in our parameter scan with $|a_1^d|/M_{\tilde
  q}<3$ this condition is satisfied almost everywhere.  Similarly, we
must exclude unphysical regions with negative soft {squared}
masses of sfermions carrying {U(1)$_{\rm em}$ or}
SU(3)$_C$-charges which can occur if $y_t$ drives the third-generation
sfermion masses to negative values at the electroweak scale. This
limits the mass splitting and thus the size of flavor-violating
effects. In the following plots the black regions are unphysical due
to $m_{\tilde f}^2<0$ or an unstable vacuum.  The actual experimental
lower bounds on the masses have no relevant effect. This is due to the
fact that close to the negative mass bound, the soft masses decrease
from typical masses of $\mathcal{O}(M_\text{SUSY})$ to zero quite
rapidly. This happens in intervals of $M_{\tilde q}$ and $a_1^d$ that
are really small as compared to the intervals we are scanning over.
Therefore we will not distinguish between the negative soft mass
bounds and the bounds resulting of sfermion masses falling below their
experimental lower bounds.

\subsection*{Mass splitting}
The CMM model specific flavor effects are crucially determined by the
mass splitting of the right-handed down squarks (see
Eq.~(\ref{eq:vmv})).  In Fig.~\ref{fig:split} the relative mass
splitting $\Delta_{\tilde{d}}^\text{rel} = 1-
m_{\tilde{d}_3}^2/m_{\tilde{d}_2}^2$ is shown. {In
  Figs.~\ref{fig:split}--\ref{fig:higgs} we depict
  the quantities of interest as contour plots in a the $M_{\tilde
    q}$--$a_1^d/M_{\tilde q}$ plane. Here the mass of the right-handed
  squarks of the first two generations, $M_{\tilde q}$, (which is
  essentially degenerate with the masses of the corresponding
  left-handed masses) and the trilinear term $a_1^d$ are defined at
  the low scale $Q=M_Z$. In the plots we further use} $m_{\tilde
  g_3}(M_Z) = 500~\text{GeV}$, $\sgn \mu = +1$, $\tan\beta = 3$ (left)
and $\tan\beta = 6$ (right).  The mass splitting increases with
$|a_1^d(M_Z)|/M_{\tilde q}(M_Z) $ and decreases as expected with
$\tan\beta$.  For a heavier gluino mass the allowed physical region
moves to larger values of $M_{\tilde q}(M_Z)$ and changing the sign of
$\mu$ does not have any significant effect.

\begin{nfigure}[!tb]
 \psfrag{amsq}{\hspace{0.0cm}\scalefont{1.3}$\frac{a_1^d}{M_{\tilde q}}$}
 \psfrag{msq}{\hspace{-0.5cm}\scalefont{1.0}$M_{\tilde q}$[GeV]}
 \psfrag{mg500argmu0tanb3}{\hspace{-2.0cm}\scalefont{1}$m_{\tilde g_3} = 500~$GeV, $ \sgn(\mu) = +1$, $\tan\beta = 3$}
 \psfrag{mg500argmu0tanb6}{\hspace{-2.0cm}\scalefont{1}$m_{\tilde g_3} = 500~$GeV, $ \sgn(\mu) = +1$, $\tan\beta = 6$}

  \includegraphics[width=.49\linewidth]{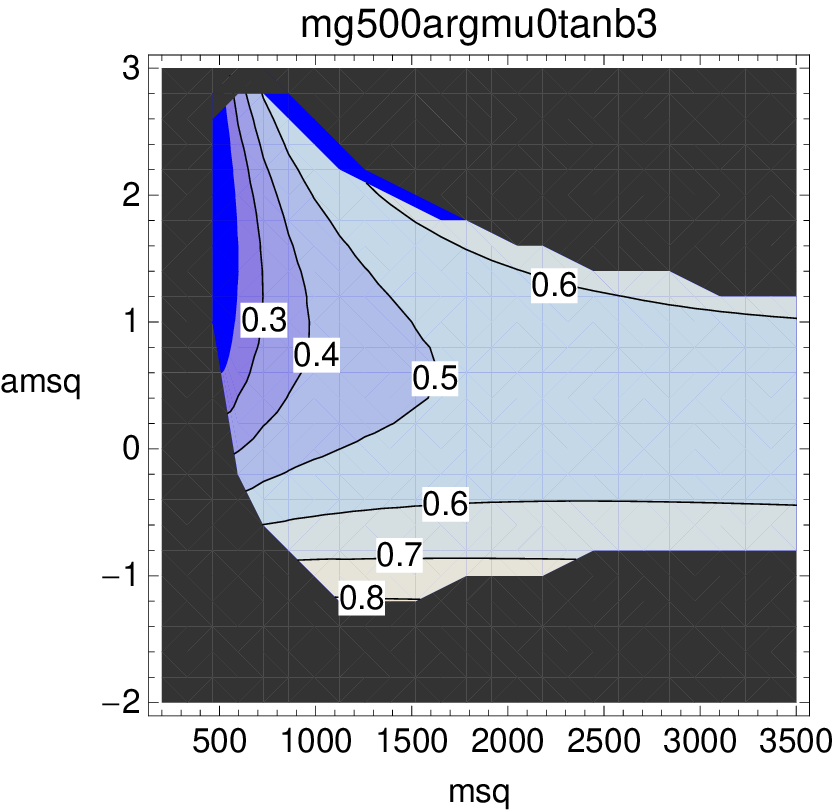}
  \hspace{.01\linewidth}
  \includegraphics[width=.49\linewidth]{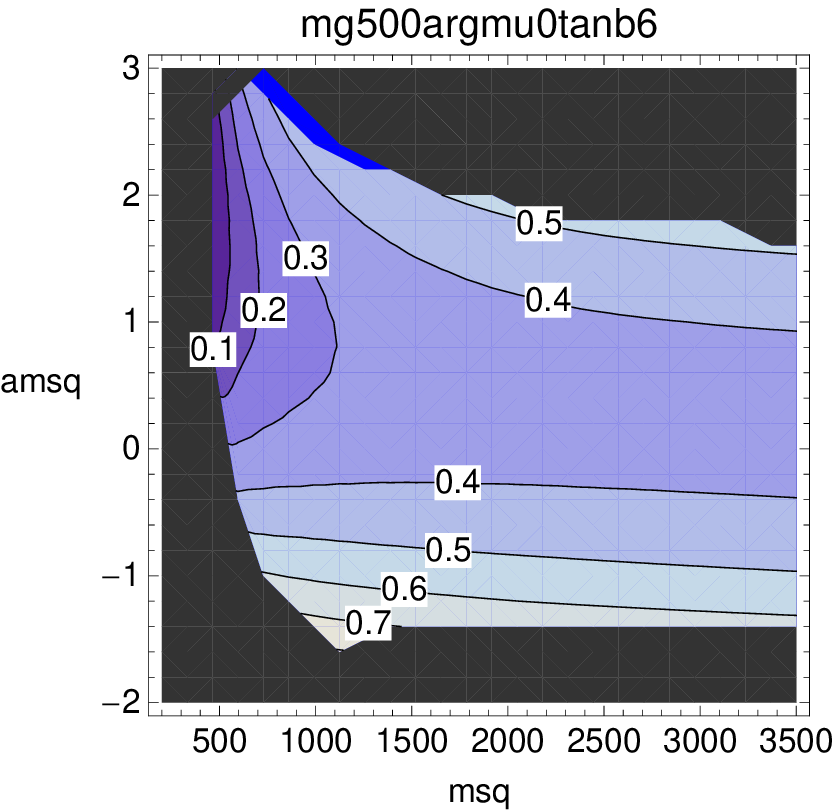}
  \caption{Relative mass splitting $\Delta_{\tilde{d}}^\text{rel} = 1-
    m_{\tilde{d}_3}^2/m_{\tilde{d}_2}^2$ among the bilinear soft
      terms for the right-handed squarks of the second and third
      generations with $\tan\beta = 3$ (left) and 6 (right) in the
    $M_{\tilde q}(M_Z)-a_1^d(M_Z)/M_{\tilde q}(M_Z)$ plane for
    $m_{\tilde g_3} = 500~\text{GeV}$ and $\sgn (\mu) = +1$.}
  \label{fig:split}
\end{nfigure}

\subsection*{Sparticle spectrum {and FCNC observables} for a specific
  parameter point}

Exemplarily, we present the output for one CMM model parameter point. 
We choose the same inputs as in Sec.~\ref{se:parMGUT} where the parameters at the GUT 
scale have been discussed: 
\begin{equation}\label{eq:bpoint2}
 M_{\tilde q} = 1500~\text{GeV},\quad m_{\tilde g_3} = 500~\text{GeV}, \quad
a_1^d/M_{\tilde q} = 1.5,\quad \arg(\mu) = 0 ,\quad \tan\beta = 6. 
\end{equation}

The sparticle spectrum at the electroweak scale is given as (mass eigenvalues):
\begin{align}
 &m_{\tilde g_1}  = 83~\text{GeV},\quad m_{\tilde g_2} = 165~\text{GeV},\\
& m_{\tilde\chi^0_i} = \left( 640,\, 632,\, 159,\,\underline{81}\right)~\text{GeV}\\
& m_{\tilde\chi^\pm_i} = \left(640,\,159\right)~\text{GeV}\\
& M_{ \tilde l_i} = \left(1427,\,1427,\, \mathbf{1074},\, 1462,\,1462,\,\mathbf{1095}\right)~\text{GeV}\\
& M_{ \tilde u_i} = \left(1519,\,1519,\,\mathbf{934},\, 1501,\,1501,\,\mathbf{485}\right)~\text{GeV} \label{ex:sup}\\
& M_{ \tilde d_i} = \left(1519,\,1519,\,\mathbf{908},\,1498,\,1498,\,\mathbf{1164}\right)~\text{GeV}. \label{ex:sdown}
\end{align}
The lightest neutralino is identified as the LSP (underlined number).
The first three entries in $M_{\tilde f_i}$, $\tilde f =\tilde l,
\tilde u, \tilde d$ correspond to sfermions with a larger left-handed
component and the last three with a larger right-handed component,
where the third generation masses are printed in bold face.  The
typical mass splitting is quite evident.  The mixing angle between the
two stop eigenstates with {$485$}~GeV and {$934$}~GeV is
$\theta_{\tilde t} = 11^\circ$  and left-right mixing in the
  down sector is negligible, owing to the small value of $\tan\beta$.
  While $M_{\tilde d_4}^2=M_{\tilde d_5}^2=m^2_{\tilde
    d_1}=m^2_{\tilde d_2}$, the flavor composition of the two
  eigenstates $\tilde d_4$ and $\tilde d_5$ is very different: $\tilde
  d_4$ is the right-handed down squark, while $\tilde d_5$ (like
  $\tilde d_6$)is a maximal mixture of right-handed sstrange and
  sbottom. We here observe a generic feature of models in which $y_t$
  is the only driver of non-universal soft squark masses: Since the
  unitary rotation transforming the squark mass matrix to diagonal
  form preserves the eigenvalues, the degeneracy in \eq{smm} persists
  in the spectrum in \eq{ex:sdown} as $M_{\tilde d_4}=M_{\tilde
    d_5}$.  The Higgs parameters read:
\begin{align}
  m^2_{H_u} = -\left(575~\text{GeV}\right)^2,\quad m^2_{H_d} =
  \left(1432~\text{GeV}\right)^2,\quad \mu = 629~\text{GeV}.
\end{align}
This fullfils the condition for electroweak symmetry breaking. The
trilinear terms are given as:
\begin{align}
  \mathsf{\hat{\tilde A}}_u = \begin{pmatrix}
 		  0& 0& 0\\	
 		  0& 0& 0\\ 
 		  0 & 0 & 46.9
 		\end{pmatrix}{\gev}, \quad
 \mathsf{\hat{\tilde{A}}}_d = \begin{pmatrix}
 		  0& 0& 0\\	
 		  0& 0& 0\\ 
 		  0 & 0.5- 0.8\,i & -14.1
 		\end{pmatrix}{\gev},\quad
 &\mathsf{\hat{\tilde{A}}}_\ell = \begin{pmatrix}
 		  0& 0& 0\\	
 		  0& 0& 0.3-0.4\,i\\ 
           0& 	0 &	-5.9
 		\end{pmatrix}{\gev}.
\end{align}
For the radiative decays we obtain
\begin{eqnarray}
\mathcal{B}\left(\tau\to\mu\gamma\right) = 1.66\cdot 10^{-8} 
\qquad\qquad  \mbox{and}\qquad 
\mathcal{B}\left(b\to s\gamma\right) = {2.89}\cdot 10^{-4} 
\end{eqnarray} 

where the latter is just above the lower end of the allowed $3\sigma$
region.  Omitting the $(3,2)$ entry of $\mathsf{\hat{\tilde{A}}}_d$
would lead to an increase of $\mathcal{B}\left(b\to s\gamma\right)$ of
about $0.03 \cdot 10^{-4}$ for this particular parameter point. Note,
however, that for smaller gluino masses and e.g. $\tan\beta=3$ effects
of up to $0.7\cdot 10^{-4}$ can be ascribed to the presence
$\mathsf{\hat{\tilde{A}}}_{d,3,2}$.  This effect was not considered in
previous analyses.

We determine the phase $\xi$ such that it leads to values of $\Delta
M_s$, $\sin\phi_s$ and $f_{B_s}\sqrt{\hat{B}_{B_s}}$ that are as close
as possible to their experimental and theoretical values, respectively,
{by minimizing the $\chi^2$ for $\Delta M_s$, $\sin\phi_s$ and
  $f_{B_s}\sqrt{\hat{B}_{B_s}}$}.  To this end we scan over the
theoretical error of $f_{B_s}^2\hat{B}_{B_s}$ (see \eq{equ:fBs}),
the experimental region for $\sin\phi_s$ (see \eq{equ:expSinphis})
and $\Delta M_s$.  As a best fit value for the chosen parameter point,
we obtain the phase $\xi = 58^\circ$, yielding $\Delta M_s =
17.68$~ps$^{-1}$ and $f_{B_s}\sqrt{\hat{B}_{B_s}} =0.260$~GeV. This
corresponds to a phase $\phi_s = -49^\circ$ {meaning 
$\sin \phi_s = -0.75$}.  Alternatively, we can also
ignore the experimental value of $\sin\phi_s$ and simply ask the
question how large $\phi_s$ can become for the parameter point in
\eq{eq:bpoint2}, given the experimental and theoretical regions
for $\Delta M_s$ and $f_{B_s}\sqrt{\hat{B}_{B_s}}$.  In this case, $\xi$
can be adjusted such that the maximally allowed (negative) phase reads
$\phi_s = -52^\circ$.  The same basic procedure is also applied in the
following analysis.

\subsection*{Correlation of observables}

A combination of the flavor observables described in Sec.~\ref{se:observables} restricts the CMM parameter space. This is illustrated in Fig.~\ref{fig:fcnc} where we distinguish again between $\tan\beta = 3$ and $\tan\beta = 6$. The green region is still compatible with $B_s-\overline B_s$, $b\to s\gamma$ and $\tau\to\mu\gamma$. For larger gluino mass the allowed area increases. Furthermore, the qualitative behavior for negative $\mu$ does not change. In this case, $\tau\to\mu\gamma$ together with $B_s-\overline B_s $ leads to the strongest constraints. Because of  decoupling  the green region increases with the SUSY scale. Furthermore we find that in the parameter space where the CMM model could be valid, the lightest supersymmetric particle is almost everywhere the lightest neutralino. 
  
\begin{nfigure}[!tb]
  \psfrag{amsq}{\hspace{+0.3cm}\scalefont{1.5}$\frac{a_1^d}{M_{\tilde q}}$}
 \psfrag{msq}{\hspace{-0.5cm}\scalefont{1.1}$M_{\tilde q}$[GeV]}
 \psfrag{mg500argmu0tanb3}{\hspace{-1.6cm}\scalefont{1}$m_{\tilde g_3} = 500~$GeV, $ \sgn(\mu) = +1$, $\tan\beta = 3$}
 \psfrag{mg500argmu0tanb6}{\hspace{-1.8cm}\scalefont{1}$m_{\tilde g_3} = 500~$GeV, $ \sgn(\mu) = +1$, $\tan\beta = 6$}

  \includegraphics[width=.49\linewidth]{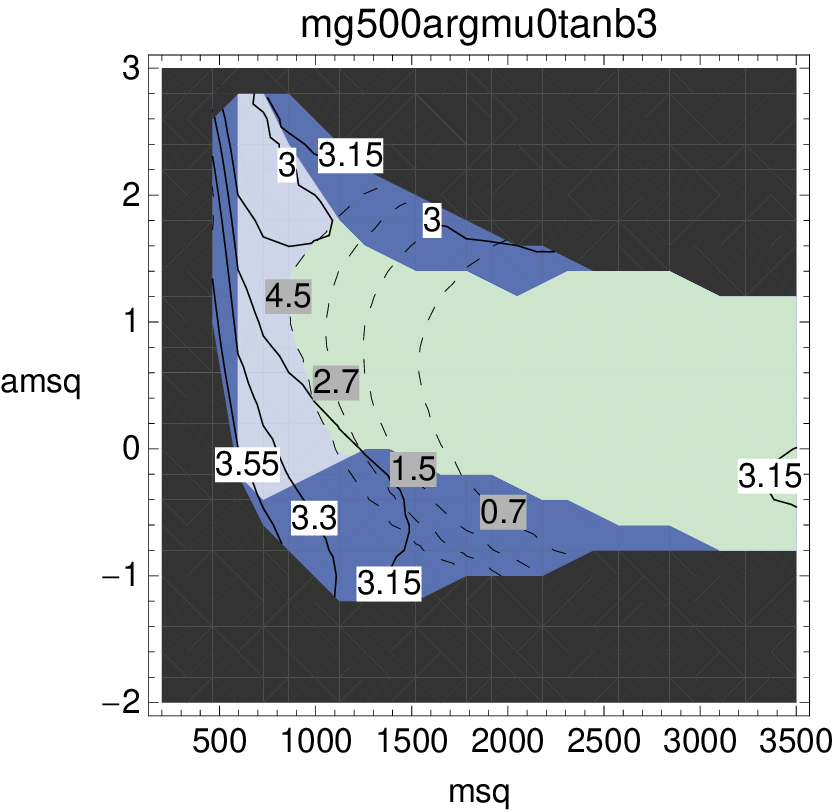}
  \hspace{.01\linewidth}
  \includegraphics[width=.49\linewidth]{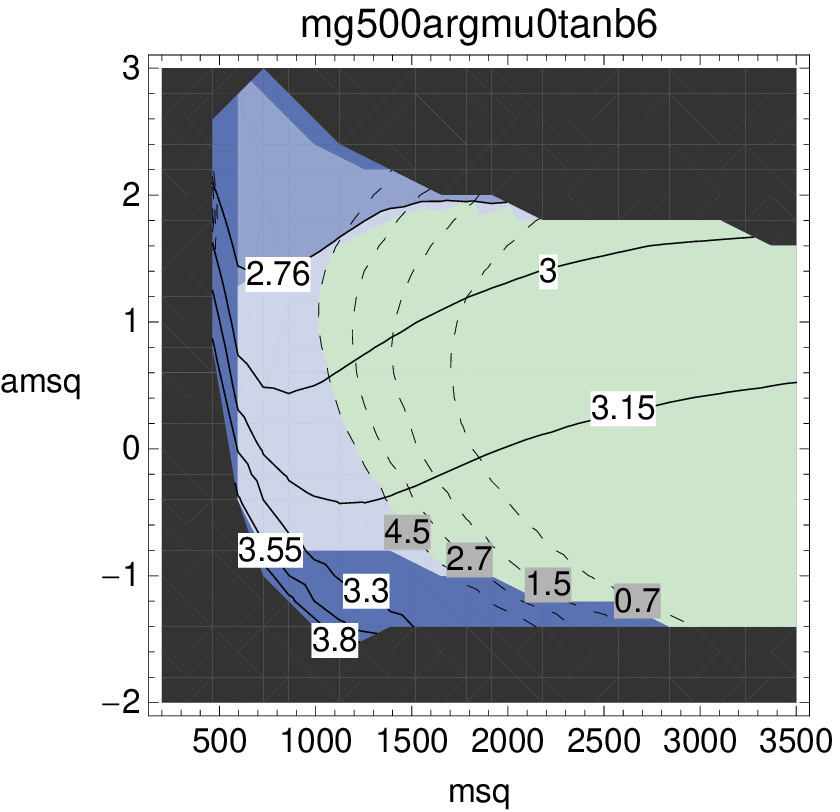}
  \caption{ Correlation of FCNC processes as a function of $M_{\tilde q}(M_Z)$ and $a_1^d(M_Z)/M_{\tilde q}(M_Z)$  
for $m_{\tilde g_3}(M_Z) = 500~\text{GeV}$ and $\sgn \mu = +1$ with $\tan\beta = 3$ (left) and $\tan\beta =6$ (right). $\mathcal{B}(b\to s\gamma)[10^{-4}]$ solid lines with white labels; $\mathcal{B}(\tau\to \mu\gamma)[10^{-8}] $ dashed lines with gray labels.  Black region: $m_{\tilde f}^2<0$ or unstable $|0\rangle$;  dark blue region: excluded due to $B_s-\overline B_s$; medium blue region: consistent with $B_s-\overline B_s$ but excluded due to $b\to s\gamma$; light blue region: consistent with $B_s-\overline B_s$ and $b\to s\gamma$ but inconsistent with $\tau\to\mu\gamma$; green region: compatible with all three FCNC constraints. }
  \label{fig:fcnc}
\end{nfigure}

What is really challenging for the CMM model is an observable not
directly related to flavor physics: the mass of the lightest neutral,
CP-even Higgs boson. As already mentioned at the end of
Sec.~\ref{se:observables}, in order to make the corrections to the
tree level Higgs mass large enough, the sfermions of the third
generation should not be too light because they enter together with
the top mass logarithmically in the radiative corrections (see
Eq.~(\ref{equ:Mh2corr})).  This is triggered by the choice of
$\tan\beta$. In Fig.~\ref{fig:higgs} one can see the same parameter
space as in Fig.~\ref{fig:fcnc} but with the predicted mass of the
lightest Higgs boson mass added (solid line with white labels). On the
left hand side for $\tan\beta = 3$ the whole green region is excluded
due to $M_{h^0}<114.4~$GeV. For negative $\mu$ the mass even tends to
smaller values.  Only for rather heavy masses, e.g. $m_{\tilde g_3} =
2500$~GeV and $M_{\tilde{q}}\gtrsim 6500~$GeV the experimental bound
can be satisfied. However, in this region of parameter space the
constraints from flavor violating processes become irrelevant. On the
right hand side of Fig.~\ref{fig:higgs} for $\tan\beta = 6$ the
situation changes such than even for light gluino masses there exist
allowed regions in the CMM parameter space.  Thus, we can summarize
this correlation between flavor violation and Higgs mass in the
CMM-model:
\begin{center}
\boxed{
 \begin{tabular}{ccccc}
small $\tan\beta$ & $\Leftrightarrow $& large flavor effects & $\Leftrightarrow$ &
(too) light  $h^0$\\
{larger} $\tan\beta$ & $\Leftrightarrow$ & 
{smaller} flavor effects & $\Leftrightarrow$
& {sufficiently} heavy $h^0$\\
  \end{tabular}}
\end{center}

\begin{nfigure}[!tb]
 \psfrag{amsq}{\hspace{0.3cm}\scalefont{1.5}$\frac{a_1^d}{M_{\tilde q}}$}
 \psfrag{msq}{\hspace{-0.5cm}\scalefont{1.1}$M_{\tilde q}$[GeV]}
 \psfrag{mg500argmu0tanb3}{\hspace{-1.6cm}\scalefont{1}$m_{\tilde g_3} = 500~$GeV, $ \sgn(\mu) = +1$, $\tan\beta = 3$}
 \psfrag{mg500argmu0tanb6}{\hspace{-1.8cm}\scalefont{1}$m_{\tilde g_3} = 500~$GeV, $ \sgn(\mu) = +1$, $\tan\beta = 6$}
  \includegraphics[width=.49\linewidth]{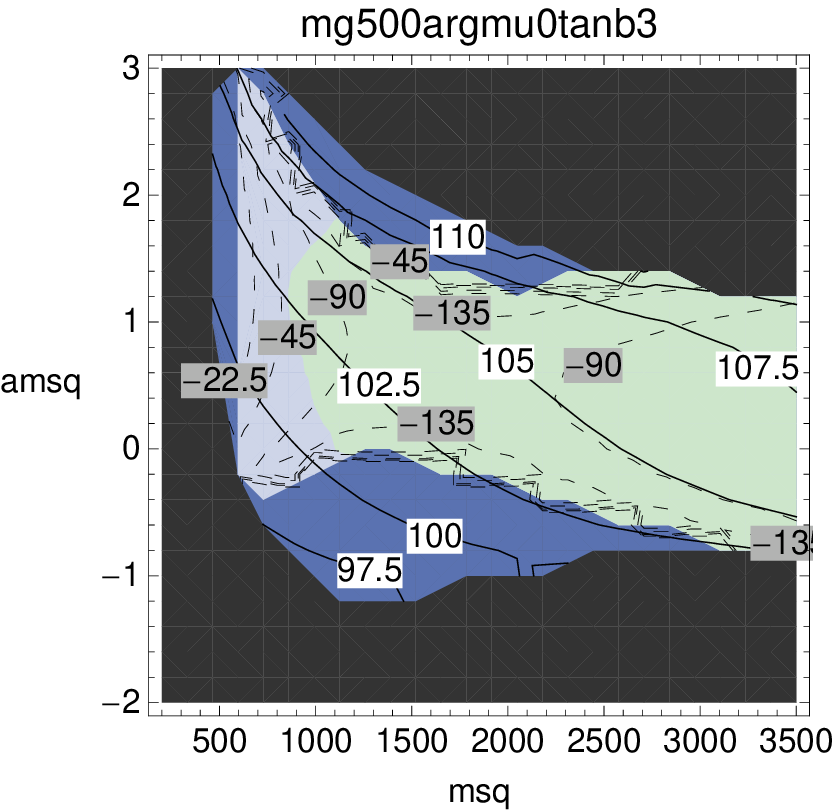}
  \hspace{.01\linewidth}
  \includegraphics[width=.49\linewidth]{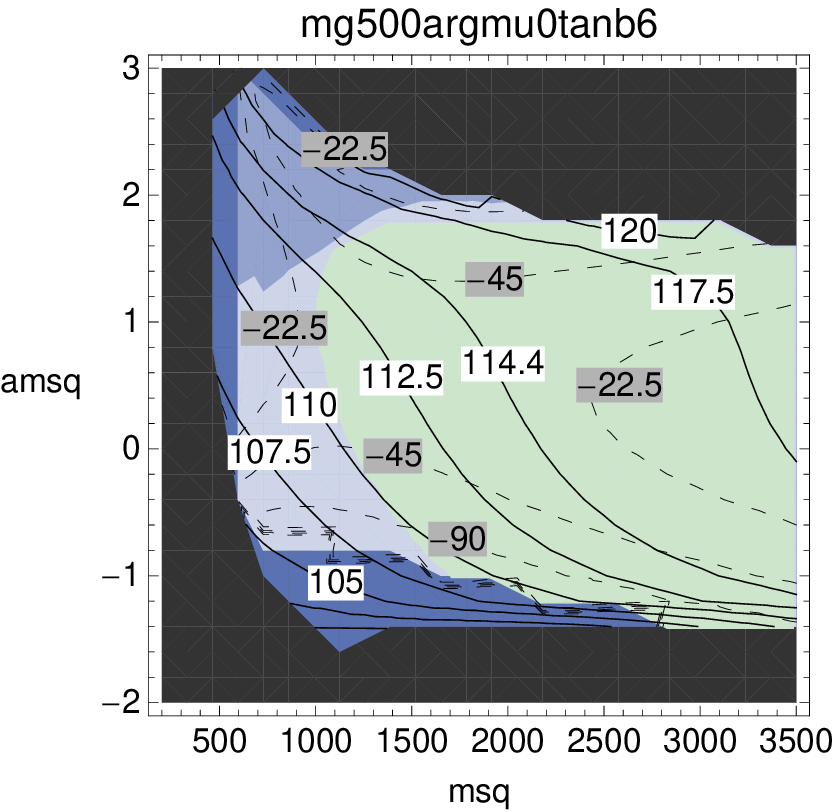}
  \caption{ Same as in Fig.~\ref{fig:fcnc}, but without labels and
    lines for $b\to s\gamma$ and $\tau\to \mu\gamma$.  {We show
      the lightest Higgs mass in GeV (solid line with white labels) and
      the phase $\phi_s$ in degrees (gray labels) for $\tan\beta = 3$
      (left) and 6 (right).   $\phi_s$ depends on the CP phase $\xi$
      of the model; the values quoted in the gray labels are the
      values of $\phi_s$ with maximal possible $|\phi_s|$.  }}
  \label{fig:higgs}
\end{nfigure}

In light of the recent result from D\O\ of the like-sign dimuon charge
asymmetry and the measured CP violation in $B_s\to J/\psi \phi$, it is
worth studying how large the CP phase $\phi_s$ can actually {be} in
the CMM model. It is related to the free phase $\xi$ defined in
Eq.~(\ref{eq:vmv}) which occurs in the Wilson coefficient (see
Eq.~(\ref{equ:WilsonBsBsbar})) of the $B_s-\overline B_s$ system.  In
Fig.~\ref{fig:higgs} we also compute the maximal (negative) phase
$\phi_s$ in the CMM model under the condition that $\Delta M_s$ lies
within its $3\sigma$-range and the hadronic matrix element within its
error bar.  

{From Fig.~\ref{fig:fcnc} we see that $\tau \to \mu \gamma$ alone
  puts a lower bound on $M_{\tilde q}$, so that the squark masses of
  the first two generations lie essentially above 1 TeV. One also
  realizes that the bound on ${\cal B}(\tau \to \mu \gamma)$ is more
  constraining than the measured value of ${\cal B}( b \to s
  \gamma)$. Fig.~\ref{fig:split} shows that the dominantly right-handed
  sbottom is about half as heavy as the down-type squarks of the first
  two generations. The sample parameter point discussed in
  \eqsto{eq:bpoint2}{ex:sdown} further shows that we can expect a
  dominantly right-handed stop with mass around 500 GeV. The sleptons
  are heavy and seemingly out of the discovery range of the LHC. On
  the other hand, the light gaugino-like chargino and neutralinos
  should permit nice signatures in the ``golden'' trilepton search 
  channels.} 
  Fig.~\ref{fig:higgs} reveals that the lower
  bound on the lightest neutral Higgs boson mass excludes the whole
  plotted region if $\tan \beta=3$. In the $\tan \beta=6$ case this
  bound has a much milder effect, essentially leading to a preference
  of the upper half of the plotted region, where
  $a_1^d>0$. Remarkably, almost all of the allowed region permits
  large effects in \bbms, with CP phases well in the range needed to
  explain the Tevatron data and quoted in \eq{equ:expSinphis}. That
  is, \bbms\ is much more sensitive to the new physics effects than
  the rare decays entering our analysis. The light gauginos are,
  of course, a consequence of our choice of $M_{\tilde g}=500\,$GeV 
  in our numerical studies. We may ask how the patterns of 
  Figs.~\ref{fig:split}--\ref{fig:higgs} change, if $M_{\tilde g}$
  is increased. In particular, one might expect that that the 
  FCNC constraints become weaker so that one could instead 
  obtain lighter squarks of the first two generations. However, this 
  is not the case, instead the lower bound on $M_{\tilde q}$ becomes 
  stronger with increasing $M_{\tilde g}$, in order to avoid too light 
  third-generation squarks and problems with $\Delta M_s$.

We conclude that in the CMM model it is indeed possible to explain the
observed discrepancies in the $B_s$ system naturally with the free
phase $\xi$ and simultaneously satisfy other FCNC bounds.  Compared to
the generic MSSM that can also describe CP violation in $B_s-\overline
B_s$ mixing, {but does not suppress FCNC elsewhere}, the CMM
model in its original formulation does not induce any dangerous
effects in e.g.\ Kaon mixing or $\mu\to e\gamma$ due to the smallness
of $(U_{\text{PMNS}})_{13}$ which translates into the particular
structure of the right-handed down squark mass matrix in
\eq{eq:vmv}.  By contrast, the generic MSSM {lacks a
  symmetry principle that governs the structure of the squark mass
  matrices in a way which suppresses $b\to d$, $s\to d$ and $c\to u$
  transitions while permitting large CP-violating effects in $b\to s$
  transitions.}  Note, that there are also some effects in $\Delta B =
1$ penguin diagrams such as $B_d\to\phi K_S$ and $B_s\to\phi\phi$,
{which triggered the early studies in \cite{Harnik:2002vs} and
  \cite{Jager:2003xv}.  The experimental value of $\Delta M_s$ restricts
  the size of the new physics contribution to $\mathsf{M}_{12}^s$ to
  smaller values than those allowed before the discovery of \bbms. In
  the portion of the CMM parameter space complying with all of today's
  experimental constraints the contribution to CP asymmetries in $b\to
  s$ penguin decays is small and typically in a range which cannot be
  resolved within present experimental errors. We will discuss 
  the impact of the CMM model on these CP asymmetries in the light 
  of future experimental uncertainties in another paper.}


\section{Comparison with other GUT analyses}\label{sec:disussion}

In the following we compare the CMM model and our results with
analyses of other authors.

{Moroi's landmark papers \cite{Moroi:2000tk} have laid out the
  basic idea of the CMM model, namely flavor violation in the soft
  squark mass terms driven by RG evolution above the GUT scale in
  conjunction with large lepton-flavor violation. The paper discusses
  the effect in an SU(5) context and focuses on the phenomenological
  effects in \bbmd\ and CP violation in $b\to s$ penguin decays.
  Written prior to the $B$ factory era, the author mentions the
  possibility of new CP phases in $b\to s$ penguin decays of order
  5$^\circ$. The consequences of a large top Yukawa
  coupling (studied in minimal SU(5) and SO(10) models) 
  in conjunction with universal soft terms 
  at $M_\text{Pl}$ for low-energy flavor observables have already been 
  studied by Barbieri et al.\ in 1995
  \cite{Barbieri:1995rs}. In this paper 
the non-degeneracy of third-generation fermions with the first and
second generation is emphasized. Since \cite{Barbieri:1995rs}
  has been written at a time at which 
 the neutrino mixing matrix was unknown, the phenomenological results 
 cannot be compared to ours in a meaningful way. 
 Harnik et al.\ have analyzed the $b\to s$ penguin
  amplitude in a framework inspired by the CMM model
  \cite{Harnik:2002vs}: Motivated by a $2.7\sigma$ discrepancy between
  the measured mixing-induced CP asymmetry in $B_d^0\to \phi K_S$ and
  the SM expectation, they have supplemented the MFV-MSSM by a $\tilde
  b_R-\tilde s_R$ mixing term determined by the atmospheric neutrino
  mixing angle. Since no RG analysis has been worked out, the authors could
  not find the correlations between the various observables and the
  sparticle spectrum, which originates from the small number of GUT
  parameters and is presented in the preceding section. However,
  correlations among \bbms, $b\to s\gamma$ and CP violation in $B_d\to
  \phi K_S$ stemming from the $\tilde b_R-\tilde s_R$ off-diagonal
  element of the squark mass matrix are already studied in
  \cite{Harnik:2002vs},  scanning over MSSM parameters.
  The authors of \cite{Harnik:2002vs} also discusses the possibility that the $\tilde
  b_R-\tilde s_R$ and $\tilde b_R-\tilde b_L$ mixings are
  simultaneously large. We do not see how this can be achieved even in
  a widely defined class of CMM-like models: A large $\tilde
  b_R-\tilde b_L$ mixing amounts to a large value of $m_b\mu
  \tan\beta$ and therefore inevitably to a sizable $\tan \beta$
  ($|\mu|$ is fixed from the condition of electroweak symmetry
  breaking) and the corresponding smaller value of $y_t$ quickly renders
  the PMNS-driven $\tilde b_R-\tilde s_R$ mixing small, see
  Fig.~\ref{fig:fp}. In \cite{Jager:2003xv} two authors of this paper
  have analyzed \bbms\ in conjunction with $\tau \to \mu \gamma$ in
  the CMM model as defined in this paper performing an RG analysis
  which has not yet included the MSSM Higgs sector (and the constraint
  from $m_{h^0}$) and $b\to s \gamma$. Both \cite{Harnik:2002vs} and
  \cite{Jager:2003xv} found order-of-magnitude enhancements of $\Delta
  M_{B_s}$ (which was unknown at the time) over the SM prediction in
  those regions of the parameter space explaining the experimental
  anomaly in $B_d\to \phi K_S$ seen at the time. This merely reflects
  the larger sensitivity of \bbms\ to $\tilde b_R-\tilde s_R$ mixing
  compared to  $b\to s$ penguin amplitudes.}

{Among the papers studying GUT flavor physics in a SUSY SU(5) context,
  \cite{Hisano:2003bd} has a significant overlap with our analysis: In
  \cite{Hisano:2003bd} also universal scalar masses and trilinear
  $A$-terms are postulated at the reduced Planck scale and the PMNS
  matrix appears in the RGE of the right-handed down squarks. Like us,
  the authors of \cite{Hisano:2003bd} study $\tau\to\mu\gamma$, $b\to s
  \gamma$ and $\phi_s$, but with focus on the mixing-induced CP
  asymmetry in $B_d\to \phi K_S$.  The study goes beyond ours by
  considering the electric dipole moment of the muon and CP violation in
  $B_d^0\to M_s^0\gamma$. Values of $\tan\beta$ up to 30 are considered,
  which are inaccessible in the CMM model.  In a recent phenomenological
  update \cite{Hisano:2008df} the authors of \cite{Hisano:2003bd} have
  calculated $|\phi_s|$ in their SU(5) model and found a maximal value
  of 9$^\circ$. This is in sharp contrast to the situation found by us
  in the CMM model.}  The work \cite{Cheung:2007pj} also studies the
possibility that in SUSY GUT models with heavy right-handed neutrinos
the large atmospheric mixing angle can affect $b\to s$ transitions due
to a large Dirac neutrino Yukawa coupling (which in our case is equal to
the top Yukawa coupling).  Using the mass insertion approximation, the
correlation of new physics effects in $B_q-\bar B_q$-mixing ($q = d,s$)
and the radiative decays $\tau\to e(\mu)\gamma$ is discussed. In
contrast to our work, the GUT model is not specified and a detailed
renormalization group analysis is missing.  Employing the approximate
GUT relation $ \delta_{RR}^{d\,ij}\approx \delta_{LL}^{\ell\,ij}$ (which
is not necessarily true with large mixing and is not invariant under the
RG) the ratio of the $B_d$ and $B_s$ mixing frequencies is correlated to
the corresponding ratio of the LFV decays $\tau\to e(\mu)\gamma$. {In
  \cite{kpy} the RG-induced flavor violation is studied in conjunction
  with dimension-5 Yukawa terms and the corresponding soft SUSY-breaking
  terms. These papers find that given the constraint from
  $\mathcal{B}(\tau\to\mu\gamma)$ the impact on \bbms\ is maximal for a
  particular value of the ratio $m_{\tilde{g}}/m_0$ around 0.3. Our
  sample point discussed in Sec.~\ref{se:parMGUT} is in qualitative
  agreement with this finding.  LFV decays are also correlated with
  quark FCNC processes in various SU(5) and SO(10) scenarios in
  \cite{dm}.  The authors discuss several ans\"atze to alleviate
  different tensions in quark FCNC data.  
  As an important difference with respect to the
  CMM model the scenarios of \cite{dm} contain relevant sources of
  flavor symmetry breaking among the first two fermion generations, so
  that e.g.~$\mathcal{B}(\mu \to e \gamma)$ places a constraint
  on the parameter spaces.} {Recently Buras et al.\ have presented a
  correlated analysis of many flavor observables in an SU(5) scenario
  with right-handed neutrinos \cite{Buras:2010pm} and mSUGRA boundary
  conditions at $M_\text{Pl}$.  The Yukawa sector is less constrained
  than in the CMM model and therefore the correlations between different
  FCNC observables are weaker. Like us and in contrast to
  \cite{Hisano:2008df}, the authors of \cite{Buras:2010pm} find that the
  current upper bound on ${\cal B}(\tau \to \mu \gamma)$ still permits a
  sizable CP phase in \bbms. In \cite{Buras:2010pm} also FCNC
  transitions among the first and second generation are studied, e.g.\
  \kkm\ and $\mu \to e \gamma$. This procedure is not in our philosophy,
  because these transitions are highly sensitive to corrections from
  higher-dimensional operators, which are moreover welcome to fix the
  poor Yukawa unification in the first two generations. Our approach,
  pursued in two previous papers, is to constrain the flavor structures
  of higher-dimensional operators from data on \kkm\ \cite{Trine:2009ns}
  and $\mu \to e \gamma$ \cite{Girrbach:2009uy}.}
 
{There are numerous papers on the MSSM with GUT boundary
  conditions placed at $M_\text{GUT}$. These papers are different in
  spirit to \cite{Barbieri:1995rs,Moroi:2000tk,Chang:2002mq,
    Harnik:2002vs,Hisano:2003bd,Jager:2003xv,Cheung:2007pj,Hisano:2008df,
    Buras:2010pm} and this paper, all of which employ RG effects above
  $M_\text{GUT}$. Here we discuss two of these papers with particular
  emphasis on flavor physics: In \cite{Ciuchini:2003rg} correlations
  of quark and lepton FCNCs are studied in SU(5) SUSY GUTs, but
  without neutrinos (and thus without the PMNS matrix).  The authors
  of \cite{Ciuchini:2003rg} assume generic flavor-violating entries to
  be present in the sfermion matrices at the GUT scale and correlate
  quark and lepton FCNCs in a general way via SU(5) symmetry.}  Using
the mass insertion approximation, an upper bound for the off-diagonal
elements of the right-handed down squark matrix of the form
$\left|\delta_{RR}^{d\,ij}\right|\leq
\frac{m_L^2}{m_D^2}\delta_{LL}^{\ell\,ij}$ has been derived.  The
authors have found that the bound on $\delta_{RR}^{d\,23}$ induced by
$\tau\to\mu\gamma$ is stronger then those from {the $B$ physics
  observables known at the time.}  The authors of
\cite{Albrecht:2007ii} have studied an SO(10) SUSY GUT model with
$D_3$ family symmetry which was proposed in
\cite{Dermisek:2005ij}. This model involves Yukawa unification of the
third generation at the GUT scale, which immediately implies large
$\tan\beta\approx 50$ at low energies. This is already in sharp
contrast to the CMM model where $y_b$ is suppressed by a factor $
\langle 45_H\rangle/M_\text{Pl}$ compared to $y_t$ {and the
  phenomenology is very different.}  With 24 input parameters, all
parameters at low energy (including SM parameters) can be calculated
with the RGE and it is possible to get realistic quark and lepton
masses as well as the PMNS and CKM matrix. The authors have a closer
look at their SUSY spectrum and study FCNC processes like
$B_s\to\mu^+\mu^-$, $B\to X_s\gamma$, $B\to X_s\ell^+\ell^-$ and
$\Delta M_{d,s}$ and the decay $B^+\to \tau^+\nu$. The combination of
this observables is {challenging for the} model, since mass
hierarchies enter loops in FC observables. The authors conclude that
this problem occurs in a wider class of SUSY GUTs with unified Yukawa
couplings of the third generation. This argument, which was further
pursued in \cite{Altmannshofer:2008vr}, is, however, not applicable to
the CMM model.

\section{Conclusions}
We {have} studied a supersymmetric SO(10) GUT model originally
proposed by Chang, Masiero and {Murayama} (CMM model) \cite{Chang:2002mq},
in which the large atmospheric {neutrino} mixing angle
{$\theta_{23}$} is transferred to $b\to s$ and $\tau\to\mu$
transitions. {At low energy the model is an MSSM whose parameters
  are highly correlated through the GUT boundary conditions.  The key
  features of the CMM model are soft SUSY-breaking terms which are
  universal near the Planck scale and a Yukawa sector with a
  non-renormalizable term in the SO(10) superpotential as the only source
  of flavor violation. Renormalization-group effects of the large top
  Yukawa coupling $y_t$ drive the sfermion masses of the third
  generation away from those of the first two generations. The
  transition from weak to mass eigenstates involves rotations with the
  atmospheric mixing angle among right-handed bottom and strange squarks
  and left-handed tau and muon sleptons. This leads to potentially large
  FCNC effects in transitions between the second and third generation,
  while other FCNC transitions are essentially unaffected. We have
  performed an extensive RGE analysis to connect Planck-scale and
  low-energy parameters, focusing on the numerically dominant effects
  associated with the large parameters $y_t$ and $\theta_{23} \simeq
  45^\circ$.}

{We have then analyzed the FCNC observables $ \mathcal
  B(\tau\to\mu\gamma)$, $\mathcal B(b\to s\gamma)$, the mass difference
  $\Delta M_s$ in \bbms, and the corresponding CP phase $\phi_s$, taking
  into account the LEP lower bound on the lightest neutral Higgs boson
  mass $m_{h^0}$.  The analysis involves only seven new parameters, so
  that the model is very predictive.  We find that $\tau\to\mu\gamma$
  constrains the sfermion masses of the first two generations to lie
  above 1~TeV, while the third-generation sfermions can be substantially
  lighter. The intergenerational sfermion mass splitting is larger than
  in models which impose universal soft terms at the GUT scale, such as
  the CMSSM. At the same time the CMM model permits light gauginos.
$b\to s\gamma$ is less constraining than $\tau\to\mu\gamma$, while
\bbms\ turns out to be most sensitive to CMM effects. One of
the model parameters is a CP-violating phase accompanying $\widetilde
b_R \to \widetilde s_R$ transitions and we can accommodate the
recent hints for new physics in \bbms\ \cite{Lenz:2010gu}.} 

{We find that the LEP bound $m_{h^0} \geq 114.4$~GeV places a
  powerful constraint on the parameter space of the CMM model: E.g.\ for
  $\tan\beta = 3$ the sfermion masses must be unnaturally high to comply
  with $m_{h^0} \geq 114.4$~GeV, which in turn does
  not permit visible effects in the FCNC observables.  However, for
  $\tan\beta =6$ we find regions of the CMM parameter space compatible
  with all data and large effects in \bbms. The pattern of sparticle
  masses is very distinctive: Sfermions are heavy, with the exception of
  a dominantly right-handed stop. Since $y_t$ is the only source of
  sfermion non-universality, eight out of twelve squarks and four out of
  six sleptons are essentially degenerate. Most importantly, two of the
  physical squarks are maximal $\widetilde b_R$--$\widetilde s_R$
  mixtures and likewise two sleptons are maximal $\widetilde
  \tau_L$--$\widetilde \mu_L$ mixtures.
  This should lead to distinctive features in the collider signatures at
  CMS and ATLAS.}  

{In summary, we have performed an RG analysis of the CMM model
  relating several observables to seven new parameters beyond those of
  the standard model. We find that the model can explain the hints for a
  large CP phase in \bbms\ seen in current data without violating other
  FCNC constraints, vacuum stability bounds or the experimental lower
  bounds on $m_{h^0}$ and supersymmetric particle masses.}

\section*{Acknowledgments}
The presented work is supported by project C6 of the DFG Research Unit
SFB--TR 9 \emph{Computergest\"utzte Theoretische Teilchenphysik} and by
the DFG grant No.~NI 1105/1-1.  J.G.\ and W.M.\ acknowledge the
financial support by \emph{Studienstiftung des deutschen Volkes}.  S.J.\
was supported by the Science and Technology Facilities Council [grant
number ST/H004661/1] {and acknowledges support from the NExT
  institute and SEPnet.}

\begin{appendix}

\section{Higgs Sector and Yukawa Couplings in the CMM Model} 
\label{se:cmm-higgs-yukawa}

The CMM model makes use of small Higgs representations: SO(10) is
broken to the standard model via the Higgs fields $\text{16}_H$,
$\ovl{\text{16}}_H$ and $\text{45}_H$.  The electroweak symmetry is
then broken when the neutral component in the doublet $H_u \in
\text{10}_H$ acquires a vev.  In addition to $H_u$, the theory
contains a second Higgs doublet, $H_d$, which couples to down quarks
as well as charged fermions.  A priori, this field can originate from
two different SO(10) representations (or a combination of the two).

As discussed in Sec.~\ref{se:model}, the adjoint Higgs field is
assumed to acquire two distinct vevs.  While the primary task of
$\text{45}_H$ is to break SU(5) to the standard model group, the SU(5)
singlet component might acquire a vev as well when SO(10) is broken
via the spinorial Higgs field.\footnote{Unfortunately, the authors of
  Ref.~\cite{Chang:2002mq} do not specify how SU(5) is broken.  They
  only mention the SU(5) singlet vev, which is necessary for the
  masses of down quark and charged lepton not to be too small.  With
  the given Higgs fields, however, $\text{45}_H$ has to break the
  SU(5) symmetry.}

If $H_d$ was contained in $\text{10}_H$ as well, the mass and weak
eigenstates would coincide and the quark mixing matrix would be the unit
matrix.  (Mixing among leptons could originate from the Majorana mass
matrix for the right-handed neutrinos.)  We therefore have to consider
an additional Higgs field in the theory, which can incorporate all or
part of $H_d$.\footnote{Note that a second ten-dimensional Higgs field
  is required for a non-vanishing coupling $\text{10}_H \text{45}_H
  \text{10}^\prime_H$ in the superpotential.  However, in order to have
  only two massless doublets, usually all components of
  $\text{10}^\prime_H$ become massive when the SU(5) symmetry is
  broken.}
This case is realized in the CMM model.  In order to allow for an
asymmetric Yukawa coupling matrix for down quarks and charged fermions,
the matter fields couple to $\text{10}_H^\prime$ via a
non-renormalizable interaction, $\text{16 16 10}_H^\prime\,
\text{45}_H$.  As already mentioned, this higher-dimensional operator
can be generated by integrating out massive SO(10) fields.  
{Depending on the representation of the massive field, four invariants can
appear} \cite{Wiesenfeldt:2005zx},\footnote{In Ref.~\cite{Wiesenfeldt:2005zx},
  Eqn.~(29c) should read
  \begin{align*}
    \widehat Y_{16} & = \frac{h_{ij}^{16}}{M_{16}} \left\{ \frac{1}{4}\,
      \epsilon_{abcde}\, \mathsf{10}_i^{ab}\, \mathsf{10}_j^{cf}\, H^d\,
      \Sigma^e_f + \overline{H}_a\, \Sigma^a_b\, \mathsf{10}_i^{bc}\,
      \mathsf{5}^*_{jc} + \overline{H}_a\, \mathsf{10}_i^{ab}\,
      \Sigma_b^c\, \mathsf{5}^*_{jc} \vphantom{frac12} \right\}_A
  \end{align*}
  such that $h^{16}$ is antisymmetric.  For a more substantiated
  approach to describe the vector-spinor 144, see
  Refs.~\cite{babu-144}.}
\begin{align} 
  & \left( \text{16 16} \right)_{10} \left( \text{10}_H\, \text{45}_H
  \right)_{10} , & & \left( \text{16 10}_H \right)_{\overline{16}}
  \left( \text{16 45}_H \right)_{16} , \nonumber
  \\
  & \left( \text{16 16} \right)_{120} \left( \text{10}_H \text{45}_H
  \right)_{120} , & & \left( \text{16 10}_H \right)_{\overline{144}}
  \left( \text{16 45}_H \right)_{144} .
  \label{eq:op-invariants} 
\end{align}
{The expressions in the right column can be expressed in terms of
  those in the first column through a Fierz transform.} 
The contributions are either symmetric or antisymmetric
\cite{Barr:2005ss}, {so that the effective Yukawa coupling 
$\mathsf{Y}_2$ has a mixed symmetry.}

The CMM model focuses on the singlet vev, $v_0 =
\VEV{S\left(\text{45}_H\right)}$, for two reasons. One, $v_0$ is an
order of magnitude larger than $\sigma$ such that the ratio
$M_\text{Pl}/v_0 \sim 10^{1}-10^{2}$ is smaller than the top-bottom mass
ratio.  Thus, $\tan\beta$ can be as large as 10 with moderate values for
$\mathsf{Y}_2$.  {Two, for $\sigma = 0$ the contributions to
  $\mathsf{Y}_d$ and $\mathsf{Y}_e$ from \eq{eq:op-invariants} satisfy
  relation \eq{eq:yukawa-unification-su5} with no further symmetry
  requirements.}

In contrast, the adjoint vev, $\sigma$, leads to different contributions
to $\mathsf{Y}_d$ and $\mathsf{Y}_e$. 
Since $\VEV{\Sigma_{24}\left(\text{45}_H\right)} \propto Y$
and the (unnormalized) hypercharge generator is given by $\sigma_2
\otimes \diag\left(2,2,2,-3,-3\right)$, we can group them into two
different effective operators, $h_1$ and $h_2$, such that
\begin{align}
  \begin{split}
    \mathsf{Y}_d \, & =\,  {\frac{v_0}{M_\text{Pl}} \mathsf{Y}_2} \, -\,  
                     3 \frac{\sigma}{M_\text{Pl}} h_1 \, +\,  2
    \frac{\sigma}{M_\text{Pl}} h_2 \,,
    \\[2pt]
    \mathsf{Y}_e^{{\top}} \, & = \,  {\frac{v_0}{M_\text{Pl}} \mathsf{Y}_2}
    \, -\,  3 \frac{\sigma}{M_\text{Pl}} h_1 \, -\,  3
    \frac{\sigma}{M_\text{Pl}} h_2 \,.
  \end{split}
\end{align}
{As a result,} 
\eq{eq:yukawa-unification-su5} will be modified.  Hence,
these operators with $\VEV{\Sigma_{24}\left(\text{45}_H\right)}$ can
naturally explain the unsuccessful Yukawa unification for the lighter
generations.
Note that we only deal with one set of operators
(\ref{eq:op-invariants}) so that they appear with both possible vevs of
the adjoint Higgs field $\text{45}_H$.

\end{appendix}
{\small
  
}%

\end{document}